\documentclass[onecolumn]{emulateapj}
\usepackage{setspace}
\usepackage{ulem}
\usepackage{color,ulem}
\usepackage{multirow}
\usepackage{amsfonts}
\usepackage{booktabs}
\usepackage{natbib}
\include{epsf}
\include{epstopdf}
\DeclareGraphicsExtensions{.png,.jpg,.pdf,.eps}
\newcommand{\beq}{\begin{equation}}
\newcommand{\eeq}{\end{equation}}

\def\la{\hbox{\raise.35ex\rlap{$<$}\lower.6ex\hbox{$\sim$}\ }}
\def\ga{\hbox{\raise.35ex\rlap{$>$}\lower.6ex\hbox{$\sim$}\ }}

\def\water{H$_2$O }

\def\beq{\begin{equation}}
\def\eeq{\end{equation}}
\def\beqa{\begin{eqnarray}}
\def\eeqa{\end{eqnarray}}
\def\sub#1{_{_{#1}}}
\def\order#1{{\cal O}\left({#1}\right)}
\newcommand{\sfrac}[2]{\small \mbox{$\frac{#1}{#2}$}}
\shorttitle{Streaming instability in turbulent protoplanetary disks}
\shortauthors{Umurhan et al.}
\begin{document}

%
%
\title{{Streaming instability in turbulent protoplanetary disks}}

\author{Orkan M. Umurhan$^{1,2}$\altaffilmark{*}, Paul R. Estrada$^{2}$ and Jeffrey N. Cuzzi$^{2}$}
\affil{$^1$ SETI Institute, 189 Bernardo Way, Mountain View, CA 94043, U.S.A.}
\affil{$^2$ NASA Ames Research Center, Moffett Field, CA 94053, U.S.A}
\altaffiltext{*}{Email: orkan.m.umurhan@nasa.gov}
\date{}


 \begin{abstract}
 {The streaming instability for solid particles in protoplanetary disks is re-examined assuming the familiar alpha ($\alpha$) model for isotropic turbulence. Turbulence always reduces the growth rates of the streaming instability relative to values calculated for globally laminar disks. While for small values of the turbulence parameter, $\alpha < 10^{-5}$, the wavelengths of the fastest-growing disturbances are small fractions of the local gas vertical scale height $H$, we find that for moderate values of the turbulence parameter, i.e., $\alpha \sim 10^{-5}-10^{-3}$, the lengthscales of maximally growing disturbances shift toward larger scales, approaching $H$.  At these moderate turbulent intensities and for local particle to gas mass density ratios $\epsilon < 0.5$, the vertical scales of the most unstable modes begin to exceed the corresponding radial scales so that the instability appears in the form of vertically oriented sheets extending well beyond the particle scale height. We find that for hydrodynamical turbulent disk models reported in the literature, with $\alpha  = 4\times 10^{-5} - 5\times 10^{-4}$, together with state of the art global evolution models of particle growth, the streaming instability is predicted to be viable within a narrow triangular patch of $\alpha$--$\tau_s$ parameter space centered on Stokes numbers, $\tau_s \sim 0.01$ and $\alpha \sim 4\times 10^{-5}$ and, further, exhibits growth rates on the order of several hundred to thousands of orbit times for disks with 1 percent ($Z= 0.01$)} cosmic solids abundance or metallicity. Our results are consistent with, and place in context, published numerical studies of streaming instabilities.
 
  \end{abstract}

\large
\section{Introduction}\label{sec:intro}
The formation of the first 100 km size planetesimals remains a poorly understood chapter in the standard story of planetary formation. A given radial zone in a protoplanetary disk will initially contain $\mu$m size particles of material which is solid under local conditions.  These grains grow by sticking until they become sub-mm to mm-sized aggregates \citep{Birnstiel_etal_2012, Estrada_etal_2016}. In the inner nebula, some (still mysterious) heating process melts these (ice-poor) aggregates and forms chondrules. These chondrules apparently can form few-cm-scale aggregates depending on local disk properties \citep{Simon_etal_2018}. The picture is less clear in the outer nebula, where chondrule formation may never occur and the material remains ice-rich. 

However, what remains elusive is understanding how these aggregate mm-to-cm-size clusters eventually coalesce into 100 km sized planetesimals, especially if the nebula is even weakly turbulent. This is because further particle evolution by ``incremental growth" (by sticking) encounters several barriers - including (but not limited to) the radial-drift barrier, the bouncing barrier, and the fragmentation barrier. 
For a more comprehensive discussion of these barrier mechanisms see the discussion found in \citet{Brauer_etal_2008,Zsom_etal_2010,Birnstiel_etal_2012}, and \citet{Estrada_etal_2016}. Additional barriers to incremental growth in turbulence reappear at 1-10km size, due to gravitational stirring of such objects by fluctuations in the {\it gas} density, much like giant molecular clouds scatter stars in our galaxy \citep{Ida_etal_2008, Yang_etal_2009,Nelson_Gressel_2010, Gressel_Nelson_2011,Yang_etal_2012}. This latter realization has led to a growing suspicion that planetesimals may have been ``born big", close to the typical 100km sizes we see today \citep{Johansen_etal_2007, Cuzzi_etal_2008, Morbidelli_etal_2009}. Moreover, recent meteroritical work suggests that a substantial amount of the accretion that formed the solar nebula's first planetesimals, some of them leaving behind only their molten Fe-Ni cores, and even the initial 20-50 M$_\oplus$ core of Jupiter, occurred in less than 0.5 Ma after the formation of the first remaining solids \citep[the so-called Calcium-Aluminum refractory oxide inclusions,][]{Kruijer_etal_2017}. Thus, formation of sizeable planetesimals seemed to have started early, well inside the snowline, and after Jupiter's initial core formed (which probably required snowline planetesimals as precursors). It is in this context that current planetesimal formation theories must be assessed. How can planetesimals be born big, starting very early (and continuing for several Myr), directly from mm-cm size objects?

\par

One popular hypothesis is that clumps of small particles are collected into 100km or larger planetesimals by the streaming instability (SI) and ultimately gravitational collapse. SI is a {\it linear instability} (grows without limit from small perturbations, under the right conditions) that can enhance the concentration
of particles in protoplanetary disks \citep{YG2005,YJ2007} (YG2005 and YJ2007 hereafter).  The dynamics involves the resonant
momentum exchange between the disk gas and its embedded particles treated as a pressure-free second fluid
\citep{Squire_Hopkins_2018a,Squire_Hopkins_2018b,Hopkins_Squire_2018a,Hopkins_Squire_2018b}
 -- see also Section \ref{broad_physical_overview}. In protoplanetary disks, the SI is strongest for axisymmetric  disturbances and
the growth rates are most rapid when the local volume mass densities in the gas ($\rho_g$) and particle components ($\rho_p$) are comparable, i.e.,
when $\epsilon \equiv \rho_p/\rho_g \geq 1$.  Linear stability analyses indicate that for
laminar Keplerian flows, the SI grows fastest for small wavelengths 
and for near-unity Stokes numbers $\tau_s \equiv \Omega t_s$, where $t_s$ is the particle gas drag stopping time and $\Omega$ is the local
disk rotation time (also see section \ref{some_physical_properties})\footnote{In fact, the inviscid calculation indicates that the growth rates asymptote to finite values
as the wavelength of the vertical disturbance approaches zero provided the radial wavelength is larger than some minimum value.  This suggests the problem may be ill-posed in the inviscid limit. However this short wavelength catastrophe is averted when viscosity is included (see our general results below).} 
{ {Note, in our discussion throughout this work we sometimes refer to $\tau_s$ by its more commonly used symbolic designator, ``St".}}

These features of the SI suggest that this process may play an important role either (a) in the late stage of a protoplanetary disk's evolution when the disk has lost most of its gas 
because of photodissociation-or-MHD-driven winds from the star or disk
{ {\citep[e.g.,][]{Ercolano_etal_2017,Carrera_etal_2017}}}
, and/or accretion onto the central object or (b) if
the disk is laminar (nonturbulent), 
in which case the particle component can settle to high $\epsilon$ near the disk midplane. 
%
Several detailed numerical simulations of laminar disks have shown that the SI rapidly enhances local particle concentrations 
\citep[e.g.,][among many others]{Bai_Stone_2010,Lyra_Kuchner_2013, Carrera_etal_2015, Yang_etal_2014,Yang_etal_2017,Schreiber_Klahr_2018, Li_etal_2018}.
Particle concentration is further helped along
if particle self-gravity is included in models -- e.g., as done in \citet{Johansen_etal_2007,Simon_etal_2016,Simon_etal_2017} 
{ {and most recently \citet{Gerbig_etal_2020}
  -- and can drive particle enhancements to the precipice of gravitational collapse and onwards \citep{Johansen_etal_2015,Schafer_etal_2017,Li_etal_2019}.  }}
\par
However,  regions of protoplanetary disks in which particle growth 
is of greatest interest (i.e., 1-100 AU) are possibly weakly-to-moderately turbulent \citep{Turner_etal_2014,Lyra_Umurhan_2019}.
That is, recent theoretical advances suggest that non-ionized regions of protoplanetary disks support several instability processes that can lead to sustained turbulence: the Vertical Shear Instability (VSI), Convective Overstability (COV) and Zombie Vortex Instability (ZVI).  

Turbulence in disks is often thought of in terms of a zero-order closure ``alpha-disk" model,
wherein gas turbulence is represented by an enhanced viscosity
quantified by a turbulent (kinematic) viscosity coefficient, $\nu_t \equiv
\alpha c_s H$,where $c_s$ and $H$ are the local isothermal soundspeed and the vertical pressure scale heights, respectively 
\citep{Shakura_Sunyaev_1973,Lynden-Bell_Pringle_1974}.\footnote{A turbulent viscosity assumes that downscale momentum transfer occurs
in the inertial range of a fully developed statistically steady turbulent fluid. The ``$\alpha$-model" scales this in terms of the typical speed- and length-scales encountered in locally rotating sections of protoplanetary disks ($c_s$ and $H$). As such, the quantity $\alpha$ is typically interpreted to be the inverse of the local turbulent Reynolds number, Re$_{{\rm t}}$ and might be thought of as a measure of the amplitude of the turbulent velocity field compared to the local sound speed.
For a more pedagogical review appropriate to astrophysical fluids see \cite{Regev_etal_2016}.}  
The alpha-disk model, notwithstanding its crudeness, does a good job in characterizing disk structure and consequent flow in a turbulent protoplanetary disk medium. 
{ {Most recently \citet{Stoll_etal_2017} examined the turbulent state of the VSI and found the emergence of large-scale meridional flow {to be well predicted by} an $\alpha$-model albeit with effectively non-isotropic diffusion stresses owing to its
characteristic large vertical motions.  However, MHD turbulence might not be as well represented by an $\alpha$-model, and we keep this in mind throughout.}}
  Numerical analyses of the three above-mentioned 
  turbulence generating mechanisms report values of $\alpha$ in the range of $10^{-5} - 10^{-3}$.
This turbulence may also, in principle, diffuse particles away from the disk midplane, reducing 
  the values of the density ratio ($\epsilon$) near the midplane, while also radially diffusing and dispersing radial concentrations of particles before they can grow appreciably 
  { {\citep{Fromang_Papaloizou_2006,Okuzumi_Hirose_2011,Zhu_etal_2015,Riols_Lesur_2018,Yang_etal_2018} 
  }}. 
  On the other hand, some numerical simulations seem to show SI manifesting even in the presence of turbulence (see below). 
  
   
    \par
So, what really happens to the SI in the presence of turbulence? YG2005 presented a brief and mostly qualitative discussion of the possibilities, but declined to pursue them on the grounds that protoplanetary disks were probably nonturbulent. There are a few numerical simulations examining the fate of the SI in either an axisymmetric or fully 3D model
  of a protoplanetary disk experiencing some level of turbulent motions, whether self-excited or driven by some other mechanism \citep[including but not limited to,][] {Johansen_etal_2007,Balsara_etal_2009,Bai_Stone_2010,Tilley_etal_2010,
  Carrera_etal_2015,Yang_etal_2017,Yang_etal_2018,Li_etal_2018}.\footnote{ {Note that \citet{Fromang_Papaloizou_2006}'s two-fluid turbulent set-up is a possible ``precursor" turbulent SI analysis, although this is not yet been demonstrated.}}
  Some of these studies 
   examined the fate of particle clumping in the presence of magnetorotational turbulence, with
  turbulent intensity $\alpha \sim 10^{-3} - 10^{-2}$, and for 
 a range of the two particle parameters $\tau_s$ and $\epsilon$ \citep{Fromang_Papaloizou_2006,Johansen_etal_2007,Balsara_etal_2009,Tilley_etal_2010},
 { {while others modeled only the self-generated ``midplane" turbulence surrounding a layer of particles that had settled toward the midplane
 of an otherwise laminar disk flow \citep[e.g.,][]{Bai_Stone_2010,Carrera_etal_2015,Yang_etal_2017,Li_etal_2018}, and most recently in a forced
 driven model of turbulence \citep{Gole_etal_2020}\footnote{This study appeared during the revision phase of this work.}.} 
 
 While the final state of the SI subject
 to these different kinds of turbulence indeed varies, a common point of agreement between all of these
 investigations is that interesting solids clumping and instability emerges for
 parameter values in which $\epsilon \sim 1$ and, most importantly, when $0.1 < \tau_s < 3$. The results reported in \citet{Johansen_etal_2007} are particularly noteworthy as they show that 
 the SI emerges, in turbulence, with a preferred radial lengthscale of about one pressure scale height 
 ($H$) and has a growth time scale of about 10 local orbital periods. 
 A question facing these and other previous numerical studies of SI in turbulence is whether such combinations of initial conditions - large particles that have somehow grown without being disrupted in such moderate levels of turbulence - are self-consistent (see section \ref{self_consistent}). Meanwhile, several other studies have shown that quite {\it small} particles can undergo SI in disks that are {\it not} turbulent at all, globally; the particles experience only a tiny amount of local turbulence, called ``midplane turbulence", generated {\it by} the densely settled particle layer  \citep{Barranco_2009,Carrera_etal_2015,Yang_etal_2017}. Our results explain these different outcomes in a unified and consistent way. 
  \par
  Well-resolved numerical experiments of two-fluid processes are expensive. A theoretical prediction for the fate of the SI under arbitrary turbulent protoplanetary disk conditions 
would be a useful tool both in developing some quantitative estimate for the expected length and timescales of growth in such a nebula, and in planning future detailed numerical experiments. 
   We present such a theory, extending the SI analysis done in YG2005 and YJ2007 with the
   addition of a simple $\alpha$-model of disk turbulence, that 
   provides an effective turbulent viscosity acting on the gas as well as a prescription for the effective turbulent particle diffusion resulting from the { {statistically steady stirring}} of
   particles by inertial scale gaseous eddies.  
   The basic assumption is that the fundamental processes driving turbulence 
   are unaffected by the 
   particles, and  lead to a statistically steady isotropic
   turbulence in the gas\footnote{This tactic was hinted at in Section 3.2.2. of YG2005. Also, an assumption similar
   to this was employed in studying Tolmien-Schlichting waves of cold (non-MHD) turbulent protoplanetary disks
  \citep{Umurhan_Shaviv_2009}.}
   { {\citep[see however][and Section \ref{turb_constraints}]{Lin_2019}.} 
   
\par
{ {
   We present a linear stability analysis of the SI in such an $\alpha$-disk model of protoplanetary disk turbulence. We determine the growth rate and wavelength of the fastest growing modes as functions
   of various properties including gas disk opening angle, particle Stokes numbers, local particle-to-gas mass density ratio, and the intensity of turbulence. { r{ During the revision phase of this manuscript, \citet{Chen_Lin_2020} released a study with similar aims and we find that our results are in mutual agreement.}}
   In Section 2 we review the basic properties of the SI, including recent theoretical developments regarding resonant drag instabilities.  
   In Section 3 we motivate our model equations, describe the steady state, introduce infinitesimal perturbations, and discuss
   various caveats with respect to our representation of turbulence.
   Section 4 is concerned with verification. 
   In Section 5 we survey the results of the stability analysis, focussing on the most rapidly growing modes.  
   Section 6 applies our theory to four recently published numerical studies of the SI.  
   In Section 7 we identify regions in the parameter space of particle Stokes numbers and disk turbulent intensity for which the SI remains a feasible path to planetesimal formation at various locations of a model disk with 1 percent metallicity.  
   We consider these in light of various known barriers to particle growth. 
 In Section 8 we summarize our findings, we discuss various issues spurred by our analysis, and point to future directions.}}}


\section{SI Mode review}
\subsection{Broad physical picture}\label{broad_physical_overview}
{ { {Because a pressure-supported gaseous disk orbits the central object at sub-Keplerian speeds, momentum exchange between gas and particles via drag forces induces a relative radial drift.}}
{ {In steady state, for example, a radially diminishing steady pressure profile 
typically causes the gas to spiral out while causing a single-size particle to spiral inward (e.g.,
see Eqns. [\ref{steady_gas_speeds}-\ref{steady_particle_speeds}]).  If multiple particle size species are considered, one or several (but not all) of their smaller components can spiral outwards with the gas \citep[e.g.,][]{Estrada_etal_2016, Benitez-Llambay_etal_2019}.}}

The SI arises from perturbations in this relative drifting steady state and how it modifies oscillatory motions in the disk:
Momentum exchange is generally modeled as a function of the
relative velocities between the two fluids multiplied by the product of the two fluid densities {\it{times}} a drag coefficient representing the type of physically appropriate drag mechanism.  
The momentum channeled from
the mean state and into perturbations through modifications of the
drag exchange term due to particle
density fluctuations is the root of the linear instability. YG2005 and YJ2007 show that these
density fluctuations draw momentum from the mean drift state and destabilize
oscillating disk inertial waves (e.g., see YJ2007).   The
insightful precursor toy model of \citet{Goodman_Pindor_2000} argues that
this sort of mean-momentum wave-phase sensitive ``tapping" via gas drag can generically lead to
instability in otherwise damped oscillating systems.
\par
A recent comprehensive theoretical study \citep{Squire_Hopkins_2018a,Squire_Hopkins_2018b} demonstrates that
the SI is a member of a particular class of resonant drag instabilities (RDI).  A two-fluid system, in which
one component is pressure free and streaming with velocity ${\bf w}_s$ with respect to the second (non-zero pressure) fluid, is potentially resonantly unstable to any wave phenomenon with wavevector  ${\bf k}$ supported
by the fluid if ${\bf w}_s \cdot {\bf k}$ equals the oscillation frequency of the wave phenomenon.  In this broad framework the SI is an RDI arising from the particle stream's resonance with
the inertial waves supported by the gas.  We apply this prescription in rationalizing the trends contained in the inviscid models discussed in the verification section \ref{verification}.
\par

Generically speaking, however, the potential for instability holds for any class of waves that the fluid system 
might support including sound waves, gravity waves, magnetosonic waves as well as Rossby waves
and potentially many others
\citep{Hopkins_Squire_2018a,Hopkins_Squire_2018b}.  For example, \cite{Schreiber_Klahr_2018} recently have shown that the SI occurs for non-axisymmetric vertically restricted disturbances in simulations of disks which means that the waves with which the particle stream becomes resonant are not axisymmetric inertial oscillations 
but, instead, either non-axisymmetric inertial oscillations or Rossby waves.  
Similar effects seem to be characterizing the particle-vortex numerical experiments recently
reported in \cite{Surville_Mayer_2018}.
Three key ingredients for resonance are that (a) there exists some means of momentum/energy 
exchange between the two fluid systems, for example, whether it be by means of classical fluid drag (as it is for the SI), or
via dynamical drag if the two fluids are self-gravitating \citep[e.g., see Chapter 13 of][]{Chandrasekhar_1961}, (b) a relative drift velocity between the particle and gas components manifests, and (c) the fluid component supports
 some kind of wave phenomenon. } 
%
\subsection{Some physical properties}\label{some_physical_properties}
We review some of the basic physical properties of the SI based on the
analysis of YJ2007.  The analysis here and throughout this paper is based
on a (nearly) point-analysis performed at
some disk cylindrical radial position $r$.  The disk position is assumed
to be locally isothermal\footnote{{ {``Local" in the sense common in protoplanetary disk literature, namely, that it is only a function of the radial coordinate.}}}
 characterized by a temperature $T$ and soundspeed $c_s \equiv \sqrt{{\cal R}T/\mu}$, where the gas constant ${\cal R} \approx 8314$ J/kg/K and $\mu$ is the gas mean molecular weight.  The local rotation rate of the disk is $\Omega$, the Keplerian rotation speed is $v\sub{K} = r\Omega$, and the effective thickness
of the disk is measured by the disk-opening angle parameter $\delta \equiv c\sub s/v\sub K = H/r$.
{ {
We assume there exists a global pressure field ($P$) whose gradient varies on 
the radial disk scale, i.e.,  $\partial P/\partial r=\order{P\big/r}$.}}  
The SI operates
on length scales $L\sub{{\rm SI}}$ given dimensionally (not quantitatively) by
\beq
\frac{L\sub{{\rm SI}}}{r} = \frac{\partial P/\partial r}{2\rho\sub g\Omega^2 r} \sim \delta^2.
\eeq
This means that $L\sub{{\rm SI}} \sim \delta H$; 
In this work, {\it we adopt the definition $L\sub{{\rm SI}} \equiv \delta H$.}
{ {We refer to $\delta$ as measuring the relative flow of the particles past the gas
at the midplane typically $\sim \delta c_s$ (see below).\footnote{The lengthscale $\delta H$
is typical also of the most unstable VSI modes (Umurhan et al., 2016), and the quantity $\delta^2$ is the same as the pressure gradient parameter $\eta$ of Nakagawa et al (1986).} 
The disk opening angle is also equal to $\delta$.
}}

The analysis of YJ2007 (and YG2005) assumes that vertical variation of gas density plays no role
and there is no momentum or mass diffusion ({\it ie.,} no turbulent viscosity or diffusivity). They show that the SI is the primary instability of axisymmetric inertial modes. 
The stability of a given mode with radial and vertical wavenumbers, $k\sub x$
and $k\sub z$ respectively, is a function of the local value of $\epsilon$
and the stopping time $t_s$ defined according to whether the particles
are in the Epstein or Stokes regimes respectively:
\beq
\displaystyle
t_s = 
\left \{\begin{array}{lr}
\displaystyle\frac{\rho\sub\bullet a}{\rho_g c_s}; & a \le (4/9)\lambda_{mfp}; \\
\displaystyle\frac{8\rho\sub\bullet a}{3\rho_g  C_d \Delta V_{pg}}; & a > (4/9)\lambda_{mfp}
\end{array}\right. ,
\eeq
where $\rho\sub\bullet$ is the density of a given particle, $a$ is the particle radius, and  $\lambda_{mfp}$ is the molecular mean-free-path.  The Epstein
regime is appropriate for particles whose radii $a \le (4/9)\lambda_{mfp}$  while the stopping times
for $a > (4/9)\lambda_{mfp}$ are for the Stokes regime, in which $C_d$ is a
particle drag coefficient and $\Delta V_{pg}$ is the relative speed between a particle 
and the gas (Weidenschilling 1977).  Depending on the size of the particle and $\Delta V_{pg}$, 
$C_d$ may itself be a function of $\Delta V_{pg}$ \citep[e.g., see][]{Estrada_etal_2016}.  
As noted previously, the stopping times are scaled by the local orbital time, giving the ``Stokes number" $\tau_s \equiv \Omega t_s$. 
\par
For a given pair of input parameters ($\tau_s, \epsilon$),  instabilities are typically expected for values of $k_x < k_z$ and growth rates are found to be maximal for values
of $L\sub{{\rm SI}} k_x \approx L\sub{{\rm SI}} k_z \sim 100$ or more (see Figure 1 of
YJ2007 and Figure 2 of YG2005). That is, the wavelengths of fastest growth are usually much smaller than $L_{\rm SI}=\delta H$. Instability is most favorable for values of $\epsilon \sim 1$.  Disturbingly,
the problem as set up appears somewhat ill-posed, in that instability appears to persist
for certain finite values of $k_x$ as $k_z \rightarrow \infty$.  In cases of most 
physical interest, the instability growth timescales must be much faster than the radial drift
rates (see Figure 8 of YG2005).  The analysis of YJ2007 assumes the gaseous component is compressible, yet they demonstrate that the contribution of gas compressibility is negligible to the instability mechanism (see also Section 4).  As such, they show that the particle fluctuations are the key ingredient of the instability.

\section{Model equations}
\subsection{Assumptions}
We build upon the inviscid model setup of YG2005.  We review the assumptions made and indicate
what is new to this paper:
\begin{enumerate}
\item The gas component is incompressible.  { {As examined in the original studies, the growth rate of the strongest inviscid SI is on the orbit timescale, while its
typical lengthscale $\sim \delta H$.  { {This means that acoustic disturbances}} propagate across those relevant lengthscales on correspondingly shorter timescales $\sim \delta P_{orb}$, where $P_{orb} \equiv 2\pi/\Omega$ is the local disk orbit time.}}
\item Spatial variations in all disk quantities are negligible except for the
mean Keplerian shear, which is assumed constant.
\item There are no disk vertical density variations (no vertical density stratification).
\item The Stokes numbers are constant.
\item  
{ {We consider only particles consisting of a single mass-dominant size and, thus, ours is a ``2-fluid" model. Particle growth evolutionary models suggest that the assumption of a mass-dominant size is not bad \citep{Zsom_etal_2010, Estrada_etal_2016}. { Because \citet{Krapp_etal_2019} 
have recently shown that the streaming instability is diminished for disk models containing most reasonable 
particle distributions
 in even weakly turbulent disks
 \footnote{\citet{Krapp_etal_2019} show that only for particle distributions with a very restricted range
in particle sizes and mass loading, i.e., $0.0001 < {\rm St} < 0.01$ with $\epsilon > 0.4$, is the growth rates of the SI enhanced
compared to the two-fluid approach typically taken.  See their Figure 5.}
 ,   
 our assumption is conservatively favorable to SI. }}}
\item Scales of interest are small enough so that it is appropriate to use the shearing box assumption, which neglects large scale effects including curvature.
\item Gas and particle perturbations are axisymmetric.. 
\item (New to this paper) Turbulence is assumed isotropic and is represented in the gas momentum equation by the standard $\alpha$-disk model in which turbulence is modeled as an enhanced kinematic viscosity, $\nu_t=\alpha c_s H$,  controlled by the non-dimensional parameter $\alpha$.  In other words, internal stresses arising from momentum exchange due to turbulence is modeled
with the term, $\nu_T \rho_g \nabla^2 {\bf u}_g$, on the righthand side of the gas momentum equation
\cite[e.g.][]{Shakura_Sunyaev_1973,Lynden-Bell_Pringle_1974}. 
{ {We also include the effect of radial accretion of the gas component due to the underlying turbulence-driven viscous evolution.}}
\item (New to this paper) Turbulence causes stirring of the particle component, represented by a turbulent diffusion term as a source term in the
particle mass conservation equation \citep{Cuzzi_etal_1993,Dobrovolskis_etal_1999,Youdin_Lithwick_2007, Carballido_etal_2011,Estrada_etal_2016}:
\beq
\frac{\alpha c_s H}{1 + \tau_s^2}\nabla^2 \rho_p,
\eeq
which captures the effect of diminished stirring of particles with large inertias ($\tau_s \gg 1$). 
\item { {(New to this paper) Since collisions between particles are not important the particle phase is typically assumed to be pressure free. However, turbulent stirring introduces an effective pressure gradient upon the particle momentum conservation in the form of an effective particle pressure term,
\[
-c_d^2 \nabla \rho_p; \qquad c_d^2 = \frac{\alpha c_s^2}{1 + \tau_s^2},
\]
\citep{Cuzzi_etal_1993,Dobrovolskis_etal_1999,Jacquet_etal_2011}.
The effects of this ``particle pressure" term are small, {\it except} for the wavelengths of the fastest growing mode.}}
\end{enumerate}

{ {We note a possible shortcoming of adopting the $\alpha$-disk model, namely the assumption that the turbulence is isotropic. It is known that numerical studies of at least one of the proposed hydrodynamical mechanisms that may drive turbulence in protoplanetary disk Ohmic (``dead") zones (i.e., the VSI) have seen turbulent stresses that are clearly non-isotropic \citep{Stoll_etal_2017}, as well as in Dead Zones whose activity is driven by sandwiching MHD turbulent layers like
in \citet{Yang_etal_2018}.
Incorporating such higher-order effects would require formulating a Reynolds averaging type of mixing scale model that takes into account the anisotropy of the shear stresses following the approaches found in \citet{Cuzzi_etal_1993,Dobrovolskis_etal_1999}.  This should be revisited in future analyses.}}

\subsection{Equations of Motion and Steady State}
We write the fundamental equations of motion in the local  frame rotating at $\Omega$. The
radial coordinate is $x$, the azimuthal coordinate $y$ and $z$ is the vertical coordinate. 
We represent the mean azimuthal shear as a departure from a mean
Keplerian state $V_0 = -(3/2)\Omega(r-r_0) {\bf {\hat y}}$ \citep{Umurhan_etal_2004}.   The disk gas  velocity relative to the mean state is ${\bf v}_ g = 
u_g {\bf {\hat x}} +  v_g {\bf {\hat y}} + w_g{\bf {\hat z}}$
while the corresponding relative 
particle velocity is  
${\bf v}_p = 
u_p {\bf {\hat x}} + v_p {\bf {\hat y}} + w_p{\bf {\hat z}}$.  
By writing
$P = \tilde P + p$,
we split the
 pressure field into a sum of the background field ($\tilde P$) -- i.e., that drives the relative mean motion between the gas and particles -- and perturbation field ($p$).
  The axisymmetric equations of motion for the gas are 
\beqa
 \rho_g\left(\partial_t + u_g\partial_x + w_g\partial_z\right)u_g -2\Omega \rho_g v_g
&=& -\partial_x p  - \partial \tilde P/\partial r  + \rho_p\rho_g \mu (u_p-u_g) +  
(\alpha c_s H \rho_g) \nabla^2 u_g, \label{gas_eqn_start} \\
\rho_g\left(\partial_t + u_g\partial_x + w_g\partial_z\right)v_g + 
\sfrac{1}{2}\Omega \rho_g u_g
&=& \rho_p\rho_g \mu (v_p-v_g) + (\alpha c_s H \rho_g) \nabla^2 v_g -(3/4)\alpha \rho_g \delta c_s \Omega, \label{gas_azimuthal_momentum} \\
\rho_g\left(\partial_t + u_g\partial_x + w_g\partial_z\right)w_g
&=&  -\partial_z p  +
\rho_p\rho_g \mu (w_p-w_g) +  (\alpha c_s H \rho_g) \nabla^2 w_g, \\
\partial_x u_g + \partial_z w_g &=& 0.
\eeqa
Momentum exchange between gas and particle phases arises in terms above of the type $\rho_p(u_g-u_p)/t_s$, where we have also defined the parameter $\mu \equiv 1/\rho_gt_s$.  
{ {It is assumed that the conditions in the disk (i.e., gas density and temperature as well as particle abundance) fall into the Epstein regime (though see Sec. \ref{combined_limits_on_particle_size}). The background disk pressure gradient, the primary driver
of the SI, is given by}}
\beq
- \partial \tilde P/\partial r = 2 \eta \Omega^2 r = 2\delta c_s \Omega.
\eeq
{ {In order to account for the viscous torque induced by the background alpha-disk model, we include on the LHS  of equation (\ref{gas_azimuthal_momentum}) the appropriate background viscous forcing in the form of an acceleration  $-\alpha c_s H  \rho_g  \nabla^2 {\bf v}_g$.}}
The equations of motion of the the particle phase are, 
\beqa
& & \rho_p\left(\partial_t + u_p\partial_x + w_p\partial_z\right)u_p -2\rho_p\Omega v_p
=   
-c_d^2\partial_x \rho_p
+ 
\rho_p\rho_g \mu (u_g-u_p),  \\
& & \rho_p\left(\partial_t + u_p\partial_x + w_p\partial_z\right)v_p + 
\sfrac{1}{2}\rho_p\Omega u_p
=  \rho_p\rho_g \mu (v_g-v_p),  \\
& & \rho_p\left(\partial_t + u_p\partial_x + w_p\partial_z\right)w_p
=  
-c_d^2\partial_z \rho_p + 
\rho_p\rho_g \mu (w_g-w_p),  \label{vertical_momentum_particles} \\
& &\partial_t \rho_p + \partial_x \rho_p u_p + \partial_z \rho_p w_p 
= \frac{\alpha c_s H}{1 + \tau_s^2}\nabla^2 \rho_p,
\label{particle_eqn_end}
\eeqa
{ {where the typical kinetic energy per unit mass of the particles induced by the turbulent stirring of the gas is 
given by $c_d^2 = \alpha c_s^2\big/\left(1+\tau_s^2\right)$.}}
Steady uniform solutions of Eqns. (\ref{gas_eqn_start}-\ref{particle_eqn_end}) are sought assuming no vertical velocities and constant steady gas and particle densities, $\overline{\rho}_g$ and
$\overline{\rho}_p$. 
Following { {the procedures of}}  Nakagawa et al. (1986) and YJ2007 (their Eqns 7-8)
we have that uniform gas velocities $U\sub {g0}, V\sub {g0}$ are
\beqa
U\sub {g0} &=& \frac{2\epsilon \tau_s - (3\alpha/2)(1+\tau_s^2 + \epsilon)}{(1+\epsilon)^2 + \tau_s^2}
\delta c\sub s, \nonumber \\
V\sub {g0} &=& -\left(\frac{1+\epsilon + \tau_s^2 + (3\alpha/4)\tau_s \epsilon}{(1+\epsilon)^2 + \tau_s^2}
\right)\delta c\sub s,
\label{steady_gas_speeds}
\eeqa
and the uniform particle velocities are
\beqa
U\sub {p0} &=& -\frac{2\tau_s + (3\alpha/2)(1+\epsilon)}{(1+\epsilon)^2 + \tau_s^2}
\delta c\sub s, \nonumber \\
V\sub {p0} &=& -\left(\frac{1+\epsilon - (3\alpha/2)\tau_s}{(1+\epsilon)^2 + \tau_s^2}
\right)\delta c\sub s,
\label{steady_particle_speeds}
\eeqa
\citep[cf.,][]{Dipierro_etal_2018}.  
{ {Given the importance of the relative radial 
velocity between the gas and particle phases in rationalizing the
SI in terms of resonant conditions \cite[a la][]{Squire_Hopkins_2018a,Squire_Hopkins_2018b,Hopkins_Squire_2018a,Hopkins_Squire_2018b}, we
find
\beq
\Delta U \equiv U\sub {g0}-U\sub {p0} 
= \tau_s\left(\frac{2+2\epsilon - 3\tau_s\alpha/2}{(1+\epsilon)^2 + \tau_s^2}
\right)\delta c\sub s .
\label{relative_radial_drift}
\eeq
}}
\
\subsection{Linearized perturbations}
We linearly perturb equations (\ref{gas_eqn_start}-\ref{particle_eqn_end})
around this steady state according to
\beq
u_g \rightarrow U\sub {g0} + u_g', \quad
v_g \rightarrow V\sub {g0} + v_g', \quad
w_g \rightarrow w_g', \quad p \rightarrow p',
\eeq
for the gas quantities and
\beqa
& & u_p \rightarrow U\sub {p0} + u_p', \quad
v_p \rightarrow V\sub {p0} + v_p', \quad w_p \rightarrow w_p',\quad
\rho_p \rightarrow \overline\rho_{p}\left(1+\Delta_d'\right),
\eeqa
for the particles.  Since the gas is incompressible,
we further write the quantities  $u_g'$ and $w_g'$ as derived
from a perturbation streamfunction $\psi'$:
\beq
u_g' = \partial_z \psi', \qquad w_g' = -\partial_x \psi'.
\eeq
One can formally define the azimuthal gas and particle ``fluid" vorticity fields as: 
\beq
\omega_g' \equiv \partial_z u_g' - \partial_x w_g',
\qquad
\omega_p' \equiv \partial_z u_p' - \partial_x w_p'.
\eeq
The gas vorticity is related to the stream function via $\omega_g' = \nabla^2 \psi_g'$.
The perturbed quantities are then Fourier decomposed. For example, 
the streamfunction is written as
\[ \psi' \rightarrow \hat\psi \exp\big(ik\sub x x + ik\sub z z - i\omega t\big) + {\rm c.c.}, \]
where $k_x$ and $k_z$ are the radial and vertical wavenumbers (respectively), the frequency
is $\omega$, and $\hat\psi$ is the normalmode amplitude.  
{ {
$\omega_i \equiv {\rm Im} (\omega)>0$ indicates growth with a corresponding e-folding timescale,
$t_g \equiv 2\pi/\omega_i$.}} 
We restrict our consideration
to positive values of $k_x$ and $k_z$.\footnote{{ {
In performing double Fourier transforms of real quantities, one is free to restrict attention
to positive wavenumber values for one chosen spatial direction, here we choose to be the radial one ($x$).
The linearized PDE's possess z-reflection symmetry since imposing $z \rightarrow -z$ together with 
$w_g \rightarrow -w_g$ and $w_p \rightarrow -w_p$ leaves the equations invariant. The eigenvalues $\omega$ are
insensitive to the sign of $k\sub z$, confirming the same reflections made in YG2005.  Thus, we may safely restrict
attention to positive values of $k_z$ as well.}}}
Since $k_x>0$, values where ${\rm Re}(\omega)>0$ indicate outwardly propagating patterns.
We are reminded that no such symmetry characteristic exists in the radial direction due
to the imposed symmetry breaking provided by both the presence of turbulence and an externally
imposed radially dependent pressure field $\tilde P$.

\par
The combined system reduces (using Mathematica) to a generalized expression for the
dispersion relation of the form 
\beq
F(\omega;k_x,k_z,\tau_s, \epsilon,\delta,\alpha) = 0,
\label{dispersion_relation}
\eeq
in which $F$ is a sixth order algebraic equation in $\omega$.  { {In the inviscid limit}}, the six temporal modes correspond to two inertial waves in the gas phase, two inertial waves in the particle phase, and two ``zero"-temperature acoustic modes
in the particle phase \citep[e.g., see discussion in Chapter 10 of][]{Chandrasekhar_1961}\footnote{{ {
In the absence of turbulent stirring a pressure-less fluid could be viewed as a
zero-temperature gas.  With $c_d^2 \neq 0 \ll c_s^2$ the particle phase effectively behaves as a low temperature 
compressible fluid.
}}}.
The algebraic equation for $\omega$ is solved using standard root-finding 
methods found in Matlab 2017a.  The eigenvalue $\omega$ depends upon five parameter
expressions:
the two wavenumbers $k\sub x, k\sub z$, the Stokes number $\tau\sub s$,
the density ratio $\epsilon = \overline{\rho}_p/\overline{\rho}_g$, and 
the ratio of the turbulent intensity parameter to the disk opening ratio squared: $\alpha/\delta^2$.  In all of our parameter scans, we assume
$\delta = 0.05$, a typical value for nominal disk temperatures and orbit velocities.
Dynamic lengthscales are normalized by $\delta c_s /\Omega = \delta H = L\sub{{\rm{SI}}}$ and growth rates normalized by 
$\Omega$.\footnote{In YG2005 and YJ2007, the lengthscales are quoted as ``$\eta r$", where $\eta$ is the radial pressure parameter of Nakagawa et al (1986), 
which expression is equivalent to $\delta H = c_s H/v_k$.}

\subsection{Turbulent dilution model}
\par
Finally, we restrict our choice of $\epsilon$ to be physically consistent with the
idea that a turbulent disk will loft particles away from the midplane and, consequently, result in dilution of $\epsilon$ near the midplane.   
Following previous authors 
\citep{Dubrulle_etal_1995,Carballido_etal_2006,Carballido_etal_2011,Estrada_etal_2016}, 
we estimate an effective particle scale height $H_p$ from the balance between particle settling toward the midplane and upward lofting
by turbulent motions.  Thus, given an initial local ratio of the particle surface mass density to gas surface mass density, $Z$, then in a region
of vertical thickness $H_p$ we
broadly assign a local particle to gas mass density ratio via the simple relationship:
\beq
\epsilon = \epsilon(\alpha, \tau_s, Z) =
Z\sqrt{\frac{\alpha + \tau_s}{\alpha}},
\label{turbulent_dilution_model}
\eeq
which, henceforth, is referred to as the {\it{turbulent dilution model}} (TDM).  In this form, $H_p = \sqrt{\alpha/(\alpha + \tau_s)} H$.
{ {A disk with a cosmic abundance of about 1 percent would correspond to $Z = 0.01$.  Even if a protoplanetary disk starts out
with a cosmic abundance of solids, any given radial location within that disk may have values of $Z$ that vary with
the disk's evolution \citep{Birnstiel_etal_2012,Estrada_etal_2016,Sengupta_etal_2019}.  As such, we allow
for values of $Z$ departing from the fiducial cosmic abundance value.}}


\section{Verification}\label{verification}
\begin{figure}
\begin{center}
\leavevmode
\includegraphics[width=9.5cm]{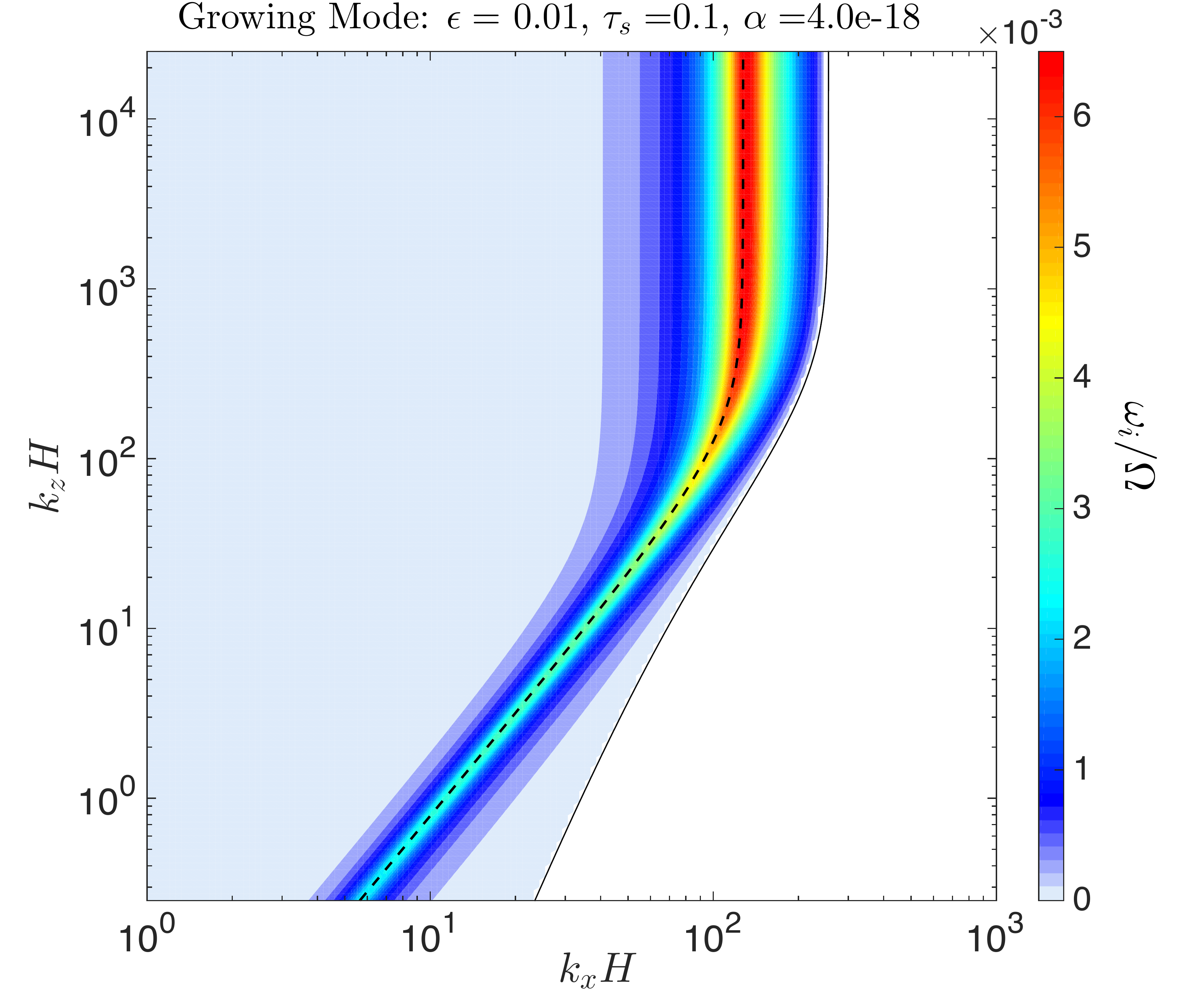}
\par
\end{center}
\caption{Maximum growth investigated in the inviscid limit ($\alpha \rightarrow 0$) for $\tau_s = 0.1$,$\epsilon = 0.01$ and $\delta = 0.05$.  The qualitative behavior of the growth rates as a function of $k_x$ and $k_z$ reported in YG2005 is recovered here.  The dashed line corresponds to the location in parameter space where the radial wavespeed of inertial gas waves equals the difference in mean radial velocities between gas and particles -- verifying predictions of \citet{Squire_Hopkins_2018a}. }
\label{Growth_rates_inviscid}
\end{figure}
As a robustness
test, we set $\alpha \rightarrow 0$ and were able to faithfully reproduce all individual eigenvalues and eigenvectors quoted in Table 1 of YJ2007 as well as the growth rate diagrams shown in their
Fig. 1. We indeed confirm that the SI is an instability of inertial modes of mixed character (i.e.,
particle-gas modes).  
{ {We note that the calculation in YJ2007 assumed a compressible gas component while our model assumes the gas
to be incompressible.  
The fact that the growth rates are essentially identical in both calculations strongly suggests that the SI (as considered in their study) is
practically insensitive to gas compressibility and, further, it would be sound to examine its evolution with the assumption of an incompressible gas.}}
\par
Figure \ref{Growth_rates_inviscid} shows the maximum growth rates as a function of $k_x$ and $k_z$ which reproduces the quality reported in YG2005.  There exists a combination of $k_x$ and $k_z$ for which the growth rates are locally maximal.  According to \citet{Squire_Hopkins_2018a},
there exists a wave-drift resonance relationship identifying this combination as
those values of $k_x$-$k_z$ for which some collective fluid mode has a projected phase speed that resonates with the relative drift velocity between the gas and the particles. 
In the case of the classic SI, one such wavemode is an inertial wave. 
 In cases for which 
both the mass-loading is weak (small $\epsilon$) and the coupling between gas and particles is strongish ($\tau_s < 1$)
inertial modes in a collective gas-particle medium can be approximated by the wave response in the gas assuming no coupling to the particles.  Figure \ref{Growth_rates_inviscid} shows the actual growth rates determined
for such a strongly coupled weakly mass-loaded model.
In this extreme case, it is elementary to show that
\beq
\omega^2 = \frac{\Omega^2 k_z^2}{k_x^2 + k_z^2},
\label{inertial_mode_gas_limit}
\eeq
\citep{Lyra_Umurhan_2019}.  Since there are no $y$ disturbances, identifying the radial component of the inertial wave to the relative drift velocities means equating
\beq
\omega = k_x \left(U\sub{p0}-U\sub{g0}\right)
\eeq
Inserting (\ref{inertial_mode_gas_limit}) and the appropriate steady radial drift expressions from 
(\ref{relative_radial_drift}) { {with $\alpha=0$}} into the above expression, we find that the desired $k_x$-$k_z$ relationship is
\beq
k_z^2 = \frac{k_x^4}{k_a^2 - k_x^2}, \qquad k_a \equiv \frac{(1+\epsilon)^2 + \tau_s^2}{2\delta (1+\epsilon) \tau_s H}.
\label{inviscid_kx_kz_relationship}
\eeq
The wave-drift resonance relationship expressed in (\ref{inviscid_kx_kz_relationship}) is shown as a dashed line over the growth rates showcased in Figure \ref{Growth_rates_inviscid}.  We see clearly that
the resonance relationship follows the maximum growth rates as one scans along $k_z$.  This lends confidence that the resonance condition is a very good predictor for identifying conditions corresponding to maximal growth of this instance of the RDI { {in the $\epsilon \ll 1$ regime}}.

\section{Results}
Our closed-form 
solutions (see Appendix \ref{appendix_1}) permit a finely-resolved sweep in parameter space varying both the Stokes number $\tau_s$ and
turbulence parameter $\alpha$.  The particle-to-gas volume mass density ratio $\epsilon$ is automatically determined as 
a function of $\alpha$ and $\tau_s$ based
on the global (unsettled) solids mass fraction $Z$ and the turbulent dilution model (TDM), eq. (\ref{turbulent_dilution_model}).  For most
parameter sweeps, we usually set $Z$ at nominal cosmic abundance, $Z=0.01$, but we also consider other
values of $Z=$0.02,0.03, 0.04, and 0.08. For these values of $Z$ but with a fixed disk opening angle $\delta = 0.05$,
we study the fastest growing mode as 
a function of $\alpha$ and $\tau_s$.
\par

\begin{figure}
\begin{center}
\leavevmode
\includegraphics[width=10.cm]{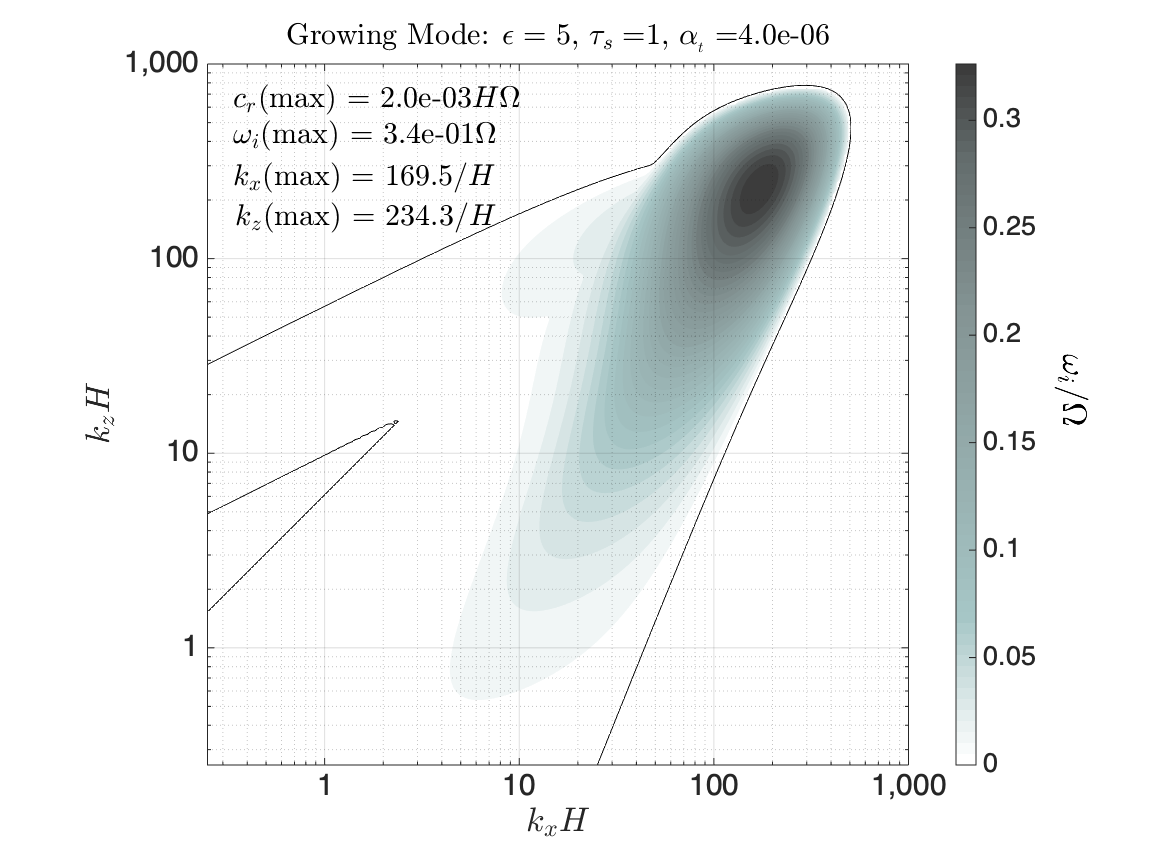}
\par
\vskip -0.25cm
\includegraphics[width=10.1cm]{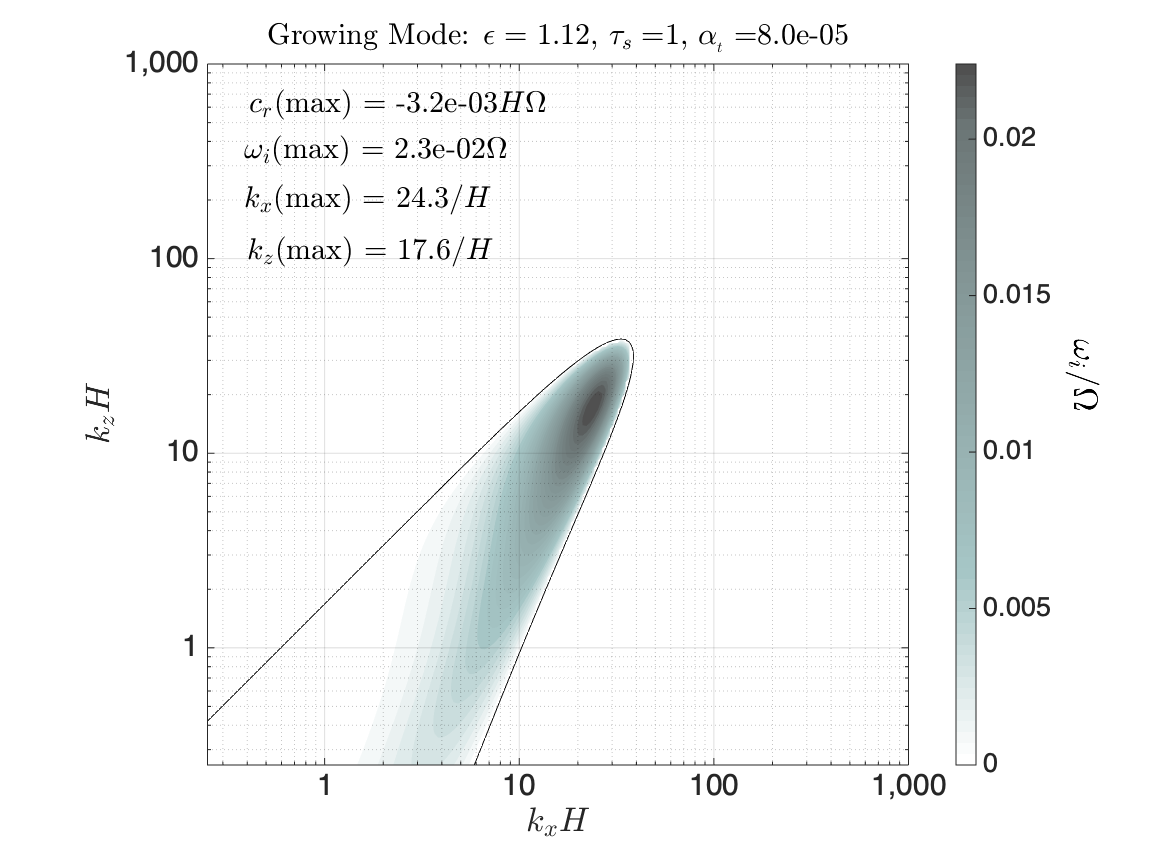}
\par
\vskip -0.25cm
\includegraphics[width=10.1cm]{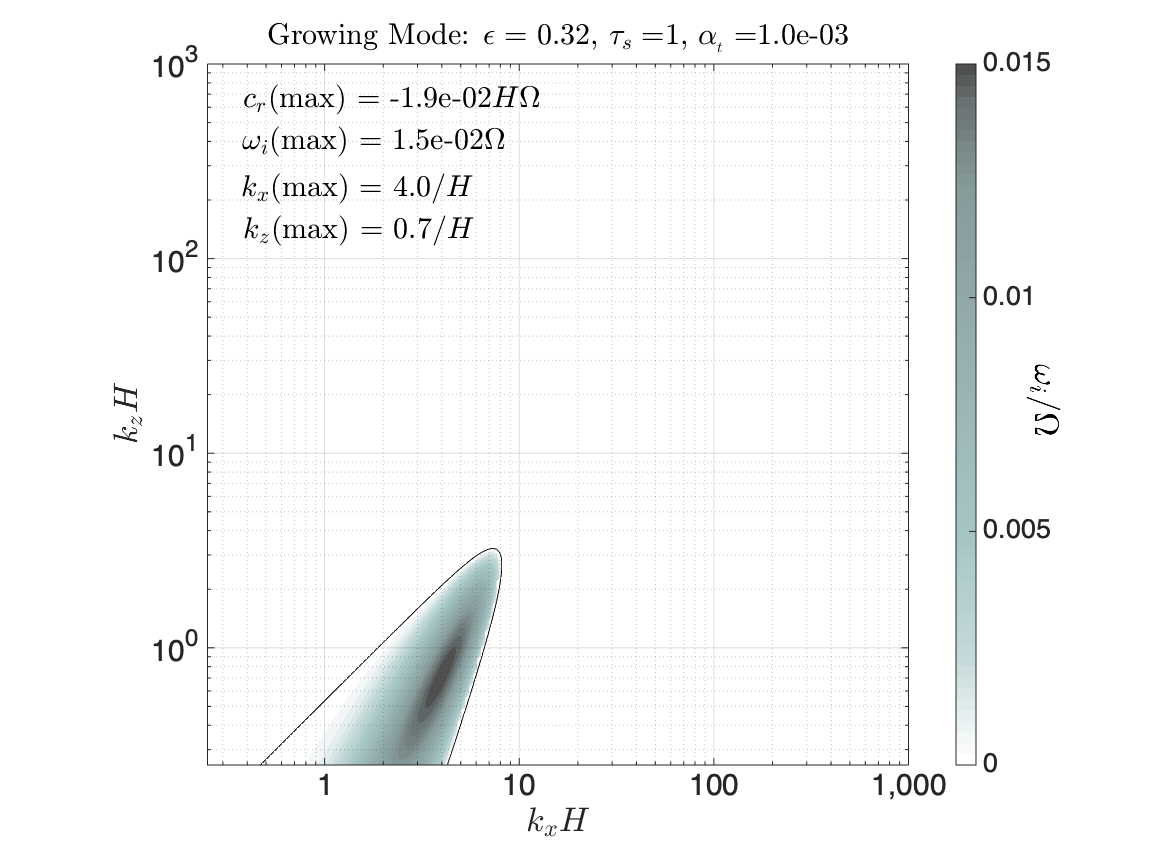}
\end{center}
\caption{Growth of the SI in a sequential progression of turbulent intensity.  Growth rates for
$k\sub x$ and $k\sub z$ are shown scaled by $H^{-1}$.  All plots assume $\delta = 0.05,Z=0.01$
and $\tau_s = 1$.  Values of $\epsilon$ follow the TDM, eq. (\ref{turbulent_dilution_model}).
The solid black line denotes zero growth.
Top panel shows $\alpha \sim 4\times 10^{-6}$ (weakly turbulent),
middle panel shows $\alpha = 8\times 10^{-5}$ (moderately turbulent)
and bottom panel shows $\alpha = 10^{-3}$ (strongly turbulent).  As $\alpha$ increases, the wavenumber
of peak growth systematically gets smaller and the spatial orientation
becomes more like vertically oriented, radially narrow sheets.  Bottom panel represents conditions
closely approximating those investigated in Johansen et al. (2007).}
\label{Growth_rates_kxkz_1}
\end{figure}

\begin{figure*}
\begin{center}
\leavevmode
\includegraphics[width=17.5cm]{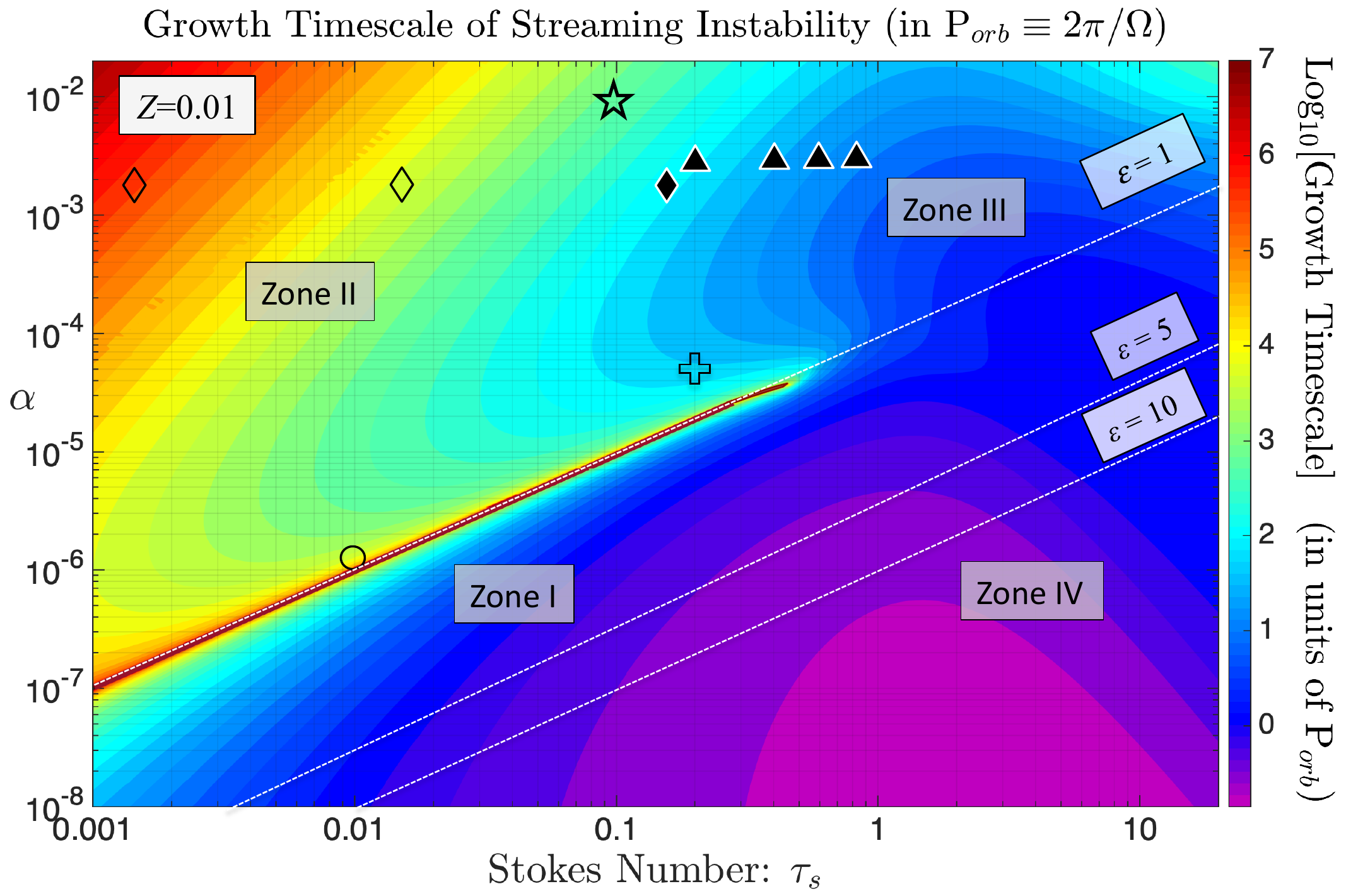}
\end{center}
\caption{Growth timescales of the fastest growing modes, $t_{gm}$, as a function of turbulent intensity $\alpha$ and Stokes number $\tau_s$, for $\delta = 0.05$ and $Z=0.01$. Colors depict e-folding growth timescales in units of local orbit times.  Published numerical simulations where SI is observed are shown as filled symbols: triangles: \citet{Johansen_etal_2007}, diamonds: \citet{Balsara_etal_2009}. Simulations where SI was not observed are shown as open symbols: circles \citet{Yang_etal_2017}, stars:
\citet{Yang_etal_2018}, crosses: \citet{Gerbig_etal_2020}, and diamonds: \citet{Balsara_etal_2009}. The $\epsilon = 1$ track leading up to the critical point ($\tau_c = 0.45$, $\alpha = \alpha_c = 3.7\times 10^{-5}$) is labeled, and effectively serves as a barrier, above which densities grow slowly.  For $\tau<\tau_c$, regions corresponding to $\epsilon < 1$ (high $\alpha$; above the track) have markedly
lower growth rates and are identified as belonging to the ``saturated turbulent"  regime (Zone II). 
while the region $\epsilon > 1$, both low and high $\tau_s$ below the track, is the ``laminar" zone
and denoted as Zone I and IV respectively.  
Zone III is the strongly turbulent region for large Stokes number. }
\label{Growth_Timescales_f01}
\end{figure*}
{ {
We caution against cavalierly linking the results of this work to the original theoretical studies (e.g.,YG2005 and YJ2007) which surveyed the inviscid linear stability calculation for $\order{1}$ values of $\epsilon$.
We observe that by adopting the TDM the inviscid limit is singular as it predicts $\epsilon \rightarrow \infty$.  Indeed the TDM assumes that some type of quasi-steady state has emerged between the particles and the surrounding turbulent state.  
The only permissible pathway linking results of this model to those of the inviscid limit is to take
the double limit $\alpha \rightarrow 0$, $\tau_s \rightarrow 0$, while setting $\tau_s/\alpha$ to a finite constant.
The latter allows for choosing an arbitrary value of $\epsilon$, but the limit corresponds to dynamics involving particles instantaneously responding to the gas motions with no relative drift.  In other words, this limit represents  regular inviscid incompressible single-fluid gas dynamics with a slightly enhanced mean density.}}

\par
\medskip
Our most basic result is this: isotropic turbulence, as measured by $\alpha$, causes the growth rates of the SI to diminish, while also increasing the wavelengths corresponding to fastest growth.  This is not surprising, because the shortest wavelength modes are eaten away by turbulent diffusion { {of momentum and particle concentration}}.  

\subsection{Individual model results}\label{individual_mode_results}
In this section we show the properties of individual models.  In particular,
Figure 
\ref{Growth_rates_kxkz_1} displays contour plots of growth rates as a function of $k_x$ and $k_z$ 
 for three values of $\alpha$ together with $\tau_s = 1$ fixed.  \par 
For {\it{weak turbulence}}
there exist wavelengths corresponding to maximum growth that are both very short, 
and have growth rates on the order of the disk rotation frequency.
For example, the top panel of Fig. \ref{Growth_rates_kxkz_1} shows results
for the very low value of $\alpha \approx 4\times 10^{-6}$ with $\tau_s = 1$, perhaps corresponding to ``midplane turbulence" around a settled particle layer in a globally laminar nebula.  The wavelengths of maximum
growth are $\lambda_x({\rm max}) = 2\pi/k_x({\rm max}) \approx 0.037 H$ and 
$\lambda_z({\rm max}) = 2\pi/k_z({\rm max}) \approx 0.027 H$; thus, only a little smaller than $\delta H$.  This fastest growing mode has an e-folding timescale $t_{gm}\approx 0.5{\rm P}_{orb}$,
where $t_{gm} \equiv t_g({\rm max}) \equiv \left[\Omega/\left(2 \pi \omega_i({\rm max})\right)\right] {\rm P}_{orb}$.  
{As the intensity of turbulence increases (middle and lower panels), the wavelengths corresponding to maximal growth get larger and
the corresponding growth rates diminish.  }
\par
The middle panel of Fig. \ref{Growth_rates_kxkz_1} shows a wavenumber survey
for $\alpha \approx 8\times 10^{-5}$ and $\tau_s = 1$ -- a so-called {\it{weak-to-moderately turbulent}} model, even if the particle layer is rather densely settled.
The fastest growing wavelengths in both directions are nearly
equal with $\lambda_x({\rm max}) \approx 0.25 H$ and $\lambda_z({\rm max}) \approx 0.34 H$,
and start becoming of the same order of magnitude as the local disk scale height.  The 
corresponding growth timescale is now considerably longer with 
$t_{gm} \approx 6.9 {\rm P}_{orb}$.\par

The bottom panel of Fig.\ \ref{Growth_rates_kxkz_1} similarly shows 
a wavenumber survey
for $\alpha \approx  10^{-3}$ and $\tau_s = 1$, a model we term {\it{strongly turbulent}}. Even here, according to the TDM, the particle layer has settled to a thickness of only $H/30$ because of the large $\tau_s$. 
The wavelengths of fastest growth 
become even larger, and the relative ($x,z$) length scales become more disparate
with
$\lambda_x({\rm max}) \approx 1.55 H$ and $\lambda_z({\rm max}) \approx 7.85 H$ (implying vertical sheet-like disturbances).
The corresponding e-folding growth timescale is  $t_{gm} \approx 10.6 {\rm P}_{orb}$. 
\par
 The pattern propagation of the turbulent SI also depends upon the 
degree of turbulence.  We measure the radial pattern speed to be given
by $c_r \equiv \omega_r/k_x$, where $\omega_r = {\rm Re}(\omega)$.  The pattern
propagation of the fastest growing mode is denoted by $c_r({\rm max})$
and equal to $\omega_r( {\rm max})/k_x({\rm max})$.  With reference to Fig. \ref{Growth_rates_kxkz_1} we see that the pattern propagation is inward
for the two largest values of $\alpha$ shown while it is outward for the nominally
weakly turbulent model. 
\par
The most strongly turbulent model shown in Fig. \ref{Growth_rates_kxkz_1} closely corresponds to the conditions modeled in Johansen et al. (2007).
 Figure 1(b-c) of Johansen et al. (2007) depicts the growth of the SI in a MRI driven turbulent
 setting in which $\alpha = \order{10^{-3}}$ and where the Stokes number $\tau_s \sim 1$.
 A cursory examination of the growing modes in those simulations { {during the early linear
 growth phase ($t<30$P$_{orb}$)}} shows that
radial wavelength of the most prominently growing structure is $\sim$1.2$H$, with a growth timescale of 10-15 P$_{orb}$.  
  Furthermore, our model predicts that the pattern speed $c_r({\rm max}) = 0.12 H/$P$_{orb}$ and inward. 
  The apparent propagation of the growing mode in Johansen et al (2007)
  is also inward with a pattern speed $~0.1 H/$P$_{orb}$ (we discuss pattern speeds further in sec. \ref{general_survey_pattern_speeds}).  While this certainly does
   not prove that our simple model is sufficiently predictive, it is encouraging that
   it predicts features that are in both qualitative and approximate quantitative agreement with previously published numerical simulations. 
\par

 \begin{figure}
\begin{center}
\leavevmode
\includegraphics[width=8.5cm]{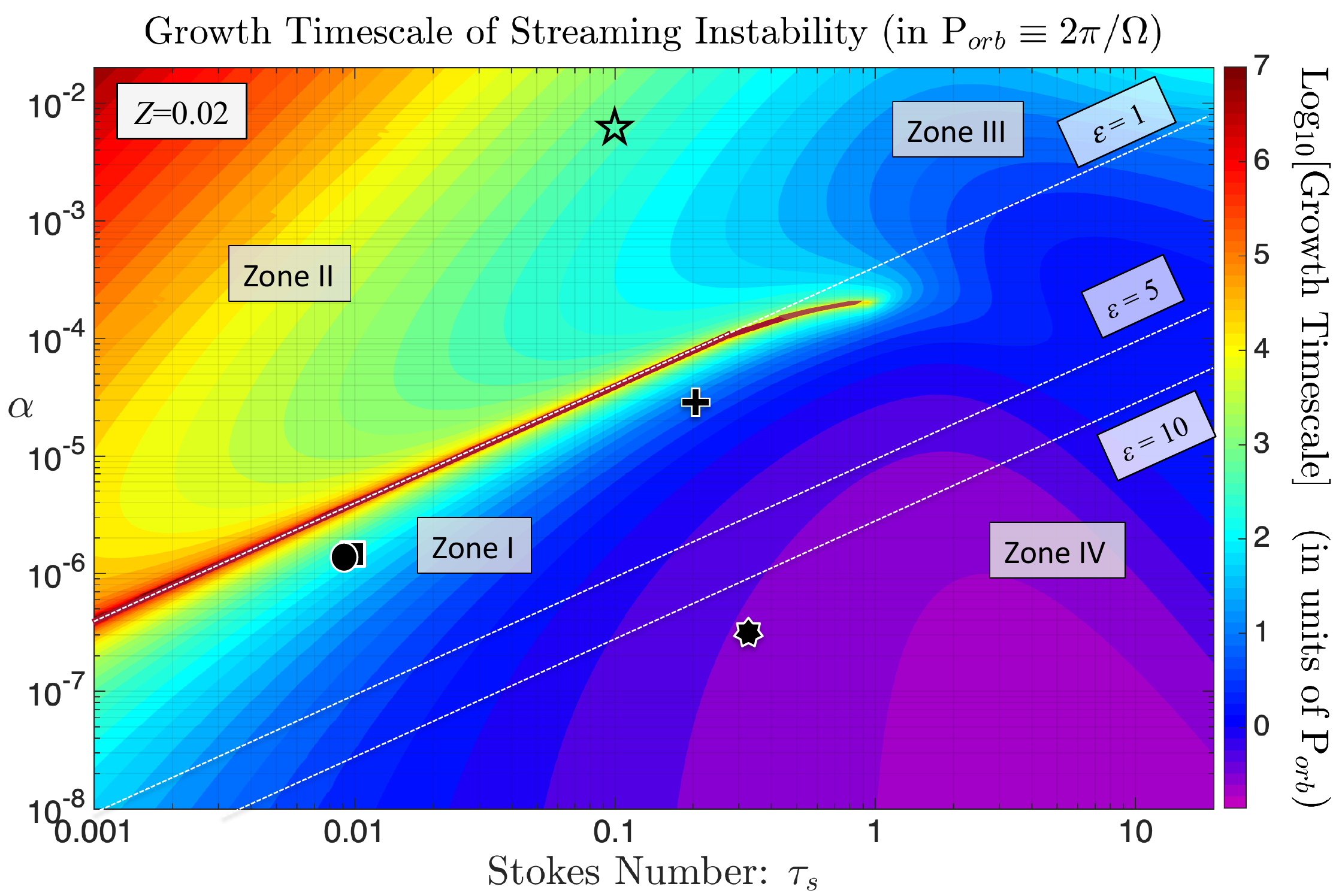}
\includegraphics[width=8.5cm]{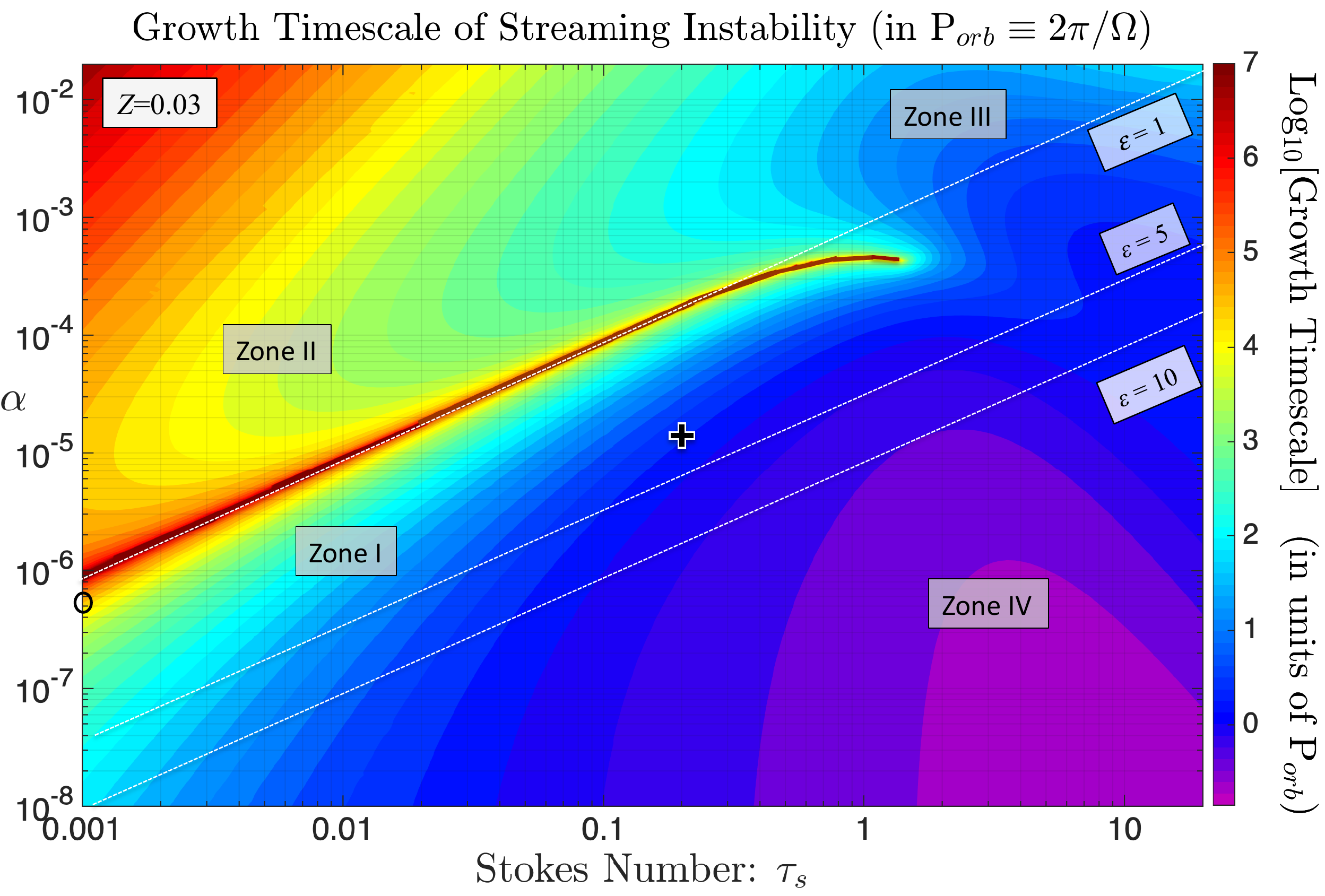}
\par
\includegraphics[width=8.5cm]{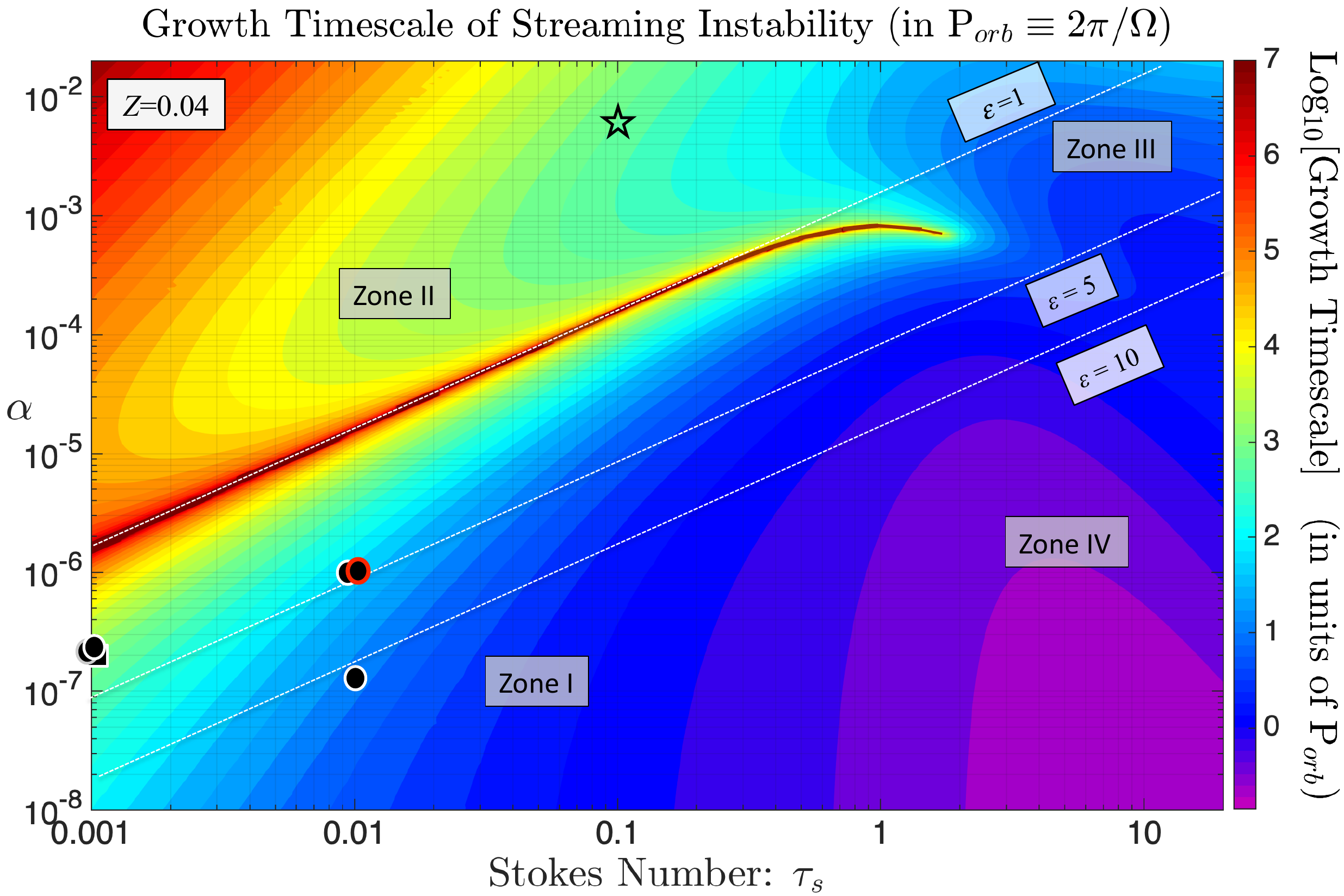}
\includegraphics[width=8.95cm]{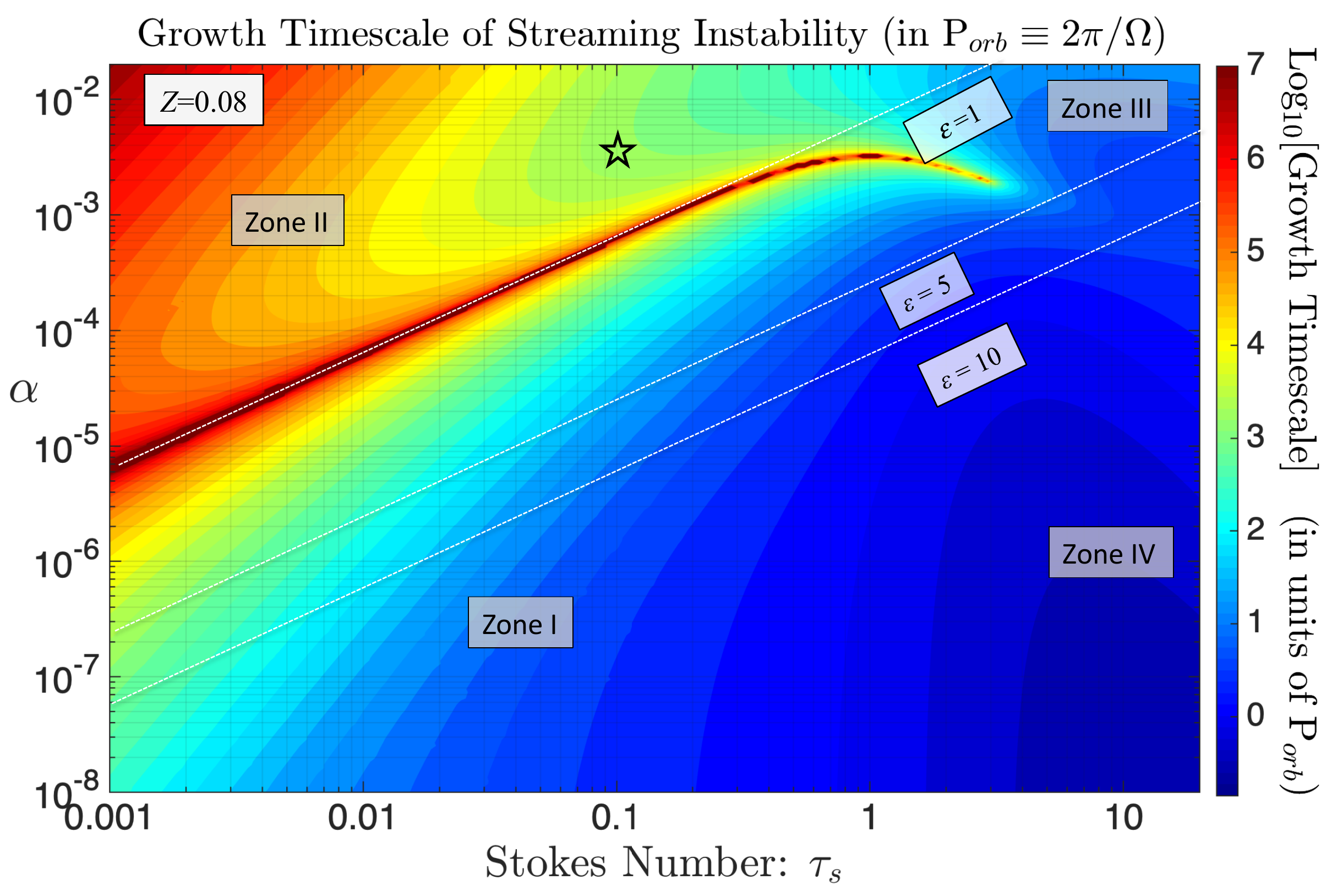}
\end{center}
\caption{Like Fig. \ref{Growth_Timescales_f01} except different values of metallicity are shown
: (top left) $Z=0.02$,  (top right) $Z=0.03$, (bottom left) $Z=0.04$, (bottom right) $Z=0.08$. 
Simulations where the SI was observed indicated by filled symbols:   7-sided star: \citet{Li_etal_2018},
circles(squares): 2D(3D) simulations \citet{Yang_etal_2017}, crosses: \citet{Gerbig_etal_2020}.  
Simulations where SI was not observed indicated by open symbols: 5-sided star: iMHD simulations of \citet{Yang_etal_2018}.
SI is only observed in numerical simulations of small-St particles 
in Zone I.  The filled circle with red outline for $Z=0.04$ corresponds to the early phase of the corresponding
simulation of \citet{Yang_etal_2017}, see also Table \ref{Yang_etal_2017_simulations}.
}
\label{Growth_Timescales_Moar}
\end{figure}


\subsection{Growth: general survey}\label{general_survey_growth_rates}
Figure \ref{Growth_Timescales_f01} shows the growth timescale of the fastest growing mode, as a function of $\tau_s$ and $\alpha$
for $Z= 0.01$.  There are several notable results.  There exists a critical branch line, defined nearly
by $\epsilon = 1$, in which the growth timescales are infinite.  This critical curve extends from $\tau_s = 0$ and terminates at a critical value of the Stokes parameter which we denote by $\tau_c$.  For the parameter combination considered here ($Z=0.01$, $\delta = 0.05$), $\tau_c \approx 0.45$ at $\alpha = \alpha_{c} \approx 3.3\times 10^{-5}$.  
{ {Although not perceptible for the $Z=0.01$ case shown in Fig. \ref{Growth_Timescales_f01}, the actual location of $\tau_c$ corresponds to a
value of $\epsilon \approx 1.1$ { {based on the turbulent dilution model}}.  Similar growth timescale plots, calculated for several
different values of $Z$, are shown in Figure
\ref{Growth_Timescales_Moar}.  The corresponding value of $\tau_c$ more clearly corresponds to increasingly larger values
of $\epsilon$ as $Z$ increases.  The critical line appears to hug the $\epsilon = 1$ line until $\tau_s$ begins to approach
$\tau_c$ from below, whereupon the curve bends downward in $\alpha$ forming a beak-like shape (this is most starkly apparent for $Z=0.08$ in Fig. \ref{Growth_Timescales_Moar}).  This critical point always corresponds to values of $\epsilon$ larger than 1.
We have summarized the observed trends in Table \ref{Critical_Values_alpha}.
}}Generally speaking, 
$\tau_c$ is some function of $Z$ and $\delta$, but a theory clarifying the meaning of this point and its mapping as a function of these and other parameters is not undertaken here. \par

\begin{table}[ht!]
\centering
\caption{Critical values of $\alpha$ and $\tau_s$ as a function of $Z$ for $\delta = 0.05$.}
\vspace{0.1in}
\begin{tabular}{c | ccc}
\hline
\hline
 $Z$ & $\tau_c$ 
  & $\alpha_c$
  &$\epsilon_c$
 \\
 \hline
0.001$^\dagger$ &0.044 &$4.4\times 10^{-8}$ & $\approx 1$	\\
0.010 &0.45 & $3.7\times 10^{-5}$ &  1.10	\\
0.020 &1.05 & $2.0\times 10^{-4}$  & 1.44	\\
0.030 &1.41  & $4.3\times 10^{-4}$  & 1.72	\\
0.040 & 1.71 & $7.1\times 10^{-4}$  & 1.96	\\
0.080 & 3.50 & $2.5\times 10^{-3}$ & 3.74	\\
\hline
\label{Critical_Values_alpha} 
\end{tabular}
\par
{\small{
$^\dagger$ General growth timescales not shown for this parameter value.}}
\\
\end{table}

The character of the turbulent SI is sensitive to whether or not  $\tau_s<\tau_c$ or $\tau_s>\tau_c$.  
In the case where $\tau_s<\tau_c$, the growth rate dramatically depends 
on which side of the branch line one is on.  
For the region below the branch line (i.e., for $\epsilon > 1$) -- the so-called laminar zone (Zone I) --
 the growth timescales are generally fairly short (less than orbit times) 
in broad accordance
with the low turbulence results depicted in the top panel of 
Fig. \ref{Growth_rates_kxkz_1} as well as in line with expectations based on
published numerical and theoretical studies examining the SI in the inviscid limit 
{ {(also see introductory discussion of section 6)}}.
\par
On the other hand, for $\tau_s<\tau_c$ and above the branch line (i.e., $\epsilon < 1$), in the so-called turbulent regime  (Zone II), 
the growth timescales are extremely long - anywhere from tens to thousands of local orbit times.  For optically thick disks in which $\tau_s = 0.01$ and where the VSI is operating $\alpha \sim 2 \times 10^{-4}$ \citep{Estrada_etal_2016, Malygin_etal_2017}, the growth timescales are just under $10^4$ orbit times (at Jupiter this corresponds to about $10^5$ years). Approaching the $\epsilon = 1$ line from either side of the branch line results in growth timescales approaching infinity.  In other words, at $\epsilon = 1$ the mode is marginal, neither growing nor decaying.  
\par
For $\tau_s>\tau_c$ the fate of the linear SI is different.  The branch line $\epsilon = 1$ ceases to have any consequence for the growth rates.  In this region the growth rates are relatively fast, anywhere between tenths and tens of orbit times.  There appears to be a boundary that separates the turbulent zone from the laminar zone (in this range around $\alpha \sim 10^{-4}$) but this is apparent only when looking at the neutral pattern propagation speed line (discussed further below).  For $\alpha > 10^{-4}$ (Zone III) the growth timescales are about tens of orbit times while for $\alpha < 10^{-4}$ (Zone IV) the growth timescales are even 10-100 times shorter.  Zone III is further distinguished from Zone IV in the character of the pattern speeds and propagation directions (next section).  Finally,
in a broad sense, we identify Zones II-III as comprising the ``turbulent regime" since they embody regions of relatively large values of $\alpha$, while we refer to Zones I,IV as comprising the ``laminar regime".
\par
 
We surveyed our results to test whether or not the incompressibility assumption remains valid.  For a given maximally growing mode characterized by wavelength $\lambda$ we find that both the growth timescale and pattern propagation timescale (across radial distance $\lambda$) are always much longer than the sound propagation time across the same lengthscale, consistent with the physical basis for neglecting compressibility in the gaseous phase.

 \begin{figure*}
\begin{center}
\leavevmode
\includegraphics[width=17.5cm]{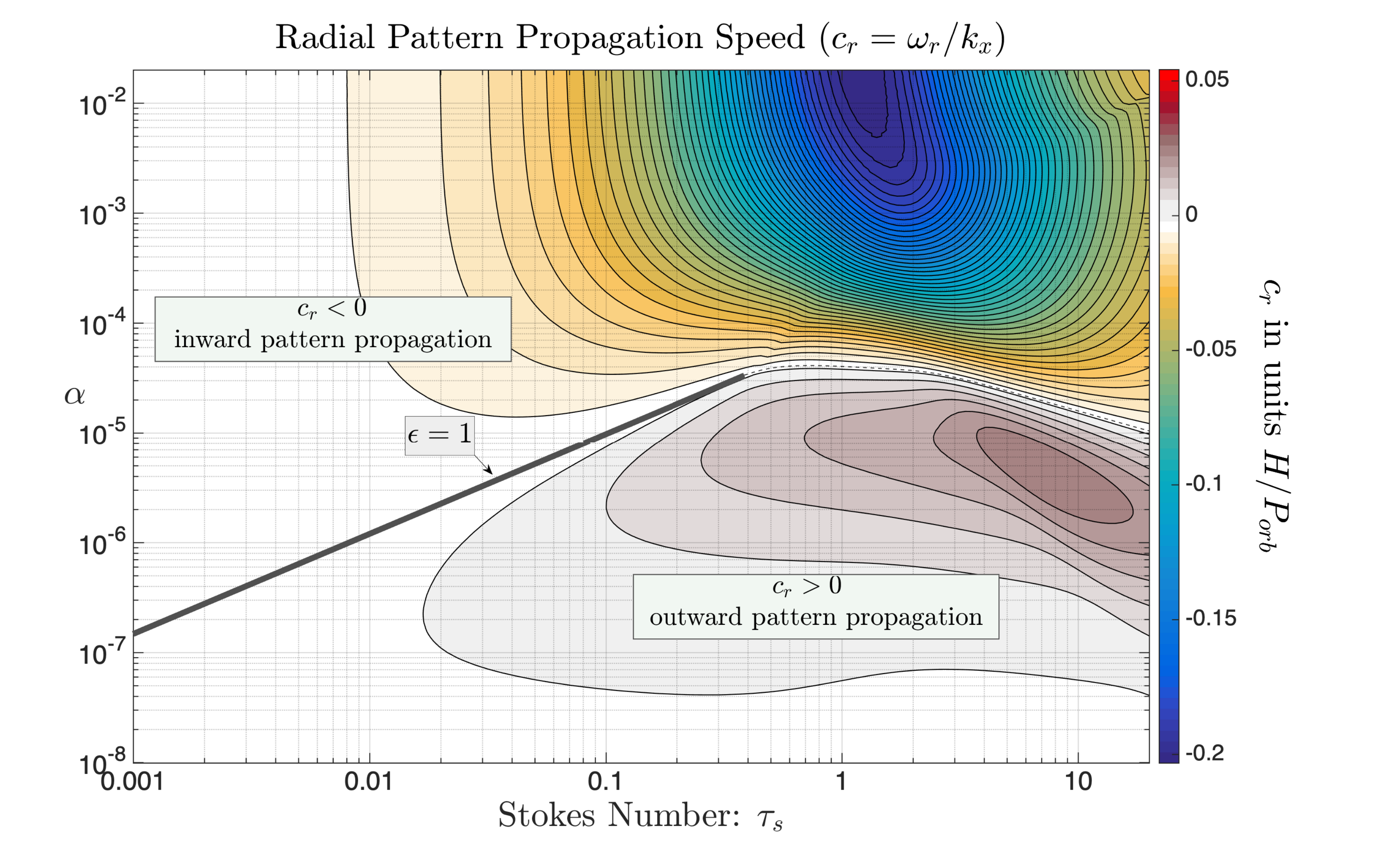}
\end{center}
\caption{Like Figure \ref{Growth_Timescales_f01} except pattern speed of fastest
growing mode shown.  Regions separating inward and outward propagation correlate with
our nominally defined turbulent and laminar regimes.  The thick and thin-dashed black lines designate locations where $c_r = 0$ in which the thick line coincides with the critical line $\epsilon =1$ subject to $\tau_s < \tau_c$ and $\alpha < \alpha_{tc}$. 
}
\label{Growth_Wavespeeds_f01}
\end{figure*}

\subsection{Pattern Speeds -- general survey}\label{general_survey_pattern_speeds}

 Figure \ref{Growth_Wavespeeds_f01} depicts the pattern propagation speed of the fastest growing
modes for the same parameter sweep discussed above.  
{ {We restrict our attention to $Z=0.01$ noting that
the qualitative character we report here carries over to the other values of $Z$.}}
We immediately observe that the pattern speed is 0 along the branch line
($\epsilon = 1$)
 separating Zone I from II.  This is consistent with predictions made for the inviscid limit (YG2005)
 wherein the pattern speed is zero for $\epsilon = 1$.  However, beyond the critical point $\tau_c$, the 0 pattern speed curve does not lie on the $\epsilon = 1$ line but, instead, follows the curve designating
 0 pattern speed which roughly separates Zone III and IV -- see dashed line of Fig. \ref{Growth_Wavespeeds_f01}.
 \par
 { {In analyzing the inviscid SI, YG2005 predict that its pattern speed depends on whether or not $\epsilon > 1$. We find that those trends carry over to this turbulent model}}:  the pattern speeds are outward ($c_r > 0$) in laminar Zone I ($\epsilon > 1$) while they are inward ($c_r < 0$) for turbulent Zone II.  Pattern speeds in this regime
 are typically less than $0.01 H/P_{{\rm orb}}$.   A stark qualitative distinction appears in examining $c_r$ in Zones III and IV.  Within the relatively turbulent
  Zone III $c_r < 0$; that is, inward, just as in Zone II.  However the pattern speeds are very high, on the order 0.1-0.2 $H/P_{{\rm orb}}$.  This high drift rate was already observed in our earlier analysis (sec. \ref{individual_mode_results}) of 
  \citet{Johansen_etal_2007}'s simulations. Meanwhile, in Zone IV the outwardly propagating patterns
  also drift with relatively high speeds (~$<0.05 H/P_{{\rm orb}}$) that are still, however, slower than those of Zone III.
  In either case, such high pattern speeds means that such structures might drift out of the region of interest on relatively short timescales, before they are able to produce planetesimals-- see further discussion in sec. \ref{realism_pp-disk_models}.2.


\subsection{Mode structure -- general survey}
{ {Similar to the previous sub-section, we restrict our discussion to $Z=0.01$ noting that
the qualitative character we report here carries over to the other values of $Z$.}}
In Figures \ref{radial_length_scale_fastest_growing_mode} and \ref{aspect_ratio_fastest_growing_mode}
we plot the structure of the fastest growing modes, where Fig. \ref{radial_length_scale_fastest_growing_mode}
shows its radial wavelength $\lambda_x\equiv 2\pi/k_x({{\rm max}})$ while 
Fig. \ref{aspect_ratio_fastest_growing_mode} displays its aspect
ratio defined to be $\lambda_z/\lambda_x \equiv k_x({{\rm max}})/k_z({{\rm max}})$.
\par
In accordance with the trends predicted in the inviscid limit, within the weakly turbulent
regions (Zones I,IV) the radial wavelengths of the fastest growing modes are on the order of $H$
and get increasingly shorter as $\alpha$ decreases.  In Fig. \ref{aspect_ratio_fastest_growing_mode} we indicate the location where $\lambda_z/\lambda_x \approx 1$; this lies mostly in Zone IV with some
spillover into Zone I.  Scanning around this region we see that $\lambda_z/\lambda_x$ indeed remains $\order{1}$ implying that the mode structure is that of axially symmetric ring structures in this rough patch of parameter space.
\par
On the other hand, in Zone II, above the branch line $\epsilon = 1$, $\lambda_x$ steadily increases with $\alpha$.  This trend continues into Zone III where the turbulence
is still relatively strong but $\tau_s > \tau_c$.  Moreover, the aspect ratio rapidly goes from flattened or tubular rings into vertically oriented sheets as $\lambda_z/\lambda_x$ steadily grows with increasing turbulent intensity.  In both Figures \ref{radial_length_scale_fastest_growing_mode} and \ref{aspect_ratio_fastest_growing_mode}, we have indicated regions for which $\lambda_z/\lambda_x > 5$; this property represents the entirety of Zones II and III.  Naturally, caution should be exercised in interpreting the results in the limit that $\lambda_z$ greatly exceeds several scale heights. { {That is, when the predicted vertical lengthscales greatly exceed $H$ throughout the majority of Zone II, the growth timescales indicated throughout Zone II of Figures \ref{Growth_Timescales_f01}-\ref{Growth_Timescales_Moar}
may be lower bounds because there may be no expression of the SI for this parameter regime.
}}
{ {It is, however, encouraging that the trends predicted here appear
to be consistent with the simulation results reported in \citet{Balsara_etal_2009}. In their Figure 7, for particle sizes
in which the SI appears to manifest, the vertical structure in the particle density appears uniform up to the scale height of the particles themselves.}}
\par



\par
\bigskip

 \begin{figure*}
\begin{center}
\leavevmode
\includegraphics[width=14.75cm]{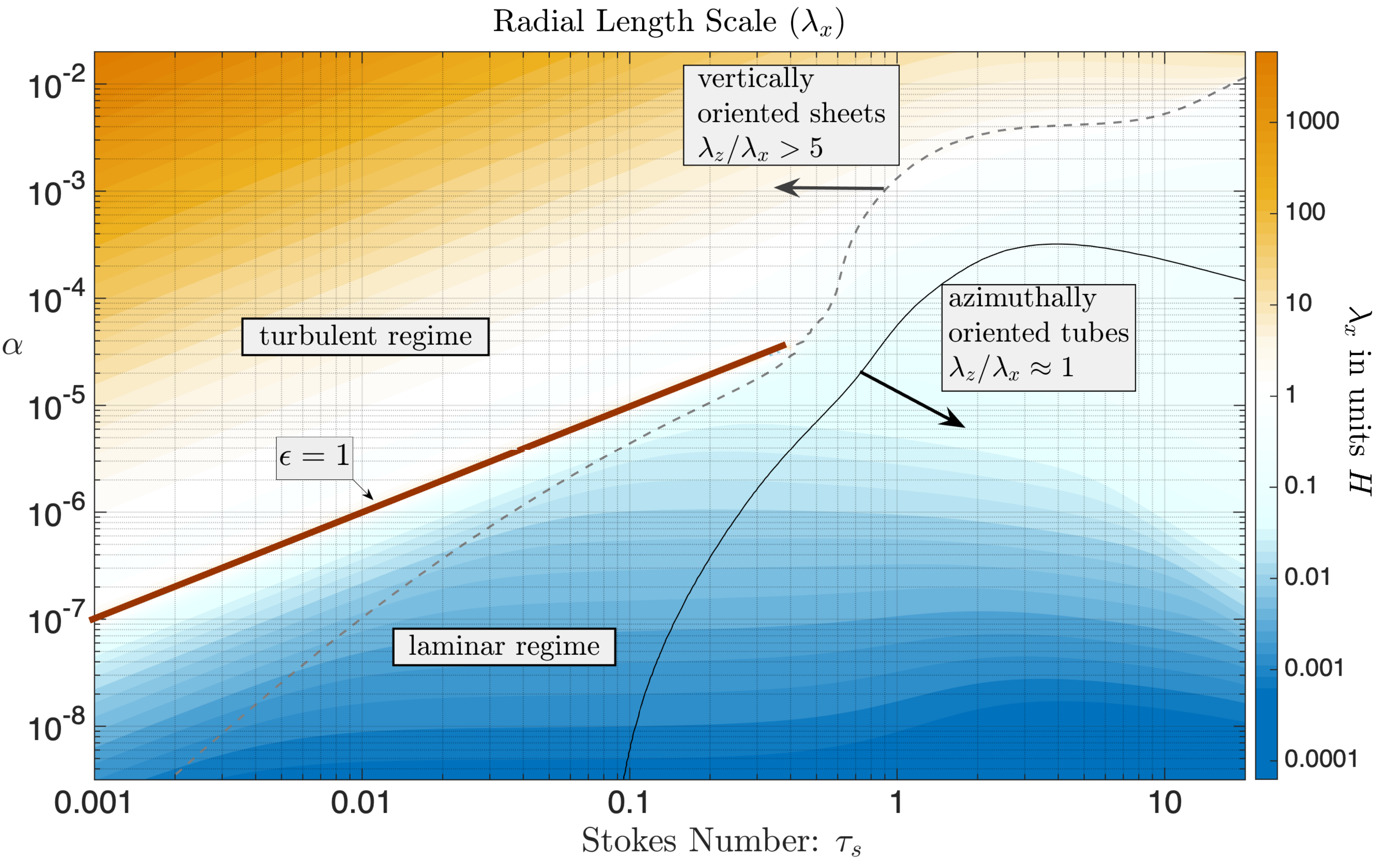}
\end{center}
\caption{Radial length scales $\lambda_x$ of fastest growing modes for parameter
values shown in Fig. \ref{Growth_Timescales_f01}.    
Vertically
oriented sheets (high aspect ratios) 
are identified for regions in parameter space where $\lambda_z/\lambda_x > 5$
(see also next figure).  
Azimuthally oriented tubular structures are identified for regions in which
$\lambda_z/\lambda_x \approx  1$ and flattened structures for $\lambda_z/\lambda_x < 1$.  For $\tau_s < \tau_ c$, note that for turbulent regimes 
of protoplanetary disk interest ($10^{-5}<\alpha < 10^{-3}$) the growing modes are 
vertically oriented sheets with radial scales
of about a scale height.}
\label{radial_length_scale_fastest_growing_mode}
\end{figure*}

 \begin{figure*}
\begin{center}
\leavevmode
\includegraphics[width=14.75cm]{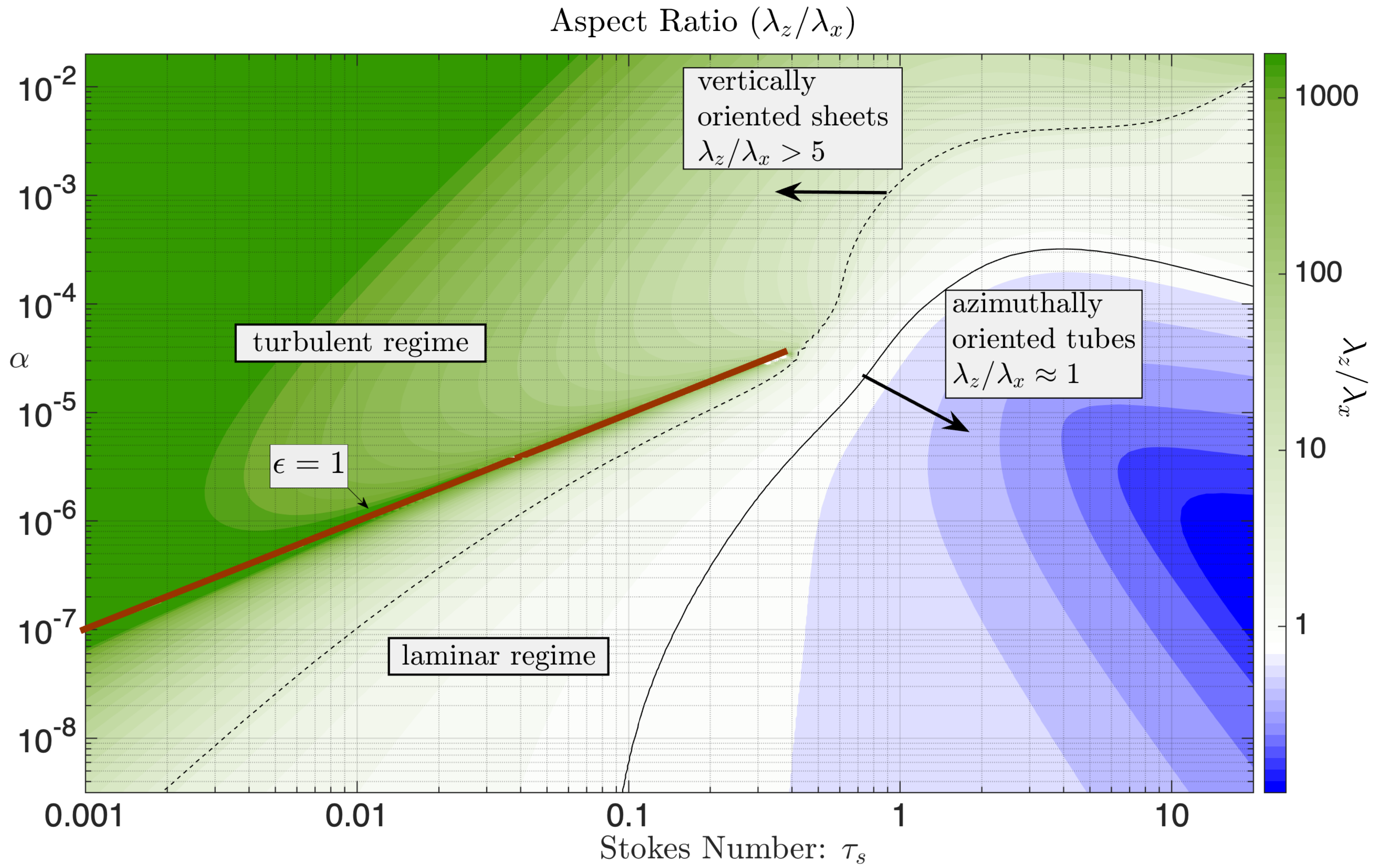}
\end{center}
\caption{Like Fig. \ref{radial_length_scale_fastest_growing_mode} 
but showing aspect ratios, $\lambda_z/\lambda_x$.}
\label{aspect_ratio_fastest_growing_mode}
\end{figure*}

\section{Comparison to some recent published studies}\label{Comparison_to_other_studies}
It is revealing to compare the predictions of the theory with published numerical simulations where SI is (filled symbols), or is not (open symbols), seen to operate. The values of $(\tau_s, \alpha)$ for  the turbulent runs are given explicitly in the papers cited \citep{Johansen_etal_2007, Balsara_etal_2009} and we have indicated their approximate
locations on the growth timescale plot Fig. \ref{Growth_Timescales_f01}.  In particular, for these simulations
(with $\delta = 0.05, Z = 0.01$)  SI is only manifest
squarely inside of Zone III, for relatively large $\tau_s$.  We note that these simulations considered several particle species with differing
 Stokes numbers while the growth rates shown in the Figure correspond to the linear theory inferred from
 assuming a single size species { {into which all mass is assigned}}.  We consider our theoretical predictions of growth rates to be upper bounds for simulations
 conducted with several species simultaneously present, based on the recent findings of \citet{Krapp_etal_2019}.
 { {In the following we examine several recent studies that provide enough information for us to compare their results with our theoretical predictions.}}
\par
{ {
\subsection{\citet{Yang_etal_2017}}
The numerical experiments run by \citet{Yang_etal_2017} are more nuanced.  They considered a single particle species
of a given Stokes number initiated in a globally nonturbulent nebula - effectively laminar flow.\footnote{The simulations of \citet{Bai_Stone_2010} are similarly set up, however we do not analyze them here because no spacetime plots of vertically
integrated density were provided.  A follow-up study examining these in more detail is warranted.}
They initialize the particles with a Gaussian vertical distribution symmetric about the midplane with a corresponding effective scale height $H_p = 0.2H$ and allow the particles to settle toward the midplane. In these simulations and others like it 
\citep{Bai_Stone_2010} the layer collapses until the particle scale height reaches a minimum ``bounce" value, out of which
the SI begins to become manifest.  
{ { We conjecture that this first bounce corresponds to the rapid transition of the fluid into an unsteady
(possibly turbulent) state. Whether or not the dynamically active fluid state during this first bounce is driven by
the SI or some other mechanism like the Kelvin-Helmholtz
instability remains to be established. \citet{Bai_Stone_2010} state that the long-time development of the unsteady particle layer under their conditions is not maintained by Kelvin-Helmholtz overturn but, rather, by the SI itself. 
 However, in a footnote, they allow that this would {\it not} be the case for smaller St.  
{ {Indeed, the recent study of \citet{Gerbig_etal_2020} shows that whether SI or the Kelvin-Helmholtz instability maintain the vertical diffusion of particles
depends upon the value of St in concert with $\delta$, with the latter process being important for values of $Z$ in the vicinity
of $0.01$ for $\delta = 0.05$, or toward higher values of $Z$ as $\delta$ increases .}}\footnote{
{ {This study was released during the revision phase of this article.  Also, the streaming parameter $\delta$ is $2\beta$ in their study.}}}
Deeper understanding of this so-called early bounce phase requires further analysis.}}
In either case, { {we conjecture that}} it is out of an { {unsteady-possibly-turbulent}} state that the subsequent SI develops in these simulations, and it may in turn drive yet another component of turbulence. 
We do not think these candidate mechanisms can be, nor need to be, distinguished at this stage. 
{ {To keep our terminology as generic and descriptive as possible, we refer to the fluid state during this early
bounce phase as ``midplane turbulence".}}

\par  
In light of the preceding discussion, for all the simulations presented \citet{Yang_etal_2017}
 record the characteristic value of $H_p/H$ at every time step.  In practice, the particles settle toward
a nominal minimum base state  value of $\left(H_p/H\right)_{{\rm s}}$ (the aforementioned ``bounce") representing the first instance
of balance between turbulent stirring and midplane directed settling,
and it is from this base state that the SI begins to emerge.  For values of $\tau_s \approx 0.01$ this state appears
to be reached by 20P$_{{\rm orb}}$  while for $\tau_s \approx 0.001$ the base state is reached at more like 80P$_{{\rm orb}}$.
}}
\par
{ {
{{ {
Thus, to place the ``detected SI" results of \citet{Yang_etal_2017} onto the theoretical growth timescale plots of Figs.
\ref{Growth_Timescales_f01}-\ref{Growth_Timescales_Moar},
we had to infer an effective value of 
the $\alpha$-parameter, henceforth $\alpha_{{\rm eff}}$, characterizing the near-midplane turbulence. }}}
Following the relationship between $H_p$ and $H$ found
after Eq. (\ref{turbulent_dilution_model}) defining the TDM, under the assumption $\tau_s \gg \alpha_{{\rm eff}}$, we approximate $\alpha_{{\rm eff}} \approx \tau_s \left({H_p}/{H}\right)_{{\rm s}}^2$,
yielding a corresponding $\epsilon_{{\rm eff}} \approx Z\sqrt{\tau_s/\alpha_{{\rm eff}}}$.  This procedure
works well for all of the simulations presented in \citet{Yang_etal_2017} that present
a time-series for ${H_p}/{H}$.  In the two cases where this information is not provided, we estimate
an average midplane value of $\epsilon_{{\rm est}} = \rho_p/\rho_g$ by matching the color of the quantity
against the provided color bar.
Equating $\epsilon_{{\rm eff}}$ to $\epsilon_{{\rm est}}$,
and subsequently inserting it into the rewritten TDM, leads to
$\alpha_{{\rm eff}} = \tau_s Z^2/\big(\epsilon_{{\rm eff}})^2$.
We have indicated the parameter locations of most of the simulations reported in \citet{Yang_etal_2017}
throughout Figs. \ref{Growth_Timescales_f01}-\ref{Growth_Timescales_Moar}.

\par
With $\alpha_{{\rm eff}}$ we computed the e-folding timescale of the fastest growing mode, $t_{gm}$, and its 
associated radial wavelength $\lambda_{_{xm}}\equiv \lambda_x({{\rm max}})$.
The {\it{observed}} e-folding growth timescales, $t_o$, 
in these simulations are difficult to assess based on the graphs provided. Thus we 
estimate $t_o$ to be less than the observed saturation timescales (hereafter, $t_{{\rm sat}}$) 
found
tabulated in Table 2 of \citet{Yang_etal_2017}.
We visually determined an effective lengthscale of structures observed to emerge during the linear phase
(hereafter, $\lambda_{_{x,{{\rm obs}}}}$) according to the following procedure: Figures 3 and 7 of \citet{Yang_etal_2017} 
show spacetime plots for the azimuthally averaged, vertically integrated, scaled particle density which
they denote by $\big<\Sigma_p\big>/\Sigma_{g,0}$.  Approximately after the time particles have settled near the midplane 
($t=20$P$_{{\rm orb}}$  or
$t=100$P$_{{\rm orb}}$ depending upon $\tau_s$, see above) but well before {$t_{{\rm sat}}$}, we count the number ($N_p$) of peaks in
$\big<\Sigma_p\big>/\Sigma_{g,0}$, across the radial size of the simulation domain, $L_x$, 
and then we estimate $\lambda_{_{x,{{\rm obs}}}} \approx L_x/N_p$.  In practice, we focused on counting resolvable peaks
within the first two predicted e-folding growth timescales because once coherent filamentary structures emerge they appear
to nonlinearly interact, resulting in a series of mergers before the 
saturated state becomes manifest.  In some cases,
it was difficult to resolve an unambiguous number of peaks and this uncertainty was noted.
In all cases we counted peaks as soon as the structures appeared coherent to the eye. }}
\par
{ {
The results of this activity are summarized in Table \ref{Yang_etal_2017_simulations}. 
Despite the crudeness of this approximate analysis we find that our theory predicts
the general properties reported in \citet{Yang_etal_2017}.
In particular, it is clear for those Yang et al (2017) initial conditions which did manifest SI, that the 
effective $\epsilon_{{\rm eff}}$'s were always greater than 1, consistent with our
predictions that growth is relatively strong for such conditions.  In the two cases discussed
by \citet{Yang_etal_2017} in which the SI was not observed, we predict that the 
growth times for those input parameter conditions are much longer than the time for which those simulations were conducted.
Of interest is the case $\tau_s = 0.001, Z=0.03$ where no structures were observed to emerge up to simulation
time t=5000P$_{{\rm orb}}$.
According to our predictions, if the simulation were run past $t=5\times 10^4$P$_{{\rm orb}}$ then the SI should become apparent.
Similar reasoning also applies to their $\tau_s = 0.01, Z=0.01$ simulation.  We also note that
their 2D simulation for the case $\tau_s = 0.01, Z=0.04$ for the box $\sim 0.4 H \times 0.4 H$ experiences a brief
dip in $H_p/H$ to $\approx 0.004$ before leveling out to $H_p/H \approx 0.01$ at $t\approx 80$P$_{{\rm orb}}$
(see red dotted time series shown in middle panel row of their Fig. 2 as well as right panel of their Fig. 3b).
During this initial bounce phase ($t<50$P$_{{\rm orb}}$) the number of filamentary structures is impossible to assess.  As such,
we examined the behavior and character of this simulation after the initial response phase passed (after $t=50$P$_{{\rm orb}}$), {when the number of filaments was first countable.}  For the corresponding
value of $H_p/H$ ($\approx 0.01$), the predicted $\lambda_{xm}$ better matches the corresponding measured 
$\lambda_{_{x,{{\rm obs}}}}$.
}}

\begin{table}[ht!] 

\caption{Simulations of \citet{Yang_etal_2017} compared to theoretical predictions}
\vspace{0.1in}
\begin{tabular}{c | c c c c  c c | c c | c c | c}
\hline
\hline
 $\displaystyle 
 \begin{array}{c}
 {\rm run} \\
 {\rm type}
 \end{array}
 $
 & $\delta$
 & $\tau_s$ 
 & $Z$ 
 &   $\displaystyle \left(\frac{H_p}{H}\right)^{\dagger}$ 
 &  $\displaystyle \frac{\alpha_{\rm eff}}{10^{-7}}$
 & $\epsilon_{{\rm eff}}$ 
 & $\displaystyle \frac{t_{_{gm}}}{{\rm P}_{{\rm orb}}}$ 
  & $\displaystyle \frac{t_{_{\rm o}}}{{\rm P}_{{\rm orb}}}^\ddagger$ 
   & $\displaystyle \frac{\lambda_{_{xm}}}{H}$ 
  & $\displaystyle \frac{\lambda_{_{x,{\rm {obs}}}}}{H}$ 
  & Identifiers$^\sharp$
 \\
 \hline
2D & 0.05 & 0.01 &  0.01 & 	--- & 	13.8$^a$  & $\sim 0.90^{b}$ 	& 11000	 & ---	& 1.2	 & --- & $\left(0.2H\right)^2,\ 2560 H^{-1}$ 	\\
 & 0.05&0.01 & 	0.02 & 	0.012   & 	14.5  &  1.66	& 	166  & $<200$ & 0.13	 & $\sim$ 0.10-0.20	& $\left(0.2H\right)^2,\ 2560 H^{-1}$	\\
  & 0.05&0.01 & 	0.02 & 	0.011   & 	12.1  &  1.81	& 	106  & $<100$ & 0.10	 & 0.08	& $\left(0.4H\right)^2,\ 2560 H^{-1}$	\\
 & 0.05&0.01 & 	0.04 & 	0.010   & 	10.0  &  4.0	& 	65 & $<50$ & 0.06 & 0.05	& $\left(0.2H\right)^2,\ 1280 H^{-1}$	\\
  & 0.05&0.01 & 	0.04 & 	0.0035   & 	1.27  &  11.2	& 	16 & $<50$ & 0.01 & n/a	$^c$& $\left(0.4H\right)^2,\ 1280 H^{-1}$\\
  & &    & 	     & 	 0.010$^d$   & 	10.0  &  4.0	& 	65 & $ $ & 0.06 & 0.04	& 	\\
 & 0.05&0.001 & 0.03 & 	--- & 	5.4$^a$  &  $\sim 1.3^b$	& 	39000  & --- & 10	 & --- 	& 	$\left(0.2H\right)^2,\ 1280 H^{-1}$\\
 & 0.05&0.001 & 0.04 & 	0.016   & 	2.25  &  2.66	& 1850	 & $<1200$ & 0.14	 & 0.10-0.20	& $\left(0.2H\right)^2,\ 1280 H^{-1}$	\\
  & 0.05&0.001 & 0.04 & 	0.015   & 	2.56  &  2.5	& 2300	 & $<2200$ & 0.17	 & 0.10-0.20	& $\left(0.4H\right)^2,\ 1280 H^{-1}$	\\
\hline
3D & 0.05&0.01 & 0.02 & 	0.010 & 	10.0  &  2.0	& 	71  & $<250$ & 0.08	 & 0.067$^e$	& $\left(0.2H\right)^3,\ 640 H^{-1}$	\\
 & 0.05&0.001 & 0.04 & 	0.011   & 	2.0  &  3.6	& 	800  & $<650$ & 	0.07 & $\sim 0.067^f$	& $\left(0.2H\right)^3,\ 640 H^{-1}$	\\
\hline
\label{Yang_etal_2017_simulations}
\end{tabular}
\par
{\small{
$^\dagger$Estimate, based on the moment the particle layer has settled but before SI grows (see text).\\
$^\ddagger$Based on the lesser of either visual identification procedure or the quoted saturation time $t_{{\rm sat}}$ from their Table 2.\\
$^\sharp$Simulation box size and resolution used in terms of grid points per scale height.\\
$^a$Based on $\epsilon_{{\rm est}}$ since no $H_p/H$ timeseries was provided for this case.\\
$^b$Based on extracting a midplane averaged  value of $\rho_p/\rho_g$ estimated from color bar for this particular run (see text). \\
$^c$No discernible wavelength structure observed in the time range ($t<50$P), perhaps because of insufficient simulation resolution.\\
$^d$Based on $t=100$P, the time after the observed initial adjustment bounce.  Input parameters are the same as above row.  See text.\\
$^e$Assessed at $t=100$P, see also discussion in section 4.2 of \citet{Yang_etal_2017}.\\
$^f$Assessed at $t=100$P before the first filament merger occurs.
}}
\\
\end{table}

\par
\subsection{\citet{Li_etal_2018}}
{ {\citet{Li_etal_2018} examine the fate of the SI under similar circumstances and initial setups
considered in \citet{Yang_etal_2014} but using three different boundary conditions -- periodic, reflecting and outflow -- 
and they show that the {nonlinear} SI state is largely insensitive to the boundary conditions employed.
Using a fixed grid resolution ($\sim H/320 \approx 0.003 H$) \citet{Li_etal_2018} checked for the emergence
of filaments for three different box sizes: $(0.2 H)^3, (0.4 H)^3, (0.8 H)^3$.  Their simulations were run with
$\tau_s = 0.314$ and they quote time series quantities similar to $H_p/H$.    During the early linear ``first-bounce" phase, all simulations collapsed into a layer with $H_p/H \sim 0.001$
within $t=1$P$_{{\rm orb}}$.  
The predicted maximum growth timescale for these input parameters are short, i.e., $t_{gm} \approx 0.5$P$_{{\rm orb}}$.
This layer then begins to develop the SI just as in the simulations reported in both \citet{Yang_etal_2014} and \citet{Yang_etal_2017}.   Coherent filaments undergoing epicyclic oscillations are clearly discernible by $t=3$P$_{{\rm orb}}$ and it is also at this
point they undergo nonlinear merging.
\par
We perform the same crude analysis as above
based on the periodic boundary condition runs \citep[see Fig. 10 of][]{Li_etal_2018} and the results
are summarized in Table \ref{Li_etal_2018_simulations}. We estimated $\lambda_{_{x,{{\rm obs}}}}$ by counting the number of filaments around $t=2$P$_{{\rm orb}}$.  We choose
this time because filaments are not easily identifiable at any time before this point, while for times after this point the filaments
begin their process of merging.  For all simulations run,  we find  
$\lambda_{_{x,{{\rm obs}}}} \approx 0.015 H$, while our theory  predicts a value approximately half of that, {$\lambda_{xm} \approx 0.007 H$}.
This means that the observed fastest growing mode encompasses about 5 grid points while
our predicted value would not be 
resolvable (at between 2 and 3 of their gridpoints).  Indeed, we note that their measured value for $H_p/H$ at
$t\approx 1$P$_{{\rm orb}}$ is much smaller than their grid resolution. We conjecture that their simulations are not sufficiently resolved 
and should be run for at least 2-3 times the resolution originally used.  Convergence would perhaps be indicated
if during this early organization phase ($t<1$P$_{{\rm orb}}$) the minimum value of $H_p/H$ remains fixed with increasing resolution.
What is not yet clear is how the response of the particles, and the turbulence they churn up as they approach the midplane,
depend upon resolution as such a systematic study remains to be done.
}}
\par

\begin{table}[ht!]

\begin{center}
\caption{Selected simulations of \citet{Li_etal_2018} compared to theoretical predictions.  Periodic boundary condition runs
shown.  Table heading are defined as in Table \ref{Yang_etal_2017_simulations}.}
\vspace{0.1in}
\begin{tabular}{c |c c c c  c c | c c | c c | c}
\toprule
\hline
 $\displaystyle 
 \begin{array}{c}
 {\rm run} \\
 {\rm type}
 \end{array}
 $
  & $\delta$
 & $\tau_s$ 
 & $Z$ 
 &   $\displaystyle \left(\frac{H_p}{H}\right)$ 
 &  $\displaystyle \frac{\alpha_{\rm eff}}{10^{-7}}$
 & $\epsilon_{{\rm eff}}$ 
 & $\displaystyle \frac{t_{_{gm}}}{{\rm P}_{{\rm orb}}}$ 
  & $\displaystyle \frac{t_{_{\rm o}}}{{\rm P}_{{\rm orb}}}^\oplus$ 
   & $\displaystyle \frac{\lambda_{_{xm}}}{H}$ 
  & $\displaystyle \frac{\lambda_{_{x,{{\rm obs}}}}}{H}^\ddagger$ 
  & Identifiers
 \\
 \hline
3D&0.05  &0.314 & 0.02 & 0.001	 & 	3.14  &  20.0	& 	0.5  & $\sim 5$ &    0.007 & $\sim 0.015$	& $\left(0.2H\right)^3,\ 320 H^{-1}$	\\
 &0.05 &0.314 & 0.02 & 	0.001   & 	3.14  &  20.0	& 	0.5  & $\sim 5$ & 	0.007 & $\sim 0.016$	 & $\left(0.4H\right)^3,\ 320 H^{-1}$	\\
  &0.05 &0.314 & 0.02 & 	0.001  & 	3.14  &  20.0	& 	0.5  & $\sim 5$ & 	0.007 & $\sim 0.016$	 & $\left(0.8H\right)^3,\ 320 H^{-1}$	
  \\
  \hline
\bottomrule
\label{Li_etal_2018_simulations} 
\end{tabular}
\end{center}
\par
{\small{
$^\oplus$ based on the first clear appearance of coherent structures.\\
$^\ddagger$ Peaks were counted at approximately t=3P, the point at which coherence was discernible to the human eye.
}}
\\
\end{table}

\subsection{\citet{Gerbig_etal_2020}}
{ {
\citet{Gerbig_etal_2020} consider the fate of SI under setups and conditions similar to those of the previous two studies.  One of their main aims is
to disentangle to what degree the midplane turbulence churned up by the settling dust particles is driven either by
the Kelvin-Helmoltz instability or the SI, and they seek to shed light on this outcome as a function of the dust streaming
parameter, $\delta$ (which is equal to $\beta/2$ in their analysis), $Z$, and for a fixed value of $\tau_s = 0.2$.
The results they uncover are subtle but it is evident from their reported simulations that when SI is not present, the dust layer
thickens consistently with what one might predict based on the Richardson criterion for Kelvin-Helmholtz overturn under the influence of gravity.  They provide the results for a suite of runs in which they quote a measured dust layer thickness plus its standard deviation, together with snapshots of the particle accumulation
as viewed from several perspectives.  For one run, $Z=0.02, \tau_s = 0.2, \delta = 0.05$, they also provide
a spacetime (radius-time) plot of the vertically integrated , azimuthally averaged particle density. 
Focusing on only a subset of simulations reported in their study -- all before self-gravity is turned on --
we compare their measured and tabulated quantities against predictions made using our theory.  
As per the procedure defined earlier, we estimate an $\alpha_{{\rm eff}}$ based on their reported value
of $H_p$.  We also apply our theory to corresponding ``high" and ``low" values of $\alpha_{{\rm eff}}$ based
on their quoted one standard deviation values of $H_p$.
The results
of this exercise
are summarized in Table \ref{Gerbig_etal_2020_simulations}.
\par
We find reasonable agreement between our theoretical predictions and their reported study.  
The predicted versus observed wavelengths are mutually consistent with one another within one standard deviation
of their reported values of $H_p$.
Most of the filament wavelength
structure that emerges in their simulations when SI is present is shown at late times ($t= 40$P), likely well after a significant
amount of filament merging has taken place.  For this reason we suppose that the observed wavelength/average-separation
($\lambda_{_{x,{{\rm obs}}}}$) should be larger than the predicted wavelength of our theory.  Given
the error bounds on their reported $H_p$, this appears to be a plausible interpretation for the simulation
with $Z=0.03, \tau_s = 0.2, \delta = 0.05$, in which our predicted wavelength ($0.05H<\lambda_{_{xm}}<0.08H$) is smaller than the observed
value ($\lambda_{_{x,{{\rm obs}}}}=0.10H$).  We also observe that for the simulation with
$Z=0.01, \tau_s = 0.2, \delta = 0.05$, in which SI is not observed, our theory predicts that the fastest
growing wavelength should be $\lambda_{_{xm}} = 1.1H$ with a growth timescale of $\sim 70$P.  Given that their simulation
was run in a box with radial length equal to $0.4H$, we predict that if the same simulation was
rerun with a radial extent of at least 1.1$H$ or longer, then SI should appear.  A similar
prediction  can be made for their $Z=0.02, \tau_s = 0.2, \delta = 0.10$ simulation, although the radial
scale of their box should be at least twice as much, if not more (see Table \ref{Gerbig_etal_2020_simulations}).
}}

\begin{table}[ht!] 
\begin{center}
\caption{Selected simulations of \citet{Gerbig_etal_2020} compared to theoretical predictions.  Table heading are defined as in Table \ref{Yang_etal_2017_simulations}.}
\vspace{0.1in}
\begin{tabular}{c c c c  c c | c c | c c | c}
\toprule
\hline
 $\displaystyle 
 \begin{array}{c}
 \delta
 \end{array}
 $
 & $\tau_s$ 
 & $Z$ 
 &   $\displaystyle \left(\frac{H_p}{H}\right)^\triangleright$ 
 &  $\displaystyle \frac{\alpha_{\rm eff}}{10^{-7}}$
 & $\epsilon_{{\rm eff}}$ 
 & $\displaystyle \frac{t_{_{gm}}}{{\rm P}_{{\rm orb}}}$ 
  & $\displaystyle \frac{t_{_{\rm o}}}{{\rm P}_{{\rm orb}}}^\oplus$ 
   & $\displaystyle \frac{\lambda_{_{xm}}}{H}$ 
  & $\displaystyle \frac{\lambda_{_{x,{{\rm obs}}}}}{H}^\ddagger$ 
  & Identifiers
 \\
 \hline
 0.05 &0.2 & 0.0002 & 0.0020	 & 	8  &  0.10	& 	7.3  & $<40$ &    0.13 & 0.13	& $\left(0.4H\right)^3,\ 160 H^{-1}$	\\
     &    &      &  $\approx 0^\flat$& 	$\infty^\varphi$  &  ---	& 	---  &  &    --- & 	& 	\\
          &    &      & 0.0055$^\sharp$& 	60.5  &  0.036	& 	17.4  &  &    0.45 & 	& 	\\
0.05 &0.2 & 0.01 & 0.0155	 & 	481  &  0.65	& 	68.0  & --- &    1.1 & ---	& $\left(0.4H\right)^3,\ 320 H^{-1}$	\\
     &    &      & 0.0140$^\flat$& 	392  &  0.71	& 	90.0  &  &    1.1 & 	& 	\\
          &    &      & 0.0170$^\sharp$& 	578  &  0.59	& 	57.6  &  &    1.1 & 	& 	\\
0.05 &0.2 & 0.02 & 	0.0125   & 	313  &  1.60	& 	13.9  & $ <10$ & 	0.18 & 0.13$^\aleph$, 0.10$^\gamma$	 & $\left(0.4H\right)^3,\ 320 H^{-1}$	\\
     &    &      & 0.0100$^\flat$& 	200  &  2.0	& 	4.72  &  &    0.09 & 	& 	\\
     &    &      & 0.0150$^\sharp$& 	450  &  1.34	& 	64.6  &  &    0.44 & 	& 	\\
0.05  &0.2 & 0.03 & 	0.0085  & 	144  &  3.55	& 	2.77  & $<40$ & 	0.06 & $0.10$	 & $\left(0.4H\right)^3,\ 320 H^{-1}$ \\
     &    &      & 0.0070$^\flat$& 	98  &  4.28	& 	2.07  &  &    0.05 & 	& \\
     &    &      & 0.0100$^\sharp$& 200  &  3.0	& 	3.84  &  &    0.08 & 	& 	\\	
0.05  &0.2 & 0.20 & 	0.0055  & 	60  &  36.4	& 	29.1  & $<40$ & 	0.04 & $0.065$	 & $\left(0.4H\right)^3,\ 160 H^{-1}$ \\
     &    &      & 0.0030$^\flat$& 	18  &  66.6	& 	16.5  &  &    0.01 & 	& \\
     &    &      & 0.0080$^\sharp$& 128  &  25.0	& 	38.2  &  &    0.09 & 	& 	\\
  0.025  &0.2 & 0.02 & 	0.0030$^\varpi$  & 	18  &  6.66	& 	2.16  & $<40$ & 	0.018 & $0.021$	 & $\left(0.4H\right)^3,\ 320 H^{-1}$ \\
    0.10  &0.2 & 0.02 & 	0.0280  & 	1568  &  0.71	& 	90.9  & --- & 	2.15 & ---	 & $\left(0.4H\right)^3,\ 160 H^{-1}$ \\
         &    &      & 0.0245$^\flat$& 	1201  &  0.82	& 	165.0  &  &    2.31 & 	& \\
         &    &      & 0.0315$^\sharp$& 	1984  &  0.64	& 	60.45  &  &   66.5 & 	& 
  \\
  \hline
\bottomrule
\label{Gerbig_etal_2020_simulations}
\end{tabular}
\end{center}
\par
{\small{
$^\dagger$Value of $\delta$ based on value of $\beta$ used in \citet{Gerbig_etal_2020}, where $\delta = 0.5\beta$.\\
$^\triangleright$Based on values found in their Figure 6.\\
$^\oplus$Based on plots shown on either their Figure 4 or Figure 5, corresponding to t=40P.\\
$^\ddagger$Unless otherwise noted, peaks were ascertained at t=40P.\\
$^\sharp$High value corresponding to 1 standard deviation.\\
$^\varphi$No reliable minimum value of $H_p$ distinguishable from zero, therefore  no theoretical prediction calculated for this entry.\\
$^\flat$Low value corresponding to 1 standard deviation.\\
$^\aleph$Peaks counted at t=40P, based on spacetime plot of their Figure 7.\\
$^\gamma$Peaks counted at t=10P, based on spacetime plot of their Figure 7.\\
$^\varpi$Error bars on scale height practically indistinguishable from plotted point, therefore high/low values not predicted.
Straight lines indicate no structure observed.
}}
\\
\end{table}

\subsection{\citet{Yang_etal_2018}}
{ {
\citet{Yang_etal_2018} examine the response of the SI in a setup that supports two scenarios: one in which the full layer is
turbulent due to the MRI (i.e., Ideal MHD, and hereafter ``iMHD"), and the other as a Dead Zone (``DZ", hereafter) in which the midplane is Ohmic and effectively MHD inactive but where the disk gas about 1 gas scale height away from the midplane is MRI active.  They consider
a particle component with $\tau_s = 0.1$ and metallicities of $Z = 0.01,0.02,0.04$ and $0.08$.  
In the following we restrict our attention to the iMHD models.  We do not consider their DZ models because the SI physics contained
in them are probably strongly influenced by coherent structures, a feature which is outside purview of our theory -- see Appendix \ref{appendix_2} for further discussion.   
\par
{ {For the iMHD models we repeat the calculations done in the previous subsections and the results are summarized
in Table \ref{Yang_etal_2018_simulations}. We estimate
$H_p$ for each simulation based on what we were best able to surmise to be the particle collective's first bounce based
on the time series found in the first column of their Figure 9.  { {The value of $\alpha_{eff}$ is different than the value quoted in their Table 1 which is based on the state of the flow at very late stages of their simulation.}}
For all of the metallicities considered we find that the fastest growing modes have wavelengths ($\sim 30$) much larger than the
radial computational domain size ($\sim 4$H).  The predicted growth timescales are several hundred orbit times (at the very least)
which is far longer than the 100 P timescales on which their simulations were conducted. 
}}
\par
However, inspection of their results shows that
the iMHD model achieves very modest increases in particle enhancements over and above the nominal base value of either the vertically integrated
particle densities, i.e., $\Sigma_p/\Sigma_{p,0}$ where enhancements range from a factors of 2 (on average) to no more than factors of 5 in extremes (see Table 3 of their study).  The particles are prevented from settling to the midplane ($H_p/H \sim 0.3$) owing to the large vertical fluid stresses arising from the MRI turbulence with effective values of $\alpha \sim 0.008$.  The top row of Figure 10 of \citet{Yang_etal_2018} depicts a spacetime diagram showing $\Sigma_{p}$ as a function of radial position and time for the four iMHD models of differing $Z$ compared against a sample run in which there is no back-reaction of the particles back onto the fluid.  Except for the $Z=0.08$ model, inspection of the remaining figures show that there is no obvious distinction between those simulations with and without back-reaction, as there is no clear nor sustained emergent radial structure beyond the periodic box scale.  In the $Z=0.08$ model (far right, top-row panel of their Figure 10), there is a brief period (between $t/P_{orb}=20$ and $40$) in which strong accumulation appears to take place but then it eventually dissipates.  
\par
We interpret the possibility of
ephemeral bursts of accumulation to be the result of concentrations driven into place by large scale coherent
structures of the flow field of the turbulent state, or perhaps even by turbulent concentration \citep{Hartlep_etal_2020}.  For example,} the driving motions of the MRI (channel modes) and its secondary turbulent transition (parasitic/Kelvin-Helmholtz overturn) are primarily axisymmetric which means that the corresponding fluid eddies and waves  of the driving scales are azimuthally aligned -- meaning that there are coherent structures in the flow and pressure field being imposed
upon the flow by the unstable MRI activity.  Such motions may be responsible for any temporary accumulations seen in these iMHD models, {{instead of SI {\it per se}}.  Moreover, our theory is unable to treat nor predict what happens to the SI in the presence of 
unsteady yet coherent structures: the simple turbulence model like we have employed here assumes the motions are uncorrelated
(implicitly placing it within the inertial range of a turbulent flow).  See more discussion in Appendix B.
}}.

\begin{table}[ht!] 

\begin{center}
\caption{Selected simulations of \citet{Yang_etal_2018} compared to theoretical predictions.  Only ideal MHD simulations
analyzed (see text).  Table heading are defined as in Table \ref{Yang_etal_2017_simulations}.}
\vspace{0.1in}
\begin{tabular}{c  c c c  c c | c c | c c | c}
\toprule
\hline
 $\delta$
 & $\tau_s$ 
 & $Z$ 
 &   $\displaystyle \left(\frac{H_p}{H}\right)^\sharp$ 
 &  $\displaystyle {\frac{\alpha_{\rm eff}}{10^{-3}}}$
 & $\epsilon_{{\rm eff}}$ 
 & $\displaystyle \frac{t_{_{gm}}}{{\rm P}_{{\rm orb}}}$ 
  & $\displaystyle \frac{t_{_{\rm o}}}{{\rm P}_{{\rm orb}}}$ 
   & $\displaystyle \frac{\lambda_{_{xm}}}{H}$ 
  & $\displaystyle \frac{\lambda_{_{x,{{\rm obs}}}}}{H}^\flat$ 
  & Identifiers$^{\dagger}$
 \\
 \hline
0.05 &0.10 & 0.01 & 0.30$^a$	 & 	9.0  &  0.07	& 	787  & --- &    38.0 & ---	& $4H\times8H\times8H,\ 16 H^{-1}$	\\
0.05   &0.10 & 0.02 & 0.26$^b$	 & 	6.8  &  0.08	& 	642  & --- &    30.2 & ---	& $4H\times8H\times8H,\ 16 H^{-1}$	\\

0.05 &0.10 & 0.04 & 	0.26$^b$   & 	6.8  &  0.16	& 	870  & --- & 	31.7 & ---	 & $4H\times8H\times8H,\ 16 H^{-1}$	\\
0.05  &0.10 & 0.08 & 	0.20$^a$  & 	4.1  &  0.41	& 	9640  & --- & 	32.6 & ---	 & $4H\times8H\times8H,\ 16 H^{-1}$	
  \\
  \hline
\bottomrule
\label{Yang_etal_2018_simulations}
\end{tabular}
\end{center}
{\small{
$^\sharp$ { {Based on the quoted values of $H_p/H$ found in the fourth column of this table and not on the asymptotic value $\alpha\sub{SS}$ quoted in Table 1 of \citet{Yang_etal_2018}.}}\\
$^\flat$ Peaks were counted at approximately t=3P, the point at which coherence was discernible to the human eye.\\
$^\dagger$Spatial computational domain sizes: (radial)$\times$(azimuthal)$\times$(vertical).\\
$^a$Read from their Figure 9 at t$\approx 25$P.\\ 
$^b$Read from their Figure 9 at t$\approx 12$P.
}}
\\
\end{table}

\section{Turbulent SI validity regime constrained by realistic protoplanetary disk models}\label{realism_pp-disk_models}
We now put these results into the context
of realistic global disk evolution and particle growth models, and other observed constraints.  
Recall that our theory assumes a single dominant, mass-carrying particle size. \citet{Benitez-Llambay_etal_2019} and \citet{Krapp_etal_2019} as
well as others note that when multiple particle species are present in the mix, 
{ {the overall growth rate of the SI may be significantly modified (probably diminished) 
as compared to single-sized particles \citep{Krapp_etal_2019}. }}



\par
Thus, we ask the questions:
for what turbulent disk conditions and particle properties does the SI provide a direct path to planetesimal formation, and, are these initial conditions realistic and self-consistent?
We consider this question for three general locations in the solar nebula, each of which nominally
representing: (i) the inner disk at $R \approx 3$AU, $T=265$K
with $c_s \approx 1050$ m/s (assuming a H$_2$ gas),
(ii)the snowline around $R \approx 5$AU, $T=140$K with
$c_s \approx 760$ m/s, and (iii)  the outer disk at $R = 30$AU, $T=73$K with $c_s \approx 550$ m/s.

\subsection{Disk Lifetime Constraints}\label{age_constraints}
For the inner disk, we rule out SI models that predict growth timescales that are significantly longer than $\sim 1$ Ma, based on the evidence that planetesimals were abundantly forming before that time \citep[][see also Section \ref{sec:intro}]{Kruijer_etal_2017,Scott_etal_2018}.  This translates to some number of equivalent P$_{orb}$ depending upon where
the turbulent SI theory is being modeled.  For example, we nominally place the snowline at $5 AU$ which means
that model parameters that predict growth time scales in excess of $(0.4-2.0)\times 10^5$ P$_{orb}$
are ruled out.

\subsection{Particle and Mode Radial Drift}\label{various_drift_limitations}
 
Growing particles will drift radially at different rates due to variable coupling with the nebula gas. The drift rate increases with Stokes number $\tau_s$, reaching a peak when $\tau_s = t_s\Omega \sim 1$, and then decreases for larger sizes. A Stokes number of unity corresponds to different size particles at different places in the protoplanetary disk, depending on the local gas surface density. For the standard Minimum Mass Solar Nebula (MMSN), meter-sized particles drift the fastest in the terrestrial planet region, whereas further out in the disk, where gas densities are low and the pressure gradients can be strong (especially if there is a strong gradient in the gas density at the disk outer edge), much smaller (mm-size) particles have $\tau_s = 1$. In fact, inward drifting particles may drift faster than they can grow in size by sticking; 
this is the so-called ``radial drift barrier" 
\citep[e.g.,][]{Brauer_etal_2008,Birnstiel_etal_2012,Estrada_etal_2016,Sengupta_etal_2019}. 

A similar criterion can be applied to the SI. Using the mean radial velocity of the particle component (eq. \ref{steady_particle_speeds}) to estimate the time $t_d$ a particle
takes to traverse the scale of the disk, 
we find (for $\tau_s \gg \alpha$) 
\beq
t\sub{d}^{-1} \equiv  \frac{\left|U\sub{p0}\right|}{r} 
= \frac{2\tau_s \delta^2 \Omega}{(1+\epsilon)^2 + \tau_s^2} .
\eeq
For a given set of parameters, the SI is considered ``viable" if the derived growth rate is faster
than the drain rate, i.e., when $\omega_i > t\sub{d}^{-1}$. Growth rates depend very
much on $\epsilon$, but sufficiently high solids-to-gas ratios that allow the solids to drive the gas motions are hard to achieve in turbulence \citep[e.g., see][]{Estrada_etal_2016}, unless one imposes arbitrary trapping mechanisms such as pressure bumps to thwart radial drift \citep[e.g.,][]{Drazkowska_etal_2013,Drazkowska_etal_2014}. Fractal particle growth leads to highly porous particles which drift radially much more slowly 
\citep{Ormel_etal_2007,Estrada_Cuzzi_2008,Okuzumi_etal_2012, Estrada_etal_2020}; this  
may provide a means to weaken the radial drift barrier and generate the necessary solids enhancements, but  decreasing $\tau_s$ also shifts the case to the left in Fig. \ref{Growth_Timescales_f01}, generally weakening SI for any $\alpha$. \par

As in the simulations reported in \citet{Johansen_etal_2007}, when $\tau_s \sim \order 1$ the pattern drift of growing modes is relatively fast and there emerges the possibility { {that the growing mode drifts in toward the star faster than the overdensity can grow sufficiently for gravitational instability to take root}}.  
{ {This concerns the notion of ``convective instability" familiar
in hydrodynamics \citep{Drazin_Reid_2004,Regev_etal_2016} and plasma physics  \citep[see Chapter 18 of][]{Bers_2016}.}}
As such, we assess the conditions in which the pattern drift timescale (to reach the star) is much shorter than the unstable
growth timescale.  The time it takes for inwardly propagating structures to drift 
into the star is $t_{w} = r/\left|c_r\right|$ (provided $c_r < 0$); so if 
$\omega_i > t\sub{w}^{-1}$,  we consider the SI as viable.

\subsection{Fragmentation Barrier}
\label
Several studies 
\citep{Brauer_etal_2008,Birnstiel_etal_2012,Estrada_etal_2016,Sengupta_etal_2019} 
show that the fragmentation barrier in a turbulent medium may be
estimated by assessing the inequality
\beq
\alpha \tau_s > {U_f^2}\Big/{2c_s^2}.
\label{fragmentation_barrier}
\eeq
The above expression
represents the condition in which the kinetic energy of colliding particles per unit mass as driven by turbulent eddies, approximately $2\tau_s \alpha c_s^2$ 
\citep{Voelk_etal_1980,Ormel_Cuzzi_2007}
exceeds their binding energy per unit mass, which is quantified by
$U_f$, the fragmentation speed for the particle (in reality a loose aggregate) in question. 
For silicate aggregates $U_f\approx 1.4$ m/s 
\citep{Zsom_etal_2010,Guettler_etal_2010}.\footnote{We note, however, that it has been suggested that the fragmentation speeds of relatively high temperature ($T\sim 200-400$ K) aggregates composed of organic covered silicate sub-micron grains may approach 100's of m/s
\citep{Homma_etal_2019}.  For the above quoted temperature range, a 0.03 $\mu$m monomer Si aggregate with about equal amount (by radius) of organic covering
has a fragmentation velocity of $\sim 100$m/s , while a 1$\mu$m sized Si aggregate with about 3\% 
organic coating (by radius) has a smaller fragmentation velocity diminish of about 10 $m/s$ (see their Figure 4), with
the assertion that this reduction in critical speeds continues for larger masses.
While it is unlikely that such sub-micron sized aggregates collide with one another with such high speeds in the turbulent nebula, this structural strengthening feature of organic covered Si grains should nonetheless be incorporated into future analyses.}

\par
The value of $U_f$ for \water ice aggregates is more nuanced. Laboratory studies of material properties suggest that sticking is significantly more effective for \water ice than for silicates \citep{Bridges_etal_1996, Wada_etal_2009}, allowing ice particles to grow larger, faster and more porous, with effective fragmentation velocity thresholds an order of magnitude or more higher than for silicates
\citep{Okuzumi_etal_2012,Wada_etal_2009,Wada_etal_2013}. On the other hand, \citet{Musiolik_Wurm_2019}
report on experimental results testing the surface energy of $\sim 1$mm \water ice spheres  
in the $175$K$<T<240$K temperature range and find that this increased strength only applies in a narrow
temperature range plateauing near 200 K, with effective sticking speeds almost 30 times lower when the grains are cold 
(i.e., $< 175$ K).   
They find that at these colder temperatures, the surface energy of \water ice grains is about the same
as the surface energy of silicate particles - suggesting that under very cold protoplanetary disk conditions, collisonal growth may not favor ice
over silicates. 

Based on the above findings, 
we consider two different values of $U_f$ for \water ice: when grains are near the ice line ($\sim 5$ AU), we adopt a 
fragmentation velocity of $U_f\approx 8.9$ m/s, which we refer to as the {\it{sticky \water ice fragmentation speed}} 
consistent with both previous studies \citep[e.g.,][]{Wada_etal_2009}, and those of \citet{Musiolik_Wurm_2019}. 
Well outside the snow line where temperatutes are low ($\sim 30$ AU), we consider both these stronger ice
particles, and also a fragmentation velocity for ice having the same value as for silicate particles
\citep[$U_f\approx 1.4$ m/s,][]{Zsom_etal_2010,Guettler_etal_2010}, which we refer to as the
{\it{cold \water ice fragmentation speed}}.


\subsection{Bouncing and drift barriers}\label{bouncing}
Numerous laboratory and theoretical studies have found that particle growth can be influenced by bouncing at much smaller sizes than the size that collides so energetically as to fragment the particles \citep{Zsom_etal_2010}.  { {Though \citet{Estrada_etal_2016} derive an expression for bouncing between similar-sized particles (their Eq. 59), this size is harder to specify rigorously because it likely also depends on material properties \citep{Guettler_etal_2010,Zsom_etal_2010}. \citet{Estrada_etal_2016} emphasize that the bouncing barrier is not impermeable but merely slows growth to a great degree. 
For this reason, it is not as instructive to include the bouncing barrier as a constraint as we do for fragmentation. A similar situation holds for the radial drift barrier alluded to earlier; it is hard to write quantitative values for this size limit \citep[e.g., see Eqns. 60-61,][]{Estrada_etal_2016}. Our fragmentation barrier upper limits on particle size (Eq. \ref{fragmentation_barrier}) are thus approximate. Under some conditions, particles may never get that large due to a combination of the bouncing and radial drift barriers \citep{Birnstiel_etal_2012,Estrada_etal_2016,Sengupta_etal_2019}. Under other conditions, growth can proceed somewhat beyond the fragmentation barrier if a distribution of collision velocities and mass transfer in collisions are allowed for \citep[e.g.,][]{Windmark_etal_2012,Drazkowska_etal_2013,Drazkowska_etal_2014,Estrada_etal_2016}.} Bouncing, drift, mass transfer, and a probability distribution for collisional velocities are included in the more realistic growth model constraints discussed in Section \ref{self_consistent}.

\subsection{Combined limits on particle size: $\tau_s$.}
\label{combined_limits_on_particle_size}
 Global numerical simulations suggest that the series of barriers to growth discussed previously are quite effective
at limiting $\tau_s$ of the particle size that dominates the mass \citep[e.g.,][]{Birnstiel_etal_2012,Estrada_etal_2016,Sengupta_etal_2019}. Recent simulations conducted for solid particle growth over a three-order-of-magnitude range of turbulent intensities ($10^{-5}<\alpha < 10^{-2}$) { {relevant to the first 0.5 Ma where planetesimal formation is thought to begin (see Sec. \ref{sec:intro})}} further indicate that
the maximum achieved Stokes numbers of the mass dominant particles\footnote{As in some of the models discussed in 
Sec. \ref{Comparison_to_other_studies}, the particle growth models of \citet{Estrada_etal_2016,Estrada_etal_2020} 
exhibit a broad size 
distribution and do not employ particles of a single size. However, in general, most of the particle mass is near the fragmentation size when drift is not important,  or near the largest particle size in the distribution when in the drift-dominated regime - either way, defining a single particle size.} 
range from $\sim 0.001-0.01$ over a wide range of disk models with initial global metallicities
of $Z\simeq 0.01$ \citep{Estrada_etal_2016,Estrada_etal_2020}. Table \ref{Stokes_numbers_alpha} summarizes these Stokes numbers, the corresponding particle radii, { {and ambient disk conditions}} at our three representative radial locations within the protoplanetary disk. We note that in these simulations, the local solids abundance { {and the nebula gas density and temperature (and thus $\delta$, as well as the pressure gradient)}} are evolving with time which means that $Z$ can have a range of values over the disk's radial extent. In particular, the snow line (and various other ``evaporation fronts'') evolve
with time leading to both local enhancements and depletions in the amounts of solids, especially in the planet forming region. The values given in 
Table \ref{Stokes_numbers_alpha} are selected at locations where the solids-to-gas ratio $Z(t)\simeq 0.01$ and { {which correspond to the ``inner disk'' and ``snow line''; they may thus lie at semi-major axes that are slightly different from our nominal definitions of $R = 3$ and $5$ AU, respectively. The ``outer disk'' location reliably corresponds to 30 AU (see caption, Table  \ref{Stokes_numbers_alpha}).}}

Despite the variable metallicity, the particle size distributions have already reached a quasi-equilibrium state at the selected times and radial locations, changing only slowly with time and depending only weakly on the instantaneous value of $Z$ \citep[see Fig. 19 of][and associated discussion]{Estrada_etal_2016}. The physics that limits the mass dominant particle $\tau_s$ in different radial regions depends on the ambient nebula conditions. In the inner disk regions 
where bulk composition is ice-free, $c_s$ and $\Omega$ are large so turbulent relative velocities are fairly large. These regions tend to be in the fragmentation regime based on previous discussion, using Eq. 
(\ref{fragmentation_barrier}) for the fragmentation equilibrium Stokes number. { {However, for $\alpha \gtrsim 10^{-3}$, the values for the fragmentation $\tau_s$ obtained from Eq. (\ref{fragmentation_barrier}) using the parameters cited in Sec. \ref{realism_pp-disk_models} are consistently smaller than the values listed in Table \ref{Stokes_numbers_alpha}, because the actual growth models include growth beyond the fragmentation barrier (see Sec. \ref{bouncing}). 
On the other hand,
for $\alpha=10^{-4}-10^{-5}$, one finds Eq. (\ref{fragmentation_barrier}) gives $\tau_s \simeq 0.01-0.1$ which is consistently larger than the corresponding values in Table \ref{Stokes_numbers_alpha}, despite the temperatures being closer to those cited in Sec. \ref{realism_pp-disk_models}, and in fact these particles have already reached the fragmentation barrier. As noted in Table \ref{Stokes_numbers_alpha}, these mass-dominant particles are in the Stokes regime, which for these relatively small $\tau_s$ means the relative velocities between them tend to be higher for a given Stokes number compared to the Epstein regime. This discrepancy does not depend on which flow regime the particles are in, but rather because eddy-crossing effects start to become important in weak turbulence, particularly when $\delta^2/\alpha \gtrsim \tau_s^{-1}$. Under these circumstances the relative velocities between the particles are higher, even for similar-sized particles, which means the fragmentation $\tau_s$ will be smaller than what Eq. (\ref{fragmentation_barrier}) would predict \citep{Ormel_Cuzzi_2007}\footnote{{ {\citet{Ormel_Cuzzi_2007} found that the transition between so-called class I and class II eddies occurs at higher values with increasing $\delta^2/\alpha$. In this case, class II eddies, where the fluctation times are shorter than the particle stopping time, dominate even for small particles resulting in the kinetic energy per unit mass between colliding particles of a given $\tau_s$ is larger.  This consequently progressively lowers the estimate from Eq. (\ref{fragmentation_barrier}) with decreasing $\alpha$. For more discussion, see \citet{Ormel_Cuzzi_2007}.}}}}. }}



In the colder, ice-rich outer disk a naive application of Eq. (\ref{fragmentation_barrier}) would give $\tau_s \gtrsim 1$ for $\alpha \lesssim 10^{-4}$ for the sticky water-ice fragmentation case. 
However, the simple constraint of Eq. (\ref{fragmentation_barrier}) assumes that the collision velocity is dominated by turbulent relative velocities, but large
values of the pressure gradient (manifested by $\delta$) in the outer nebula can drive systematic drift- and headwind-related velocities that can 
significantly exceed those due to turbulence for these weak values of $\alpha$. Under these
circumstances, bouncing plays an even more influential role by slowing growth, enhancing the importance of the drift barrier in precluding the fragmentation barrier from being reached. 
Simulations of the outer disk where particles drift faster than they can grow (Figs. \ref{Growth_Timescales_f01_with_constraints_5AU_d0p07} and \ref{Growth_Timescales_f01_with_constraints_30AU_d0p1}) are characterized by Stokes numbers $\tau_s \lesssim 0.01$ that are nearly 
constant or modestly varying with radius, which suggests that the decrease in particle radii with distance simply mirrors the 
decrease in gas density \citep{Birnstiel_etal_2012,Estrada_etal_2016,Estrada_etal_2020}. Even when the bouncing barrier was not considered, $\tau_s \lesssim 0.1$ in these regions. 
On the other hand, for large values of $\alpha \gtrsim 10^{-2}$ one 
finds that even in the outer disk particles can be in the fragmentation-dominated regime even if their Stokes numbers remain comparable to the lower 
$\alpha$ cases, because the fragmentation barrier occurs
at much smaller sizes there (see Table \ref{Stokes_numbers_alpha}). 

{ {Overall then, Eq. (\ref{fragmentation_barrier}) is a handy but crude approximation in general.}} \citet{Estrada_etal_2016} give a more in-depth discussion and derive estimates of the limits of $\tau_s$ as it pertains to their 
models. The variation in $\tau_s$ and particle size over different radial locations and with time listed in Table \ref{Stokes_numbers_alpha} { {is subtle  
but secondary to our focus here. After looking at many models, we believe the values presented are representative for the purpose of constraining regions of parameter
space that may permit SI to form planetesimals over a wide range of conditions (see Sec. \ref{self_consistent}). 
Additional description of the colored-symbol models will be addressed in \citet{Estrada_etal_2020}. 
Even with their uncertainties,}} the message of the realistic models is that plausible, self-consistent combinations of nebula turbulence and particle size typically lie in ``Zone II'', the ``moderately turbulent" regime above the $\epsilon=1$ line, where SI is only ``incipient". Moreover, any degree of particle porosity leads to even smaller $\tau_s$ \citep{Estrada_etal_2020}.


\begin{table}[ht!]
\begin{center}
\caption{Maximum$^*$ achievable Stokes numbers ($\tau_s$) in global disk evolution models for $Z=0.01$}
\vspace{0.1in}
\begin{tabular}{@{} c | cc|cc|cc  @{}}
\toprule
\hline
 
& \multicolumn{2}{|c|}{$\displaystyle \begin{array}{c} {\rm Inner \ Disk} \\ (R=3 {\rm{AU}})^\triangle \end{array}$ }
& \multicolumn{2}{|c|}{{$\displaystyle \begin{array}{c} {\rm Snow Line} \\ (R=5 {\rm{AU}})^\triangle \end{array}$ }}
& \multicolumn{2}{|c}{$\displaystyle \begin{array}{c} {\rm Outer \ Disk} \\ (R=30 {\rm{AU}})^\triangle \end{array}$ }
 \\
 $\displaystyle {\frac{\alpha}{10^{-4}}}$
&  \multicolumn{2}{|c|}{------}
& \multicolumn{2}{|c|}{------}
& \multicolumn{2}{|c}{------}
 \\
  & $\tau_s$ & $a$(cm)$^*$ & $\tau_s$ & $a$(cm)$^*$ & $\tau_s$ & $a$(cm)$^*$ \\
 \hline
100$^\dagger$ &$7.0\times 10^{-4}$ & 0.03    & $9.3\times 10^{-3}$      & 0.43 & $6.9\times 10^{-3}$  & 0.07	\\
$\;\;\;$40$^{\dagger\dagger}$ &$1.8\times 10^{-3}$ & 0.38     & $1.1\times 10^{-2}$      & 1.6  & $6.2\times 10^{-3}$  & 0.09 \\
$\;\;$10$^\dagger$ &$2.5\times 10^{-3}$ & 0.76     &$1.9\times 10^{-2}$       & 4.6  &$4.8\times 10^{-3}$	& 0.11\\
$\;\;\;\;\;$4$^{\dagger\dagger}$ &$3.4\times 10^{-3}$ & 1.9      &$1.3\times 10^{-2}$       & 5.3  &$7.7\times 10^{-3}$   & 0.12 \\
$\;\;\;\;$1$^\dagger$ &$5.7\times 10^{-3}$	  & 5.8$^\diamond$  &$1.2\times 10^{-2}$       & 13  &$1.1\times 10^{-2}$	& 0.30 \\
$\;$0.1$^\dagger$ & $2.2\times 10^{-2}$			 & 12$^\diamond$ &$4.9\times 10^{-2}$  & 39$^\diamond$  &$1.6\times 10^{-2}$	& 0.37\\
\hline
$\;\,$10$^\ddagger$ & $3.1\times 10^{-3}$ &0.79 & $4.5\times 10^{-3}$ & 0.89 & $1.9\times 10^{-3}$	& 0.05\\
\hline
\bottomrule
\end{tabular}

\vspace{0.25in}

\begin{tabular}{@{} c | ccc|ccc|ccc  @{}}
\toprule
\hline
 
& \multicolumn{3}{|c|}{$\displaystyle \begin{array}{c} {\rm Inner \ Disk} \\ (R=3 {\rm{AU}})^\triangle \end{array}$ }
& \multicolumn{3}{|c|}{{$\displaystyle \begin{array}{c} {\rm Snow Line} \\ (R=5 {\rm{AU}})^\triangle \end{array}$ }}
& \multicolumn{3}{|c}{$\displaystyle \begin{array}{c} {\rm Outer \ Disk} \\ (R=30 {\rm{AU}})^\triangle \end{array}$ }
 \\
 $\displaystyle {\frac{\alpha}{10^{-4}}}$
&  \multicolumn{3}{|c|}{------}
& \multicolumn{3}{|c|}{------}
& \multicolumn{3}{|c}{------}
 \\
  & $T$(K) & $\rho_g$ (g cm$^{-3}$) & $\delta$ & $T$(K) & $\rho_g$ (g cm$^{-3}$) & $\delta$ & $T$(K) & $\rho_g$ (g cm$^{-3}$) & $\delta$ \\
 \hline
100$^\dagger$ & 283 & $2.4\times 10^{-11}$ & 0.07 & 156 & $1.2\times 10^{-11}$ & 0.07 & 73 & $3.2\times 10^{-13}$ & 0.11	\\
$\;\;\;$40$^{\dagger\dagger}$ & 290 & $3.7\times 10^{-11}$ & 0.09 & 148 & $2.3\times 10^{-11}$ & 0.08 & 71 & $3.6\times 10^{-13}$ & 0.12 \\
$\;\;$10$^\dagger$ & 275 & $1.2\times 10^{-10}$ & 0.08 & 130 & $4.4\times 10^{-11}$ & 0.07 & 75 & $8.6\times 10^{-13}$ & 0.11 \\
$\;\;\;\;\;$4$^{\dagger\dagger}$ & 265 & $3.0\times 10^{-10}$ & 0.08 & 140 & $7.0\times 10^{-11}$ & 0.08 & 75 & $4.8\times 10^{-13}$ & 0.11 \\
$\;\;\;\;$1$^\dagger$ & 225  & $8.6\times 10^{-10}$ & 0.06 & 152 & $2.2\times 10^{-10}$ & 0.06 & 70 & $9.1\times 10^{-13}$ & 0.11 \\
$\;$0.1$^\dagger$ & 204 & $6.7\times 10^{-10}$ & 0.06 & 148 & $2.7\times 10^{-10}$ & 0.06 & 73 & $6.5\times 10^{-13}$ & 0.11 \\
\hline
$\;\,$10$^\ddagger$ & 227 & $1.1\times 10^{-10}$ & 0.07 & 135 & $3.2\times 10^{-11}$ & 0.08 & 70 & $8.8\times 10^{-13}$ & 0.11 \\
\hline
\bottomrule
\label{Stokes_numbers_alpha}
\end{tabular}

\end{center}
\par
{\small{
$^*$Corresponds to the size that dominates the mass in the particle size distribution. $\;\;$
\newline $^\triangle$Approximate radial locations. { {For this exercise, the models were evaluated for local values about $Z = 0.01 \pm 0.001$ at a mean time of $\sim 2\times 10^5$ yrs, where ``3 AU'' corresponds to radial distances where $200\lesssim T\lesssim 290$ K, the snow line to $130\lesssim T\lesssim 160$ K, and ``30 AU'' to $70\lesssim T\lesssim 75$ K, respectively.}}
\newline $^\dagger$Data shown with yellow symbols in Figs. \ref{Growth_Timescales_f01_with_constraints_3AU_d0p07}-\ref{Growth_Timescales_f01_with_constraints_30AU_d0p1} are from \citet{Estrada_etal_2020}. 
\newline $^{\dagger\dagger}$Values from MTBF models of \citet{Estrada_etal_2016} and shown with black symbols in Figs. \ref{Growth_Timescales_f01_with_constraints_3AU_d0p07}-\ref{Growth_Timescales_f01_with_constraints_30AU_d0p1}.
\newline $^\ddagger$Using \citet{Musiolik_Wurm_2019} cold ice fragmentation findings for $U_f$ and shown with orange symbols in Figs. \ref{Growth_Timescales_f01_with_constraints_3AU_d0p07}-\ref{Growth_Timescales_f01_with_constraints_30AU_d0p1}.
\newline $^\diamond${ {These particles are in the Stokes regime with particle Reynolds numbers near unity.}}
}}
\\
\end{table}


\begin{figure*}
\begin{center}
\leavevmode
\includegraphics[width=17.5cm]{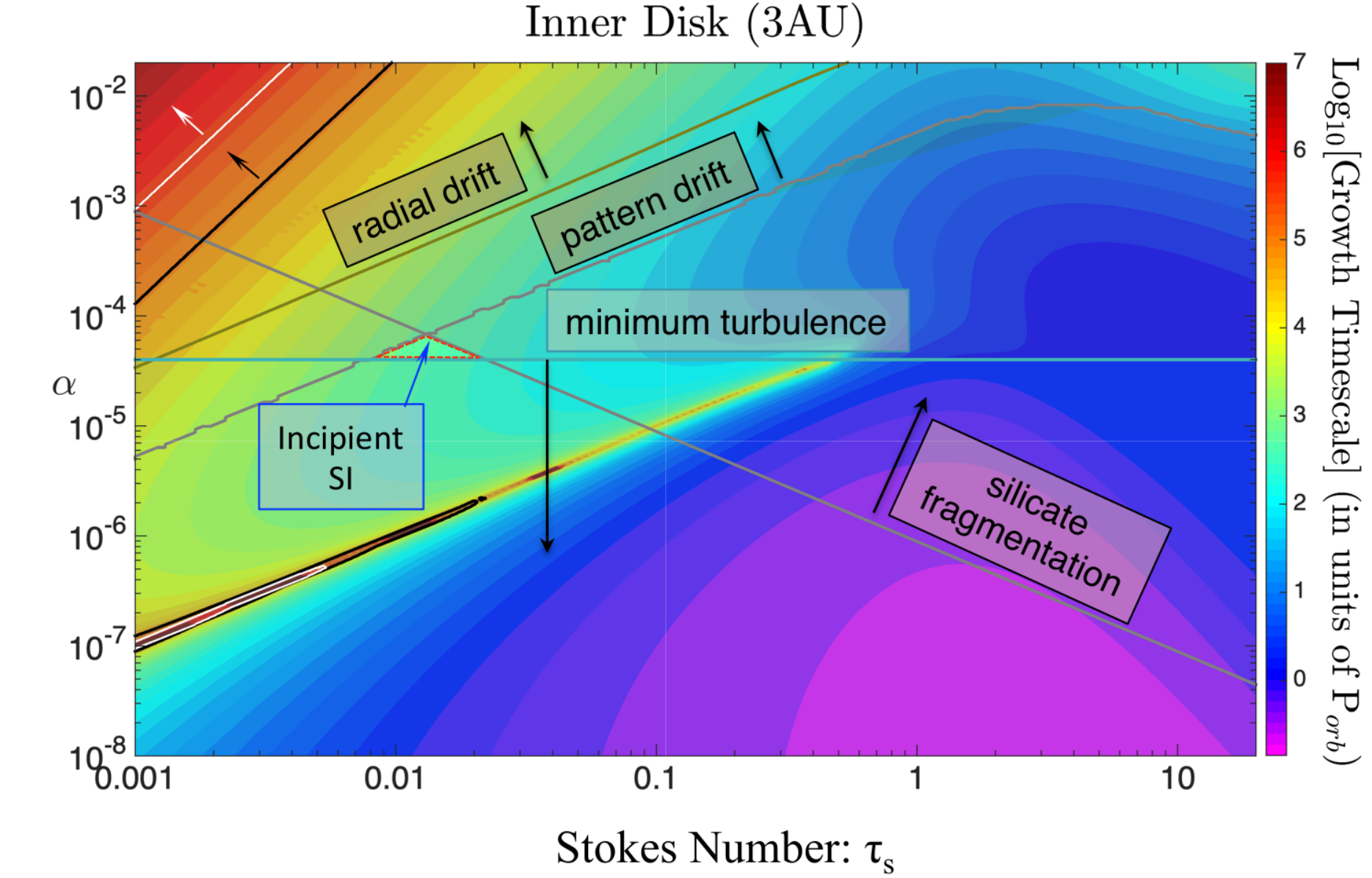}
\end{center}
\caption{\noindent 
{{Regions of $\alpha$--$\tau_s$ parameter space accessible to the SI based on the same input disk model values as shown in Fig. \ref{Growth_Timescales_f01}, $Z=0.01$ and at the nominal location of the asteroid belt, $R = 3$AU (inside the snowline for these models). Regions of parameter space that exclude the SI -- due to various particle and fluid constraints -- are indicated by various transparent color shadings associated with one of the constraining process discussed in Section \ref{realism_pp-disk_models} are represented.  Process constraints are appropriate at a disk location $R = 3$AU with $c_s = 950$m/s.  Arrows indicate regions excluded by given process.  White (black) curve indicates 2.5 (0.5) Myr disk lifetime. The SI incipient region is indicated by red-hatched triangle. }}}
\label{Growth_Timescales_f01_with_constraints_3AU}
\vspace{0.2in}
\end{figure*}

\subsection{Turbulence Constraints}\label{turb_constraints}
It is widely believed that the region where the first planetesimals were  assembled was a ``Dead Zone" extending from 1AU to $\le$ 80 AU, so-called
because the temperature and transparency of protoplanetary disk material to ionizing photons
are too low to allow magnetically driven (MHD) turbulence to arise \citep{Turner_etal_2014}. {These areas of protoplanetary disks are now more commonly referred to as Ohmic Zones \citep[e.g.][]{Lyra_Umurhan_2019}, owing to the dominance of Ohmic dissipation that suppresses 
self-generated activity like the linear MRI process.
There remains substantial debate over the true ionization state and consequent effective magnetic resistivity of disk
material, depending on the abundance of grains of sufficiently small size \citep{Okuzumi_etal_2012,Ormel_Okuzumi_2013,Simon_etal_2018}.  Even if insufficiently ionized for MRI, }
the disk remains susceptible to nonideal MHD processes 
including wind launching mechanisms  \citep{Bai_2016,Bai_etal_2016}.
Nonetheless, this debate leaves some uncertainty as to whether any regions may be {susceptible to the MRI, which numerical experiments show} to induce very strong turbulent intensities of $\alpha \sim 10^{-3} - 10^{-2}$ \citep{Balbus_Hawley_1998,Armitage_2011}.
\par
Therefore, for the purposes of this study we shall assume that any turbulence that arises in the near-midplane regions of protoplanetary disks, which are of greatest interest to planetesimal formation, stems from any one of the three purely hydrodynamical mechanisms noted in section \ref{sec:intro} that have recently been discussed in the literature
-- for a review see \citet{Turner_etal_2014} or \citet{Lyra_Umurhan_2019}. The operation of the three mechanisms depends
upon the thermal relaxation or cooling time scale $t_{{\rm th}}$ of the particle-gas mixture, which in turn depends on
the disturbance lengthscales if these lengthscales are in the optically thick regime.  Specifically, 
the [VSI/COV/ZVI] (section \ref{sec:intro})  is
expected to operate in a disk
when [$\Omega t_{{\rm th}} \ll 1, \Omega t_{{\rm th}} \approx 1, \Omega t_{{\rm th}} \gg 1$] (respectively).  Exactly which regions of protoplanetary disks are most susceptible to which mechanism
remains under discussion \citep{Malygin_etal_2017,Umurhan_etal_2017,Barranco_etal_2018}.  However, because of their complementary instability criteria, it is plausible that the full extent of protoplanetary disks that is of interest to planetesimal formation will be turbulent, due to at least one of the three
mechanisms listed.   Recently published numerical studies of the three processes show that
the turbulent intensities arising from these mechanisms lie somewhere in the range $10^{-5} < \alpha < 5\times 10^{-4}$
\citep{Lyra_Umurhan_2019}. 
Of the various published studies of the VSI, \citet{Flock_etal_2017} predict the lowest level of 
$\alpha \sim 4\times 10^{-5}$, and we use this as our minimum adopted value.
Our understanding of the nature of cold protoplanetary disk turbulence will surely continue to evolve. 
\par
{Caveat:
We have assumed the mechanism driving turbulence in disks is unaffected by the degree of particle loading. This is valid {\it a posteriori} in regions covered by all of the detailed particle growth models (figures \ref{Growth_Timescales_f01_with_constraints_3AU_d0p07} - \ref{Growth_Timescales_f01_with_constraints_30AU_d0p1}) which imply $\epsilon < 1$, and limited or negligible particle feedback. This assumption should be regarded with caution in parameter combinations where $\epsilon > 1$, but these combinations (corresponding to Zone I) are already known to lead to SI in numerical experiments.}\footnote{{For example, using 2D axisymmetric simulations
of a single-fluid {\it terminal velocity approximation} model \citep{Lin_Youdin_2017}
 as well as physical arguments, \citet{Lin_2019} argues that a feedback loop is set up wherein particle-loading and subsequent settling {\it drives buoyancy oscillations that weaken the VSI} \citep{Lin_Youdin_2015} which, in turn, further enhances particle settling.  
VSI stabilization by this process was argued for values of disk metallicity $Z \ge 0.01$ and appears effective for $\tau_s > 0.003$ in the published 2D simulations. However, 
whether or how much particle loading would damp turbulence in full 3D, more highly resolved simulations remains to be seen, as 3D energy cascade pathways and turbulent eddy motion couplings 
are different \citep{Richard_etal_2016,Lyra_Umurhan_2019}. 
Additionally, whether or not this tendency toward stabilization of the VSI persists in multi-particle component 3D disk simulations remains to be seen \citep[e.g., see discussion in][]{Benitez-Llambay_etal_2019}.}}

\begin{figure*}
\begin{center}
\leavevmode
\includegraphics[width=17.5cm]{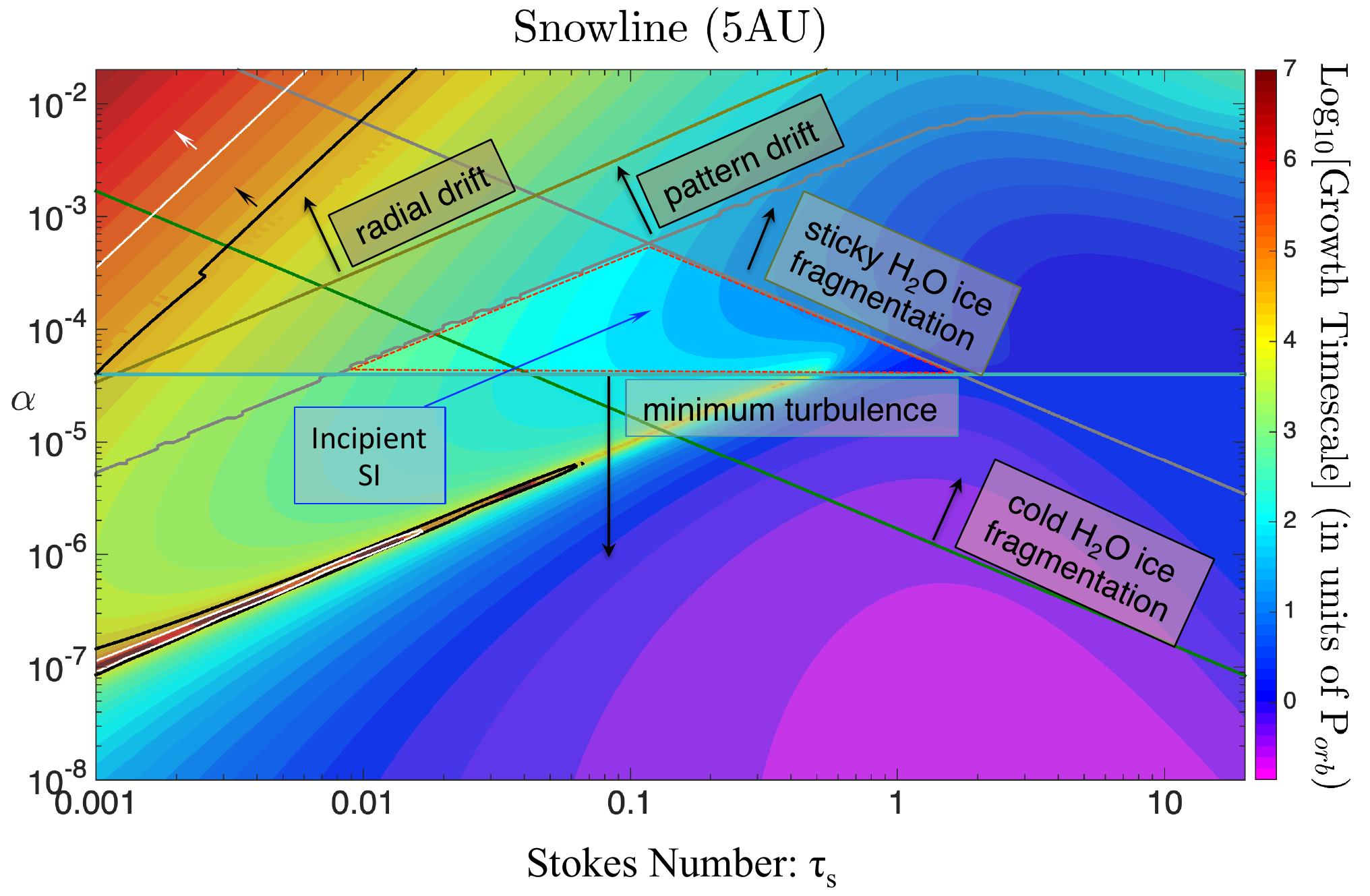}
\end{center}
\caption{\noindent 
Like Fig. \ref{Growth_Timescales_f01_with_constraints_3AU} except $R = 5$ AU, nominally the snowline.  
The SI incipient region is indicated by a red-bordered triangle { {and corresponds to water ice being stickier close to
the snow line (and thus bounded on the right by the sticky H$_2$O ice fragmentation line). The incipient region can be significantly decreased if ice is not sticky (cold H$_2$O ice fragmentation line) or loses its stickiness as one moves away from the snowline into colder regions \citep{Musiolik_Wurm_2019}.}}}
\label{Growth_Timescales_f01_with_constraints_5AU}
\vspace{0.4in}
\end{figure*}

\begin{figure*}
\begin{center}
\leavevmode
\includegraphics[width=17.5cm]{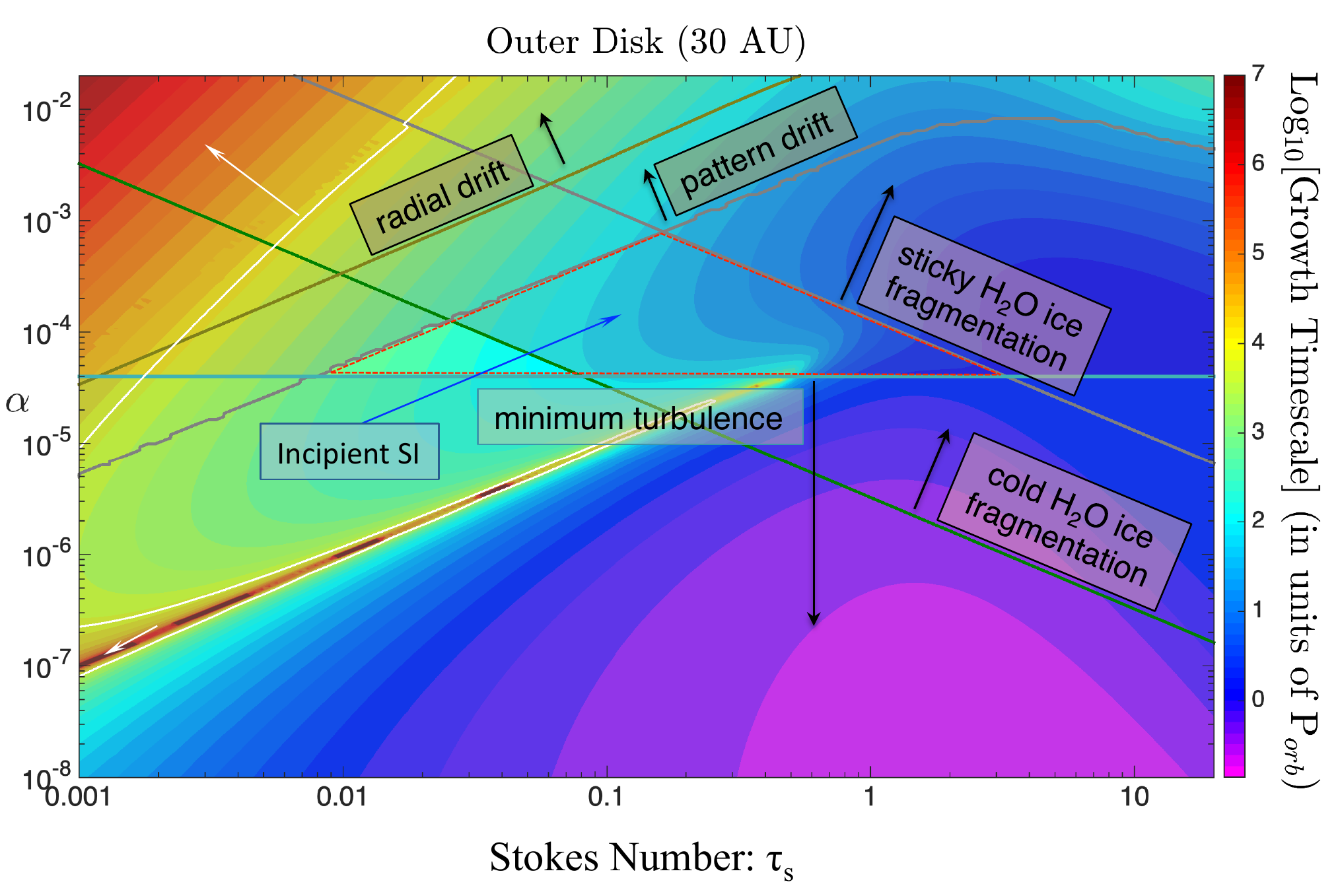} 
\end{center}
\caption{\noindent 
Like Fig. \ref{Growth_Timescales_f01_with_constraints_3AU} except $R = 30$ AU, nominally the outer disk.  Note, only
2.5 Ma line (white) shown. The SI incipient regime shown with red-hatched triangle { {and corresponds to water ice being stickier beyond
the snow line (and thus bounded on the right by the sticky H$_2$O ice fragmentation line). The incipient region would be significantly decreased if one adopts the ice cold H$_2$O ice fragmentation line which
may be the case away from the snowline in the colder regions of the disk \citep{Musiolik_Wurm_2019}.}}}
\label{Growth_Timescales_f01_with_constraints_30AU}
\vspace{0.2in}
\end{figure*}

\begin{figure*}
\begin{center}
\leavevmode
\includegraphics[width=17.5cm]{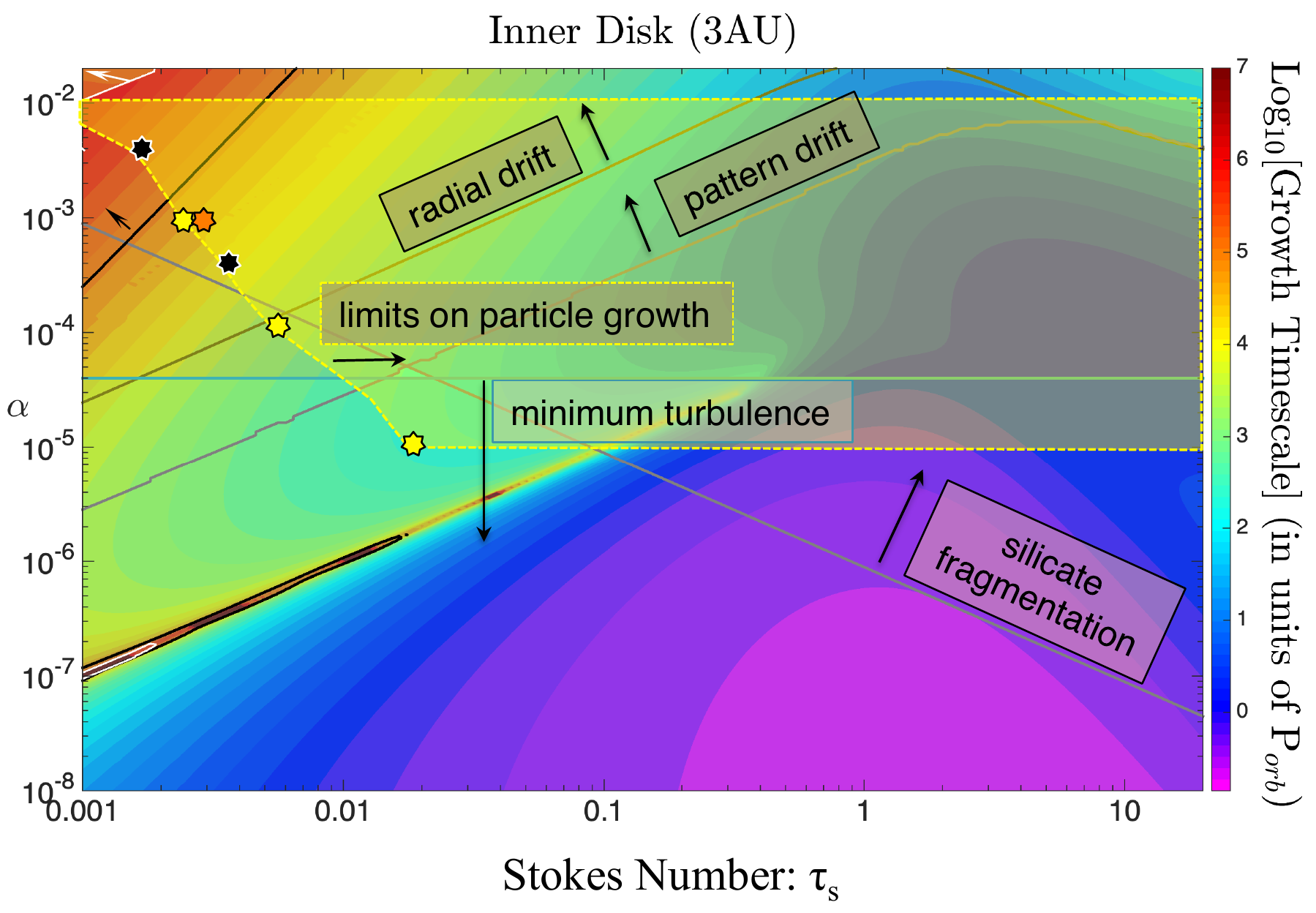}
\end{center}
\caption{\noindent 
Regions of $\alpha$--$\tau_s$ parameter space accessible to the SI based on the same input disk model values as shown in Fig. \ref{Growth_Timescales_f01} { {(except with $\delta = 0.07$)}}, $Z=0.01$ and at the nominal location of the asteroid belt, $R = 3$AU (inside the snowline for these models). { {A consequence of higher $\delta$ values is that the radial and mode pattern drift lines shift downwards relative the our nominal $\delta = 0.05$ case, further limiting the SI incipient region.}} The indicated (upper) limits on particle size (yellow and black stars) are based on the predicted mass-bearing Stokes numbers for evolutionary models of \citet{Estrada_etal_2016,Estrada_etal_2020} as listed in Table \ref{Stokes_numbers_alpha}.  The yellow patch demarcated by the dashed line identifies the parameter region ruled out based on interpolation of the yellow stars. The orange star is for a model in which water ice is only
sticky near the snow line (last row of Table \ref{Stokes_numbers_alpha}). { {The latter model's different $\tau_s$ value is attributable to less material drifting in to the inner disk relative to the sticky ice model where particles can grow larger.}} 
Note also these evolution models do not examine values of $\alpha$ beyond the vertical extent of the yellow shading. 
{ {There is no SI incipient region accessible in this case.}}}
\label{Growth_Timescales_f01_with_constraints_3AU_d0p07}
\end{figure*}

\begin{figure*}
\begin{center}
\leavevmode
\includegraphics[width=17.5cm]{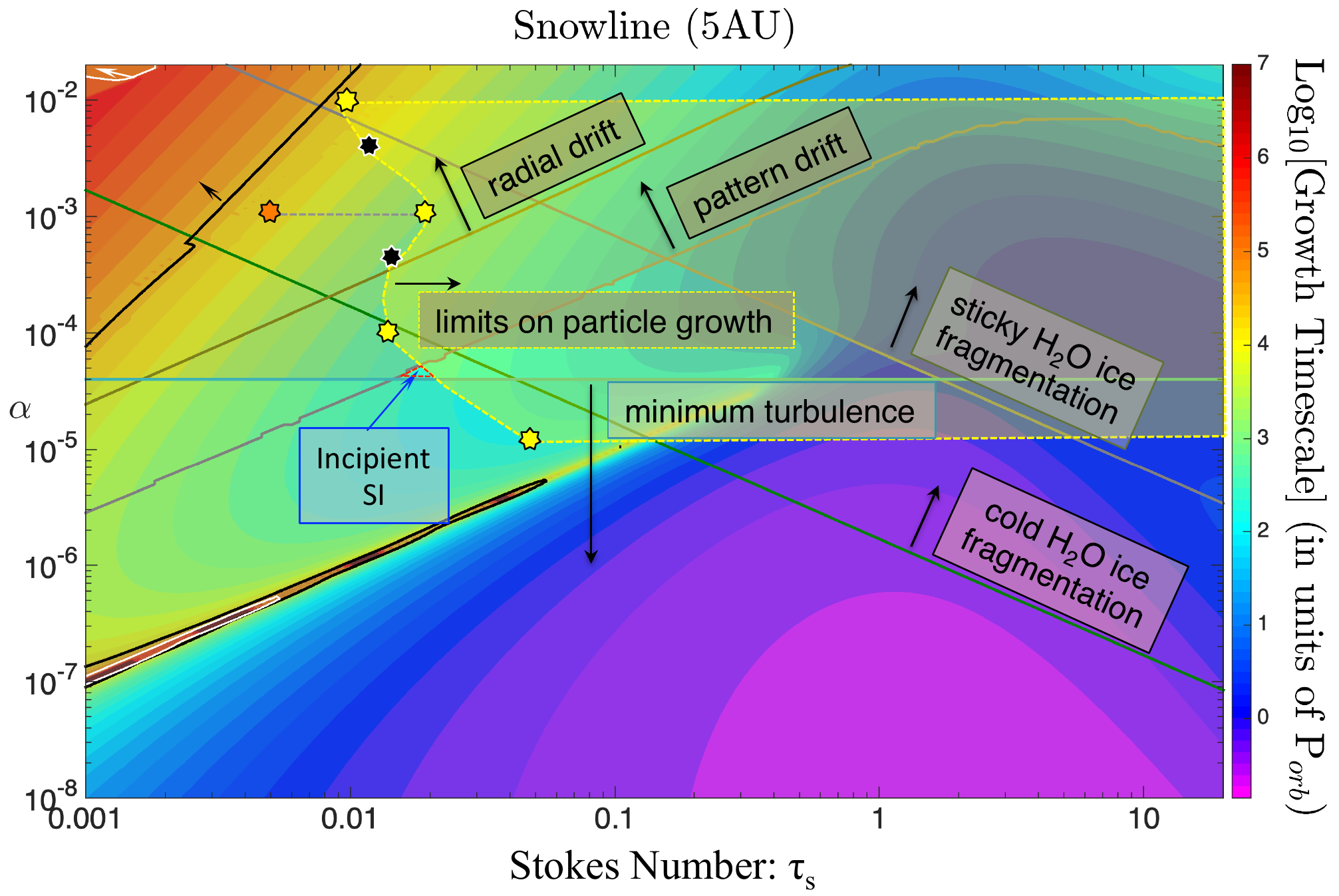}
\end{center}
\caption{\noindent 
Like Figure \ref{Growth_Timescales_f01_with_constraints_3AU_d0p07} except $R = 5$ AU, nominally the snowline. { {The orange star is for a model in which water ice has the low stickiness appropriate for cold temperatures (last row of Table \ref{Stokes_numbers_alpha}), and thus leads to smaller particles.}}
{ {We note that a tiny SI incipient regime survives, indicated by the red-hatched triangle.}}}
\label{Growth_Timescales_f01_with_constraints_5AU_d0p07}
\end{figure*}

\begin{figure*}
\begin{center}
\leavevmode
\includegraphics[width=17.5cm]{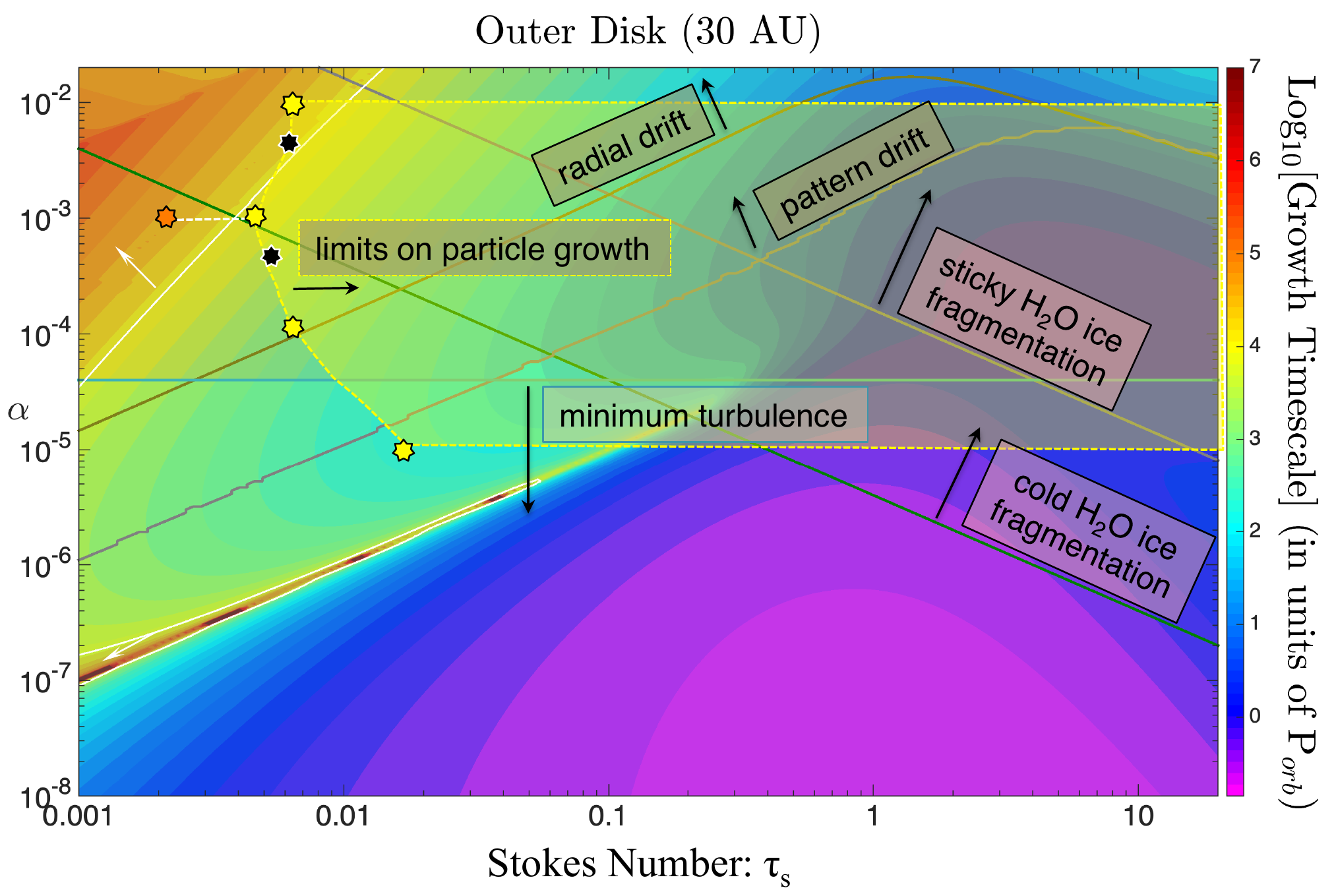}
\end{center}
\caption{\noindent 
Like Figure \ref{Growth_Timescales_f01_with_constraints_3AU_d0p07} except $R = 30$ AU and $\delta = 0.1$, nominally the outer disk.  { {The orange star is for a model in which water ice has the low stickiness appropriate for cold temperatures (last row of Table \ref{Stokes_numbers_alpha}), and thus leads to smaller particles. All of the yellow stars shown are thus likely overestimating $\tau_s$ for the 30AU case if one adopted the cold ice model for them as well, but even at their current values they preclude the tiny incipient region allowed by the simpler, straight-line constraints of pattern drift and minimum turbulence.}}
 Note, only 2.5 Ma line (white) shown. { {Like the inner disk, there is no SI incipient region accessible in this case.}}}
\label{Growth_Timescales_f01_with_constraints_30AU_d0p1}
\vspace{0.2in}
\end{figure*}

  \subsection{Regions of parameter space that permit the SI to operate for $Z=0.01$}\label{self_consistent}
  Given the physical constraints outlined in the previous six subsections, we can now assess in what parts of the $\alpha$--$\tau_s$ parameter space the SI is likely to operate.  We examine this for three locations: $R=3$AU (``inner disk"), $R=5$AU (``snowline") and $R=30$ AU (``outer disk"), in a $Z=0.01$ disk with constant $\delta = 0.05$.  By implication, we imagine the following survey to be predictive
  for an early phase of the solar nebula with uniform Z ($=0.01$) and well before wind loss substantially evaporates the disk's gas \citep[say, $< 2.5$ Ma;][]{Carrera_etal_2017}. { {It is increasingly thought that planetesimal formation had to proceed in the inner nebula before the 0.5 Ma time marker (relative to CAIs), and that even Jupiter's core may have formed in less than 1 Ma \citep[e.g.,][]{Kruijer_etal_2017,Simon_etal_2018}.}}
For higher, perhaps local, values of $Z$, { {to which particle size is insensitive}}, the reader is referred to Figure \ref{Growth_Timescales_Moar} for comparison of how Zones I and II shift.
  \par
We approach this task by first comparing the SI's predicted growth timescales with constraining timescales based on the various disk processes and particle growth barriers discussed in sections \ref{age_constraints}-\ref{turb_constraints} { {for our nominal disk model with $\delta = 0.05$. We then delineate regimes of ($\alpha,\tau_s$) space deemed implausible by the best current models of growth-by-sticking \citep[sections \ref{bouncing} and \ref{combined_limits_on_particle_size}]{Estrada_etal_2016,Estrada_etal_2020} over the first 0.5 Ma, which are typically characterized by larger $\delta \gtrsim 0.05-0.1$.}}  We declare the SI to be ``incipient", or capable of leading to some degree of enhancement, in those regions of ($\alpha,\tau_s$) space which are both reasonable from the standpoint of particle growth, {\it and} where the SI's growth timescales are shorter than the aforementioned constraining timescales.  
 
{ {{\it Nominal fragmentation constraints:}}} The results { {for our nominal models}} are shown in Figures \ref{Growth_Timescales_f01_with_constraints_3AU}-\ref{Growth_Timescales_f01_with_constraints_30AU}, { {and those with the added constraints of particle growth models are shown in Figures \ref{Growth_Timescales_f01_with_constraints_3AU_d0p07}-\ref{Growth_Timescales_f01_with_constraints_30AU_d0p1}}}. All these plots overlay various excluded regions on the growth timescale plot similar to Figure \ref{Growth_Timescales_f01} for the three disk zones of interest. { {The timescales in Figures \ref{Growth_Timescales_f01_with_constraints_3AU_d0p07}-\ref{Growth_Timescales_f01_with_constraints_30AU_d0p1} however were recalculated using the higher $\delta$ values associated with the particle growth models. If a zone of incipient SI is allowed, it is indicated by a red-hatched triangle in these figures.}} 

    Figure \ref{Growth_Timescales_f01_with_constraints_3AU} shows
  the situation for the inner disk where the temperature is too large to support water ice particles.  We find that the SI is
  incipient in a relatively small triangular patch of parameter space centered about a value of $\alpha \sim 5\times 10^{-5}$ and $\tau_s \approx 0.01-0.02$. The region of accessible parameter space is bounded on the small side of $\tau_s$ lines showing conditions where the growing SI pattern drifts into the star faster than it can achieve an $e$-folding level of growth (see discussion in sec. \ref{various_drift_limitations}).  { {The region is also bounded on the higher side of $\tau_s$ by the constraint imposed by the silicate fragmentation line. Within this SI incipient region, the growth timescales are relatively long - thousands of orbit times.}} 

{ {The situation at the snow line can vary significantly depending on whether one adopts the sticky or cold H$_2$O fragmentation condition, as can be seen in Figure \ref{Growth_Timescales_f01_with_constraints_5AU}. Taking the sticky-ice constraint opens  up a much larger region for incipient growth ranging from $\tau_s \approx 0.01-1$, and for turbulent intensities as high as $\alpha \sim 5\times 10^{-4}$ that extends into Zone III, and even Zone I for the largest $\tau_s$ (see Fig. \ref{Growth_Timescales_f01}), and growth timescales are as little as 10's of orbits. However, the sticky ice condition may only apply to regions that are close to the water ice evaporation temperature \citep{Musiolik_Wurm_2019}. Adopting the cold water ice fragmentation condition returns the situation to the one seen in Fig. \ref{Growth_Timescales_f01_with_constraints_3AU} with only a tiny incipient zone.  Likewise at 30 AU, the region of SI incipient growth depends strongly on how sticky water ice is. Sticky water ice, as has been adopted in most growth models to date, leads to an even larger incipient region where turbulence can be as high as $\alpha \sim 10^{-3}$, and encompasses a range from $\tau_s \approx 0.008-4$ again extending as far as Zones III and I. On the other hand, the cold water ice fragmentation condition \citep{Musiolik_Wurm_2019} restricts the incipient region to a tiny region in Zone II, centered about $\alpha \sim 7\times 10^{-5}$ and bounded by $\tau_s \approx 0.008-0.1$.}}

{ {{\it Modeled particle growth constraints:}}} { {We now examine how the constraints imposed by the evolutionary growth model results of \citet{Estrada_etal_2016,Estrada_etal_2020} in the epoch of interest affect the regions of SI incipient growth. In Figure \ref{Growth_Timescales_f01_with_constraints_3AU_d0p07}, we return to the inner disk, but now
plot the maximum achieved Stokes numbers from Table \ref{Stokes_numbers_alpha} (colored stars) which correspond to the particle sizes that carry most of the mass (see Sec. \ref{combined_limits_on_particle_size}). Note that Fig. \ref{Growth_Timescales_f01_with_constraints_3AU_d0p07} contains new SI mode growth rates, recalculated for $\delta = 0.07$ which represents a characteristic value for the particle growth and drift models. The main effect of this is that the radial and pattern drift constraint lines have shifted downwards, with the latter further restricting the SI incipient region to a vanishingly small range even before particle constraints are folded in.
However, when one considers the growth models -- even for the minimum $\alpha$ case -- this region becomes inaccessible.}}
As was discussed in Sec. \ref{combined_limits_on_particle_size}, not all the plotted points fall on the representative silicate fragmentation line { {which assumes that turbulence dominates their relative velocities, and that eddy-crossing effects are unimportant.}} 
The points defined by $\alpha \gtrsim 10^{-4}$ 
broadly lie above their respective fragmentation lines { {(as can be determined from the temperatures listed in Table \ref{Stokes_numbers_alpha})}} because there has been significant
growth beyond the fragmentation barrier (section \ref{combined_limits_on_particle_size}}) due to both ``mass transfer" \citep[e.g.,][]{Wurm_etal_2005,Windmark_etal_2012} and ``lucky particle"  growth in those models
\citep[e.g.,][]{Garaud_etal_2013,Drazkowska_etal_2014}, whereas for $\alpha \lesssim 10^{-4}$ { {the mass-dominant particles are entering a low turbulence regime in which they are subject to eddy-crossing effects  
so that their fragmentation Stokes numbers are smaller than what Eq. (\ref{fragmentation_barrier}) would predict. 
Thus, all the particle evolution models discussed here are in fact in the fragmentation regime.}}
Overall then, { {we find that}} in the silicate-dominated region illustrated by Figure \ref{Growth_Timescales_f01_with_constraints_3AU_d0p07}, all plausible, self-consistent combinations of turbulent intensity and particle size lie { {well within Zone II, and do not overlap any of the region in $\alpha$--$\tau_s$ parameter space in which SI incipient growth is permissible.}}


For both the nominal snowline and the outer disk, Figures \ref{Growth_Timescales_f01_with_constraints_5AU_d0p07} and \ref{Growth_Timescales_f01_with_constraints_30AU_d0p1} show that { {the situation is also strongly limited by realistic growth models,}} with respect to regions of incipient growth. { {As was the case in Fig. \ref{Growth_Timescales_f01_with_constraints_3AU_d0p07}, Figure \ref{Growth_Timescales_f01_with_constraints_5AU_d0p07} corresponds to a value of $\delta = 0.07$ which represents a characteristic value across all models listed in Table \ref{Stokes_numbers_alpha} for the snow line. Like before, the higher $\delta$ further restricts the incipient growth region, though not as dramatically as for the inner disk. In this case we see that}} the SI appears to be incipient only within a tiny triangular patch of parameter space, again around $\alpha \sim 5\times 10^{-5}$ and $\tau_s \approx 0.01$ indicated in the figure by the red-hatched triangular boundary.
The constraining processes of { {the minimum level of turbulence, and the mode pattern drift timescales, bound the region from the bottom and left. However the evolutionary growth models between $\alpha \sim 10^{-5}-10^{-4}$ cut off access to the regions of $\alpha$--$\tau_s$ parameter space that would be permissible from either water ice fragmentation constraint. 
Figure \ref{Growth_Timescales_f01_with_constraints_30AU_d0p1} shows the case for the outer nebula at 30AU, in which $\delta = 0.1$. Here the mode pattern drift line moves even further downward, increasing the lower bound of $\tau_s \simeq 0.04$ a factor of 5 more than for the $\delta = 0.05$ case (Fig. \ref{Growth_Timescales_f01_with_constraints_30AU}) and shrinking the SI incipient zone. Meanwhile, particle Stokes numbers remain about $\tau_s \sim 0.01$, further to the left and eliminating the small incipient zone.}}  

In Figure \ref{Growth_Timescales_f01_with_constraints_5AU_d0p07}, { {almost}} all models with $\alpha \lesssim 4\times 10^{-3}$ are in the drift-dominated regime, { {whereas in Figure \ref{Growth_Timescales_f01_with_constraints_30AU_d0p1} the fragmentation limit has only been reached for $\alpha = 10^{-2}$ (recall the fragmentation limit for the black and yellow symbols are more closely associated with the sticky H$_2$O ice fragmentation line).}} { {In Fig. \ref{Growth_Timescales_f01_with_constraints_5AU_d0p07}, the models for $\alpha = 10^{-4}-10^{-5}$ are again in a regime where eddy-crossing is starting to become important, so their fragmentation $\tau_s$ will be smaller than what the sticky ice line predicts, but still larger than their current values. Here, only the model for $\alpha = 10^{-5}$ is in the Stokes flow regime.}} The remaining models in Figs. \ref{Growth_Timescales_f01_with_constraints_5AU_d0p07} and \ref{Growth_Timescales_f01_with_constraints_30AU_d0p1} (orange stars) which are associated with the cold H$_2$O ice model, have reached the fragmentation size 
(though significant growth beyond it has occurred in Fig. \ref{Growth_Timescales_f01_with_constraints_5AU_d0p07} at the snow line). The curious inflection that leads to peak values for $\tau_s$ for 
$\alpha \sim 10^{-3}$ are real and due in part to enhanced growth about the snow line, whereas the trend towards larger $\tau_s$ seen for $\alpha \sim 10^{-2}$ at 30 AU is due to both reaching the fragmentation barrier and the lower gas surface density as a result of the more rapid viscous evolution compared to the other models. Overall though, { {as was the case}} for the silicate-dominated { {inner disk}} region, all self-consistent combinations of turbulent intensity and particle size lie within Zone II, { {from the standpoint of SI}}.

The interplay between particle drift, bouncing and fragmentation, and the manner in which particle growth proceeds,
especially around evaporation fronts like the snow line, highlight the complexity of modeling particle growth and gas evolution with time, and call for more
in-depth analyses of this type in the context of the theory presented herein.
 \vspace{0.1 in}

\section{Summary and Conclusions}

Conducting high resolution numerical simulations of the interaction of gas and particles to shed light upon the streaming instability is a computationally expensive undertaking.  It should be
of value to have some kind of theoretical guide -- however approximate -- to constrain the parameter range of its validity from the standpoint of planetesimal formation. 
The purpose of this study is to provide a theoretical framework to address the question of how and to what extent the streaming instability might be effective for planetesimal formation under globally turbulent disk conditions.

{ {Our model extends and generalizes behavior initially studied by YG2005 by representing the effects of local turbulence by an $\alpha$-model, and makes additional predictions that are consistent with previously reported numerical studies of the presence or absence of the streaming instability in turbulent disk simulations which include the following: 
\citet{Johansen_etal_2007, Balsara_etal_2009, Yang_etal_2017,Li_etal_2018,Yang_etal_2018,Gerbig_etal_2020}.\footnote{{ {For the one set of simulations
we examined where the correspondence was weakest \citep[i.e.,][]{Li_etal_2018} we conjecture that those simulations
were not run with sufficient resolution to see the short wavelengths predicted by our model. }}}
The $\alpha$ model representation of turbulence -- that characterizes its effect in the form of an enhanced isotropic turbulent viscosity and diffusion -- acts locally both to stir particles and to exchange momentum.  Underpinning its use here is the assumption that the 
processes leading to turbulence, especially in protoplanetary disk Ohmic zones, do so independent of the presence of particles with the realistic sizes and abundances treated here. }}
\par
We have examined the normal mode response of the streaming instability as a function of the disk turbulence parameter $\alpha$ and particle Stokes numbers $\tau_s$ by identifying the wavelengths, growth times, and pattern speeds of the fastest growing modes.  For given  values of $\alpha$ and $\tau_s$, the  particle to gas mass density ratio ($\epsilon =\rho_p/\rho_g$) is calculated using a turbulent dilution model (eq. \ref{turbulent_dilution_model}),  to represent the balance between the gravitational settling of particles toward the disk midplane and the vertical diffusion of the same particles due to turbulence
\citep{Dubrulle_etal_1995, Youdin_Lithwick_2007, Estrada_etal_2016}.
We hope the theoretical framework proposed and examined in this study is useful in similar future studies.  

\par
{ {
Simple turbulence models like ours involve taking higher order moments of the equations of motion which are then truncated with
some a closure relationship. In general, such resulting model equations do not conserve momentum
\citep[see extended discussion in][]{Davidson_2004}.  
\citet{Tominaga_etal_2019} highlighted the lack of angular momentum conservation in a 2D non-axisymmetric disk setting with a turbulence model similar to ours, and proposed a solution to the problem.
Together with the authors of \citet{Chen_Lin_2020}, we have conducted a proper re-analysis of the matter in the framework of these equations (not shown here) and find that this non-conservation has negligible effect on the growth rates determined for all values of $\alpha$
pertinent for realistic protoplanetary disks.  A future follow-up study explicitly detailing this result is warranted.}}
\par
The study conducted here has revealed several interesting trends for a nominal (minimum mass) solar nebular model with global solids-to-gas mass ratio of $Z=0.01$ and disk opening angle $\delta = H/r = 0.05$ (results are also given for values of $Z$ as large as 0.08).
{ {While the specific values summarized below primarily pertain to disk models with $Z=0.01$, the conclusions might be extended to higher $Z$ by comparison with figure  \ref{Growth_Timescales_Moar} (although further analysis should be done
to verify this assertion as well): }}
\begin{enumerate}

\item As turbulent intensity increases, the wavelengths of the maximally growing SI modes increase.
while the growth rates of the maximally growing SI modes diminish.

\item 
{ {The combination of ($\tau_s,\alpha$) that leads to initial $\epsilon = 1$ according to the turbulent dilution model traces a critical curve with important implications for SI behavior.
This curve terminates at a critical point $(\tau_c,\alpha_c)$ corresponding to
values of $\epsilon$ that monotonically increase with $Z$ for given $\delta$. For 
cosmic abundance $Z=0.01$, this critical point occurs at $\tau_c = 0.45$ and 
$\alpha_c \approx 3.7 \times 10^{-5}$, corresponding to $\epsilon = \epsilon_c = 1.1$.  Selected critical
point values for other $Z$ are summarized in Table \ref{Critical_Values_alpha}.}}

\item For values of $\tau_s < \tau_c$,  and for parameter combinations of $\alpha$ and $\tau_s$ that
lead to $\epsilon = 1$ according to the turbulent dilution model (eq. \ref{turbulent_dilution_model}) - that is, along the critical $\epsilon=1$ curve - the least stable SI mode neither grows nor decays and the growth timescale is effectively infinite. 

\item Provided $\tau_s < \tau_c$ we identify two regimes which straddle the above-mentioned critical line as being either ``laminar/unstable" (Zone I) or ``turbulent/saturated" (Zone II; Figure \ref{Growth_Timescales_f01}). 

\item The spatial structure of the fastest growing mode in Zone II (the turbulent/saturated regime) typically corresponds to vertically oriented sheets with radial scale of about a pressure scale height $H$.  
{ {In practice, the sheet's vertical extent should follow the particle scale height and this appears
to be consistent with the simulation results reported in \citet{Balsara_etal_2009} (see their Fig. 7). 
}} The mode structure in Zone I (the laminar/unstable regime) exhibits narrow, azimuthally oriented tubes with lengthscales much less than $H$, consistent with previous predictions and simulations 
\citep[e.g., see the recent results of][]{Carrera_etal_2015,Yang_etal_2017,Li_etal_2018}.

\item { {The theory developed here appears to reasonably predict the onset or absence of the SI in several recently published simulations,
We have plotted these correspondences in our $\alpha-$St growth timescale plots in Figures 3 and 4.
For laminar disk models in which particles settling to the midplane generate their own midplane turbulence from either the
from Kelvin-Helmholtz overturn and/or the SI itself, we find that our theory (i) reasonably predicts the radial spacing of emerging filaments 
when SI is active especially during its early onset phases well before filament-filament merging occurs and (ii)
predicts when the SI does not appear.  We have come to this conclusion after careful analysis of the three recent studies
by \citet{Yang_etal_2017,Li_etal_2018,Gerbig_etal_2020}.  \par
In simulations where turbulence is externally driven throughout the disk model by the MRI 
like \citet{Johansen_etal_2007,Balsara_etal_2009,Yang_etal_2018}, the theory does a reasonable job at predicting whether or not SI should or should not be present and, in some cases, predicts the radial wavelength structure and growth timescale of filaments
like in \citet{Johansen_etal_2007}.  In particular, the theory predicts
the absence of the SI in the simulations of \citet{Yang_etal_2018} for their ideal MHD model in which the MRI
is rampant throughout the computational domain.  On the other hand, \citet{Yang_etal_2018} also present results of Dead Zone disk models
 driven by waves coming from sandwiching MRI active layers.  In the absence of backreaction particles accumulate along filaments
 likely because the waveforcing emplaces azimuthally oriented pressure fluctuations with coherent pressure maxima toward which
 particles naturally drift.  When backreaction is turned on, the particle densities within these filaments become enhanced.
 Our theory does not explicitly
 handle SI physics in the presence of coherent structures: it implicitly assumes the scales of the system take place well inside the inertial range of the turbulent forcing where
 coherence is lost.  This matter must be addressed in
 future work.
 }} 
\item We believe that realistic protoplanetary disk conditions relevant to the early solar system { {($\lesssim 0.5$ Ma)}} in which $Z= 0.01$ { {and where typically $\delta \gtrsim 0.05-0.1$}} -- taking into account various barriers to particle growth, age constraints, and the disk's likely degree of turbulence as quantified by $\alpha$ --
restrict the SI to Zone II where it is ``incipient", or allowed, for a narrow range of self-consistent disk and particle properties. For example, { {we find}} at a disk location nominally representative
of the ice line of Jupiter's early core ($r= 5$AU) { {that the incipient range of parameter space as indicated by the red-hatched triangle in Fig. \ref{Growth_Timescales_f01_with_constraints_5AU_d0p07} is restricted about a turbulent intensity $\alpha \sim 4\times 10^{-5}$, and Stokes numbers in a narrow range between $\sim 0.01<\tau_s< 0.02$.}}
We treat this vanishingly small allowable region as only suggestive - given uncertainties in all the constraints, it may be larger. { {On the other hand, no such permissible regions are found in the inner and outer disks cases for the given models. Thus achieving growth sufficient to breach the incipient regions in the inner and outer disk appears even much more challenging.}} 

\item { {In Zone I, SI is robust and plausibly proceeds to planetesimal formation, as routinely observed in numerical simulations. In Zone II, we identify a new kind of behavior in which SI is only ``incipient", the growth timescales are very long ($10^2-10^4$ local orbit times) and the growth of particle density is limited.}}

\item { {
It should be kept in mind that while the analysis in Section 7.5 suggest that it may be difficult for the SI to trigger sufficient particle enhancements in turbulent disks with spatially uniform values of $Z=0.01$, that this prediction is also likely to be applicable only 
for the most earliest phases of a model solar disk in which there is negligible particle enrichment.  
As a disk evolves, the disk's gas content will evolve both globally due to wholesale loss via winds and locally variations induced
by transport.  Over time, then, regions will emerge with substantially enhanced values of $Z$.  The regions which are ``SI incipient" can, in principle, slip from being in Zone II to Zone I, wherein the growth is expected to be more rapid even in the face of turbulence.
A counterargument here might be that decreasing the gas density through winds might not necessarily lead to higher values of St
nor enhanced local values of $Z$ because the efficiency of radial drift can efficiently deplete a local region of its particles
while particles might encounter their fragmentation barriers at smaller sizes \citep{Birnstiel_etal_2012,Estrada_etal_2016,Carrera_etal_2017}.
  We therefore expect the 
situation to be very different for conditions corresponding to the latter stages of the disk's evolution beyond its
thick gas phase, and we caution against applying our prediction of limited to no growth for
the whole of the disk's lifetime.  Further follow up analysis are therefore necessary and warranted.
}}

\end{enumerate}
\par
\medskip
%

{\it Other Future Work:} There is much to understand about the implications of this new physics.  Included in any list of future work 
should be (a) a better physical understanding of the critical line and critical point corresponding to the special condition $\epsilon = 1$ and, especially, why this special combination of parameters
leads to exactly marginal modes, (b) the role of a particle size distribution; (c) the possible role of particle loading on damping turbulence; (d) a self-consistent analytical model of particle layers and self-generated turbulence in the limit of vanishingly small global turbulence, (e) the behavior when the predicted vertical wavelength exceeds the thickness of the particle layer, and of course (f) the regime where purely hydrodynamical turbulence is operative, and its intensity.
{ {It is  worthwhile to re-do this analysis within the framework of (i) a turbulent MHD model, and/or (ii)
a model with multiple particle sizes as in \citet{Krapp_etal_2019}, (iii) and to expand the theory to account for the presence of
coherent flow structures, e.g., like those present in \citet{Yang_etal_2018}.}}
\par
{ {
Finally, understanding the physical mechanism of the turbulent SI remains a priority.  While the mathematical 
description of the onset of the inviscid SI in terms of resonant drag energy exchange between gas waves and particles 
is satisfying \citep[i.e., the RDI,][]{Squire_Hopkins_2018a}, it might need modification to explain the onset of instability in this kind of turbulent model setting.  The mechanistic explanation of the inviscid SI presented by \citet{Jacquet_etal_2011} 
offers a framework upon which an explanation for the viscous case might be built.
}}

\section*{Acknowledgements}
{ {
{ { We thank our two reviewers who provided very thorough readings and suggestions that improved both the rigor of our analysis and the quality of this manuscript}}.
We are indebted to Allan Treiman for organizing ``Accretion:  Building New Worlds Conference" held at the Lunar and Planetary Institute (August 15-18, 2017), where the motivation for this study first germinated via in-depth conversations with A. Carballido, J. Simon and D. Carrera.  We thank Phil Armitage for organizing the ``Theoretical and Computational Challenges in Planet Formation Workshop" at the Flatiron Institute (May 20-22, 2019) where mature results of this study were presented.
We are grateful for extended conversations with K. Shariff and D. Sengupta and others at NASA Ames Research Center's  Origins Group as { {well as input and perspectives shared with us especially by M-K. Lin and K. Chen who developed very
similar results independently of -- but in parallel with -- us}}, as well as X. Bai, H. Klahr, W. Lyra, J. Drazkowska.
O.M. Umurhan acknowledges support from NASA Astrophysics Theory Program grant NNX17AK59G and The New Horizons Kuiper Belt Extended Mission. P.R. Estrada and J. N. Cuzzi 
acknowledge NASA Emerging Worlds grant NNX17AL60A for providing resources and support for this project.
Finally, we also acknowledge 
NASA's ``Internal Scientist Funding Program" grant to Ames Research Center on Origins of Planetary Systems as well as NASA's Senior NPP program.}}

\appendix

\section{A. Linearized matrix equation}\label{appendix_1}
{ {
The linearized equations of motion are non-dimensionalized on $\Omega$ and $H$.  Furthermore, in order to insure incompressibility, the radial and vertical components of the perturbation gas velocity (respectively, $u_g', w_g'$) is expressed in terms of the azimuthal perturbation streamfunction $\psi'$, which is related to the gas vorticity via,
 $\omega_g' = -K^2 \psi'$) to insure incompressibility. The result may be cast into the following matrix form }}
\beq
-i\omega \left(\begin{array}{c}
\psi' \\
v_g' \\
u_p' \\
v_p' \\
w_p' \\
\Delta_p'
\end{array}\right) +
\mathtt{M}\left(
\begin{array}{c}
\psi' \\
v_g' \\
u_p' \\
v_p' \\
w_p' \\
\Delta_p'
\end{array}
\right) = 0,
\eeq
where $\mathtt{M}$ is given by
\beq
\mathtt{M} \equiv
\left(
\begin{array}{cccccc}
a & \frac{2 i k_z}{K^2} & \frac{i k_z \epsilon }{K^2 \tau _s} & 0 & -\frac{i k_x \epsilon }{K^2 \tau _s} & -\frac{2 i k_z \epsilon  (\epsilon +1)}{K^2 \left((\epsilon +1)^2+\tau _s^2\right)} \\
 \frac{i k_z}{2} & a & 0 & -\frac{\epsilon }{\tau _s} & 0 & -\frac{\epsilon  \tau _s}{(\epsilon +1)^2+\tau _s^2} \\
 -\frac{i k_z}{\tau _s} & 0 & \frac{1}{\tau _s}-\frac{2 i k_x \tau _s}{(\epsilon +1)^2+\tau _s^2} & -2 & 0 & i k_z c_d^2 \\
 0 & -\frac{1}{\tau _s} & \frac{1}{2} & \frac{1}{\tau _s}-\frac{2 i k_x \tau _s}{(\epsilon +1)^2+\tau _s^2} & 0 & 0 \\
 \frac{i k_x}{\tau _s} & 0 & 0 & 0 & \frac{1}{\tau _s}-\frac{2 i k_x \tau _s}{(\epsilon +1)^2+\tau _s^2} & i k_z c_d^2  \\
 0 & 0 & i k_x & 0 & i k_z & \frac{K^2 \alpha /\delta^2}{\tau _s^2+1}-\frac{2 i k_x \tau _s}{(\epsilon +1)^2+\tau _s^2} \\
\end{array}
\right),
\eeq
where $K^2 \equiv k_x^2 + k_z^2$ and for notational convenience we define
\[
a \equiv \alpha  K^2/\delta^2+\frac{\epsilon }{\tau _s}+\frac{2 i k_x \epsilon  \tau _s}{(\epsilon +1)^2+\tau _s^2}.
\]  
We could approach solutions to this by solving for $\omega$ from the sixth order dispersion relation arising from
\beq
{\det\left(-i \omega \mathtt{I} + \mathtt{M} \right) = 0,}
\eeq
where $\mathtt{I}$ is the $6\times 6$ identity matrix.
However, the expression resulting from the above operation is extremely unwieldy and offers no insight toward the mechanisms operating in the instability.
Instead,
we choose to solve this directly by determining the eigenvalues of $\mathtt{M}$ using standard numerical
techniques found in Matlab.   We arrange the solutions in descending order of $\omega_i$.  The solutions plotted
throughout the text are the least stable mode.  In some instances there are two unstable modes but the second mode is usually dwarfed in magnitude by the first mode.  A detailed examination of the second unstable mode and its interpretation remains to be determined, but may be of interest.  
{ {We are reminded that in the inviscid theory the characteristic lengthscales of the SI are
$H\delta$ while lengthscales in the $\alpha$-disk model are measured on $H$.}}  Hence, a disparity of lengthscales
between the large-scale global turbulence and the SI appears with the ratio $\alpha^{1/2}/\delta$.

\section{B. On the Dead Zone models of Yang et al. (2018)}\label{appendix_2}
{ {
By contrast, the DZ model results (especially $Z=0.04,0.08$) exhibit particle accumulation along azimuthally elongated filaments which grow tighter, become increasingly coherent and achieve high densities, plausibly due to SI. However, we think that the physics in this case
is more complicated than our turbulence model can address.
\citet{Yang_etal_2018} depict the particle behavior in DZ models with and without backreaction \citep[see bottom two rows of Figure 10 of][]{Yang_etal_2018}.  Even in the models without backreaction (where SI is inactive), particles can be seen to accumulate into azimuthally elongated filamentary structures almost from the very beginning of the model runs.  In these models the value of $\epsilon$ in the densest regions appears to exceed unity in many places. In comparison, all DZ models with backreaction show the same early accumulation as in the non-backreacting case.  However, {\it with} backreaction these density enhancements  continue to evolve because the SI drives further density increases. }} 
\par
{ {
\citet{Yang_etal_2018} characterize the vertical distribution of particles in terms of an effective particle diffusion parametrized
by an analagous $\alpha$-parameter called $\alpha_{g,z}$.  This diffusion parameter achieves high values ($\alpha_{g,z}\sim 4\times 10^{-3}$)
similar to the values of $\alpha$ in the iMHD case, even though the value of $\alpha$ in the midplane regions of the DZ model is quite low ($\sim 2\times 10^{-4}$).  
Nonetheless, particles appear to concentrate into filaments which become reinforced and further enhanced
as the SI takes root, this is especially apparent in the simulations involving $Z=0.02, 0.04,0.08$.  Such filaments are likely pressure extrema
with concomitant zonal flows are known to be characteristic of MRI driven turbulence at the large scales 
of a simulation \citep{Johansen_etal_2009,Simon_etal_2012}.  

These scales
are, at best, at the top of the inertial range of the turbulent flow and are not contained inside the turbulence's inertial range. 
To what extent the midplane layers in their DZ models may or may not be characterized as a uniform turbulent flow \citep[see section 4.1 of][]{Yang_etal_2018} and what effect external wave forcing has on inducing
a priori high density azimuthally aligned filaments remains to be fully examined. 
By examining the development of filaments in models with backreaction against those without backreaction , it can be seen in the latter that filamentary particle enhancements emerge as a matter of course in response to the large scale wave-forcing coming from the magnetically active layers \citep[see also Figure 8 of][]{Yang_etal_2018}.\footnote{Similar structuring can be seen in the simulations of \citet{Johansen_etal_2007}.}
With backreaction turned on, these particle enhancements readily continue
their condensation into narrow coherent structures.  It is difficult to disentangle the causal sequence of events: are the observed structures growing due solely attributed and intrinsic to the SI or does the pre-existence of overdense filaments aid in triggering the SI in these cases?  
It would seem that the large scale azimuthally elongated structural forcing imprinted by the waves emanating from the overlying turbulent layers predisposes the midplane particles into undergoing the SI because of the a-priori axisymmetric clumping caused by
the radial pressure maxima induced by the waves.
The coherent large scale structure of the non-uniform wave-forcing is a physical complexity not reflected in the simple turbulence model of our theory,
nor of that of \citet{Chen_Lin_2020}.
We therefore view the physical structures observed in the midplane regions of their DZ model as outside the scope of the theoretical construct
discussed here and could be addressed in future work. Finally, the relevance of models with such deep MRI-active layers is itself open to debate \citep[e.g.,][]{Bai_Stone_2010}. 
}}

\end{document}